\documentclass[useAMS,usenatbib,usegraphicx,onecolumn]{mn2e} 

\title[Photometric scaling relations of NIRS0S and OSUBSGS samples]{{\bf Photometric scaling relations of lenticular and spiral galaxies}}
          
\author[E. Laurikainen, H. Salo, R. Buta, J.H. Knapen and S. Comer\'on]{E. Laurikainen$^{1}$\thanks{E-mail:
eija.laurikainen@oulu.fi}, H. Salo$^{1}$, R. Buta$^{2}$, J. H. Knapen$^{3,4}$ and S. Comer\'on$^{3,4}$ \\ 
$^{1}$Division of Astronomy, Department of Physical Sciences, University of Oulu, FIN-90014, Finland \\
$^{2}$Department of Physics and Astronomy, University of Alabama, Box 870324, Tuscaloosa, AL 35487 \\
$^{3}$Instituto de Astrof\'isica de Canarias, E-38200 La Laguna, Tenerife, Spain \\
$^{4}$Departamento de Astrof\'isica de Canarias, E-38205 La Laguna, Tenerife, Spain}

\begin{document}

\date{Accepted: Received:}


\maketitle

\label{firstpage}
 
\begin{abstract}

 Photometric scaling relations are studied for S0 galaxies and
  compared with those obtained for spirals.  New two-dimensional
  multi-component decompositions are presented for 122 early-type disk
  galaxies, using deep $K_s$-band images. Combining them with our
  previous decompositions, the final sample consists of 175 galaxies
  (Near-Infrared Survey of S0s, NIRS0S: 117 S0s + 22 S0/a and 36 Sa
  galaxies). As a comparison sample we use the Ohio State University
  Bright Spiral Galaxy Survey (OSUBSGS) of nearly 200 spirals, for
  which similar multi-component decompositions have previously been
  made by us. The improved statistics, deep images, and the
  homogeneous decomposition method used, allows us to re-evaluate the
  parameters of the bulges and disks. For spirals we largely confirm previous results,
which are compared with those obtained for S0s.
Our main results are as
  follows. (1) Important scaling relations are present, indicating
  that the formative processes of bulges and disks in S0s are coupled
  (e.g. $M_K^o(disk)$= 0.63 $M_K^o(bulge)$ $-$9.3), like has been
  found previously for spirals (for OSUBSGS spirals $M_K^o(disk)$=
  0.38 $M_K^o(bulge)$ $-$15.5; the rms deviation from these relations is
  0.5 mag, for S0s and spirals).  
(2) We
  obtain median $r_{eff}/h_r$ $\sim$ 0.20, 0.15 and 0.10 for S0,
  S0/a-Sa and Sab-Sc galaxies, respectively: these values are smaller
  than predicted by simulation models in which bulges are formed by
  galaxy mergers. (3) The properties of bulges of S0s are
  different from the elliptical galaxies, which is
  manifested in the $M_K^o(bulge)$ vs $r_{eff}$ relation, in the
  photometric plane ($\mu_0$, $n$, $r_{eff}$), and to some extent also
  in the Kormendy relation ($<\mu>_{eff}$ vs $r_{eff}$). The bulges of S0s 
are similar to bulges of spirals with $M_K^o$(bulge) $<$ $-$20 mag.
Some S0s
  have small bulges, but their properties are not compatible with the
  idea that they could evolve to dwarfs by galaxy harassment.  (4) The
  relative bulge flux ($B/T$) for S0s covers the full range 
  found in the Hubble sequence, even with 13$\%$ having $B/T$
  $<$ 0.15, typical for late-type spirals. (5) The values and
  relations of the parameters of the disks ($h_r^o$, $M_K^o(disk)$,
  $\mu_o(0)$) of the S0 galaxies in NIRS0S are similar to those
  obtained for spirals in the OSUBSGS.  
  Overall, our results support the view that
    spiral galaxies with bulges brighter than $-$20 mag in the
    $K$-band can evolve directly into S0s, due to stripping of gas
    followed by truncated star formation.

\end{abstract}


\section{INTRODUCTION}

The position of S0 galaxies between ellipticals and spirals in galaxy
classification schemes \citep{hubble1926,devauc1959,sandage1961} has
made them of particular interest in any scenario of galaxy formation
and evolution. Yet the debate on their origin is open.  In the current
paradigm, the hierarchical Lambda Cold Dark Matter ($\Lambda$CDM)
cosmology \citep{somerville1999,stein2002}, the disks are 
formed first by cooling of gas inside rotating dark matter halos,
whereas both the elliptical galaxies and the bulges of disk galaxies
are suggested to have formed in major or minor mergers, respectively
\citep{burk2006}. 
  The bulges formed in this manner are dynamically hot and their
  basic properties were established already in the merger process
  \citep{burk2005}, not affected by the subsequently growing
  disks. Within $\Lambda$CDM, S0s are formed in galaxy mergers in a
  similar manner as the elliptical galaxies, or they are transformed from 
  spirals which have lost their disk gas by some stripping mechanism
  \citep{gun1972,moore1996,bekki2002}.  Therefore, it is important to
  study whether the S0 galaxies are more tightly related to
  ellipticals or to spirals.  However, even if the bulges in S0s were
  found to be similar to those in spirals, this does not yet answer
  the question of what the formative processes of bulges in S0s are.
  
It has been suggested that two types of bulges appear: 1) classical
merger-built bulges, and 2) disk-like bulges formed by star formation
in the disk (Kormendy 1982; see also review by Kormendy $\&$ Kennicutt
2004). Boxy/peanut bulges are often listed as a separate class, but as
they are assumed to be part of a bar (Athanassoula, Lambert $\&$ Dehnen 2005), they are
basically disk-related structures.  Bulges in late-type spirals
are typically photometrically disk-like (Andredakis $\&$ Sanders 1994;
Carollo et al., 1997, 1998) rotationally supported structures (Cappellari et
al. 2007), whereas for the bulges of early-type spirals contradictory
results have been obtained.  In particular, the fairly large masses of
the bulges in the early-type galaxies are difficult to explain by
secular evolution alone, related to bar-induced gas infall and
subsequent star formation in the central regions of the galaxies (see
Kormendy $\&$ Kennicutt 2004), at least if no external intergalactic
material is added to the bulge.  It needs to be re-investigated what
is the nature of bulges of S0s in the nearby Universe, e.g. are they
disk-like or more likely have properties of merger built structures.
Answering this question would set important constraints on models of
galaxy formation and evolution. 

Scaling relations have been used for theoretical modeling of
elliptical galaxies (see de Zeeuw $\&$ Franx 1991), and for evaluating
the formation of galactic disks in spiral galaxies
\citep{dalcanton1997,firmani2000}. Therefore, the scaling relations
can be used as a tool to study the origin of S0s, which
morphologically appear between the two main types of galaxies.
One such scaling relation  
was introduced by \citet{kormendy1977}, who showed, using R$^{1/4}$
models, that the effective radius ($r_{eff}$) is
connected to the central surface brightness ($\mu_0$), both for elliptical
galaxies and for bulges of early-type disk galaxies.  The dispersion in
this so-called Kormendy relation is reduced by adding a third
parameter, a S\'ersic index $n$, leading to the photometric plane
(e.g. Khosroshahi, Wadadekar $\&$ Kembhavi 2000). Alternatively, if
the central velocity dispersion is used, the fundamental plane is
obtained (e.g. Djorgovski $\&$ Davis 1987; Dressler et
al. 1987). Other important scaling relations appear between the
brightnesses and the scale parameters of the bulge and the disk
\citep{courteau1996}, and between the brightness of the bulge
with the total galaxy brightness \citep{yoshizawa1975,carollo2007}.

The scaling relations studied for the S0 galaxies so far are open to
different interpretations: while the fundamental plane and the
Kormendy relation have associated their bulges with the elliptical
galaxies \citep{pahre1998,pierini2002,aguerri2005}, the scale
parameters of the bulge and the disk hint to a spiral origin
\citep{aguerri2005,lauri2009}. In a broader context, scaling
  relations for spiral galaxy samples have been extensively studied
  revealing several fundamental relations (see Ravikumar et al. 2006;
  Graham $\&$ Worley 2008 as some of the latest works). However, the
  spiral samples contain just a small number of S0s, which makes it
  difficult to draw conclusions about S0 properties.  In addition,
except for the first attempts by Aguerri et al. (2005a) and Laurikainen
et al. (2009), the scaling relations for the disk galaxies have not
yet been studied using a 2D multi-component approach, which is
important in accounting for structure, particularly in barred disk
galaxies (Peng 2002; Laurikainen et al. 2004, 2006; Gadotti 2008).  

In this study the photometric scaling relations are studied for a
sample of 175 early-type disk galaxies, mainly S0s in NIRS0S. This is
the largest sample of S0s studied in detail up to now.  We use deep
$K_s$-band images for decomposing the two-dimensional surface
brightness image to structure components, including bars, ovals and
lenses. We present new decompositions for 122 galaxies, and use our
previously published decompositions for the rest of the NIRS0S
sample. As a comparison sample we use the Ohio State University
  Bright Spiral Galaxy survey (OSUBSGS; Eskridge et al. 2002) of
  nearly 200 spirals, for which similar multi-component decompositions
  have been previously made by us (Laurikainen et al. 2004). Our main
  emphasis is to compare whether the photometric properties of bulges
  in S0s are more similar to those of elliptical galaxies or
  bulges in spirals.  Also, implications of these scaling relations
  for the formative processes of bulges in S0s are discussed.  Our
  uniform decomposition approach, applied for a statistically
  significant sample of galaxies, using deep $K_s$-band images, allows
  us to re-evaluate the properties of bulges and disks in S0s, and to
  make an unbiased comparison with spirals. 

\section{SAMPLE AND OBSERVATIONS}

Our primary sample consists of 175 early-type disk galaxies, mainly
S0s (117 S0s, 22 S0/a-Sa, 36 Sa galaxies).  It is part of the
Near-IR S0 galaxy Survey (NIRS0S) of 184 galaxies, with the following
selection criteria from the Third Reference Catalog of Bright Spiral
Galaxies (de Vaucouleurs et al. 1991; hereafter RC3): $B_{\rm T}$
$\le$ 12.5 mag, inclination $\le$ 65$^\circ$, and Hubble type $-$3
$\le$ $T$ $\le$1 \footnote{Our current NIRS0S sample differs from that
  specified by \citet{lauri2005} and \citet{buta2006} in that we use
  $B_T$ or the photographic value $m_B$, or the average of these two
  when both are available. This was done to eliminate contamination of
  the original sample by total $V$ magnitudes in RC3, which occupy the
  same column as $B_T$ in that catalog.}. Some of the observed
galaxies lie outside the original sample limits for various reasons,
usually due to specific weather conditions, e.g. high wind speed, that
prevented us from accessing the original sample galaxies. In such
circumstances we often observed a somewhat fainter galaxy ( $B_T$ $<$
12.8) than in our original sample.  In order not to miss the S0s that
might be erroneously classified as E galaxies in the RC3, our sample
includes also E galaxies classified as S0s in the Revised Shapley Ames
Catalog of Bright Galaxies (Sandage $\&$ Tammann 1981, hereafter
RSA). 
We emphasize that in the current study it is not in our interest to re-classify
galaxies in the near-IR, instead optical classifications are used.
The observations and the data reductions are described by
Salo et al. (2009, in preparation; hereafter SLBK).  The non-rebinned 2D
images typically reach the surface brightness level of 21-22 mag
arcsec$^{-2}$ in the $K_s$-band. The flux calibrations use
multiaperture flux measurements in the $K$-band, obtained from the Two
Micron All-Sky Survey (2MASS)
\footnote{The Two Micron All Sky Survey, which is a joint project of
the University of Massachusetts and the Infrared Processing and
Analysis Center/California Institute of Technology, is funded by the
National Aeronautics and Space Administration and the National Science
Foundation}
(Skrutskie et al. 2006). The orientation parameters are from SLBK, where deep
$V$-band images were used when available, otherwise $K_s$-band images
were used. 

 As a comparison sample we use the Ohio State University Bright
  Spiral Galaxy Sample (OSUBSGS; Eskridge et al. 2002), which consists
  of nearly 200 spiral galaxies, covering the Hubble types
  S0/a-Sm. The selection criteria in OSUBSGS are similar as in NIRS0S
  except that the apparent magnitude limit is $B_T$=12.0 mag. For 180
  of these galaxies we have previously made multi-component
  decompositions in the $H$-band using the same method as applied in
  this study for NIRS0S galaxies (see Laurikainen et al. 2004).  In
  order to convert the parameters of the bulge and the disk of these
  decompositions to magnitude units, the images were flux-calibrated
  using the total $K$-band magnitudes given in 2MASS. We thus
  implicitly assume that the $H-K$ color is constant throughout the
  galaxy.  As a consistency check we compared and found a good
  agreement of the photometric parameters for those Sa spirals which
  are common in the NIRS0S and OSUBSGS samples.  It is worth noticing
  that NIRS0S covers the Sa type bin better than OSUBSGS which is why
  in this study OSUBSGS is used for the Sab-Sm galaxies only.

\section{2D MULTI-COMPONENT DECOMPOSITIONS}

\subsection{Algorithm}

We use a 2D multi-component code, BDBAR \citep{lauri2004,lauri2005}, for
decomposing the 2D light
distribution of galaxies into bulges, disks, bars, ovals and lenses.
This approach has turned out to be important, not only for barred
galaxies \citep{peng2002,lauri2004},  
but also for galaxies with ovals and lenses (Prieto et
al. 2001; Laurikainen et al. 2005, 2009), since these components can
have a 
significant contribution to the total light. Bars in the
decompositions have been modeled also by \citet{prieto1997},
\citet{aguerri2005}, 
\citet{reese2007}, \citet{gadotti2008} and \citet{weinzirl2008}. 
Quantitatively the effect of bars on the fitted model was studied by
\citet{lauri2006}, who compared 2D bulge/disk decompositions with
bulge/disk/bar decompositions for a sample of 15 S0-S0/a galaxies. They
showed that the mean bulge-to-total ($B/T$) flux-ratio changed
from 0.55 to 0.30 and the mean S\'ersic index $n$ from 2.6 to
2.1 when bars were taken into account.  Including also nuclear bars
reduced the mean $B/T$ flux ratio even further to 0.25.  Obviously, bulge/disk 
decomposition and multi-component approaches represent different views of the
concept of the bulge: in simple bulge-disk decompositions for strongly
barred and lens-dominated galaxies the $B/T$ flux ratio is more like a
concentration parameter than a measure of the relative amount of light in the
spheroidal component. In the multi-component approach,
$B/T$ allows us to measure the relative amount of light of the bulge in a more reliable
manner.  This is expected to have implications for the parameters
of the bulge and disk, which
has to be taken into account when comparing our results with those
obtained previously in the literature. 
As many S0 galaxies have weak outer disks, it is sometimes useful to
apply the 1D decomposition approach \citep{lauri2000} first, which, due to
the azimuthal averaging, gives the scale length of the disk in a more
robust manner. 

In our 2D decomposition code all components use generalized elliptical 
isophotes, the isophotal radius being described as:
\vskip 0.25cm 
$r=(|x|^{c+2} + |y/q|^{c+2})^{1/(c+2)}$.
\vskip 0.25cm 
\noindent The isophote is boxy when the shape parameter $c > $ 0,
disky when $c < $ 0, and purely elliptical when $c$ = 0. Circular
isophotes correspond to $c$ = 0 and minor-to-major axial ratio $q$ =
1. Here the x-axis is along the apparent major axis of the component,
having a position angle $\phi$ at the sky. The origin is at the galaxy
center, taken to be the same for all components.

The disk is described by an exponential function:
\vskip 0.25cm 
$I_d(r_{\rmn{d}}) \ = \ I_{\rmn{o}}  \exp[-(r_{\rmn{d}}/h_{\rmn{r}})]$,
\vskip 0.25cm 
\noindent where $I_{o}$ is the central surface density,
and $h_r$ is the scale length of the disk. The radius $r_d$
is calculated along the disk plane defined by the assumed position 
angle $\phi_d$ and axial ratio $q_d$ of the disk.
$I_{o}$ is in flux units, and when converted to magnitudes it is denoted by $\mu_0$.  

The surface brightness profile of the bulge is described by a 
S\'ersic's function (S\'ersic 1963, 1968):
\vskip 0.25cm 
$I_b(r_{\rmn{b}}) \ = \ I_{\rmn{ob}} \exp[-(r_{\rmn{b}}/h_{\rmn{b}})^{\beta}]$,
\vskip 0.25cm 

\noindent where $I_{ob}$ is the central surface density, $h_b$ is the
scale parameter of the bulge, and $\beta$ = $1/n$ determines the slope
of the projected surface brightness distribution of the bulge ($n$ = 1
for an exponential, and $n$ = 4 for the $R^{1/4}$-law).
$I_{ob}$ expressed in magnitudes is denoted by $\mu_{ob}$.
The parameter
$r_b$ is the isophotal radius defined via the parameters $q_b$, $c_b$
and $\phi_b$, where the subscript stands for bulge. The effective 
radius of the bulge, $r_{eff}$, is the half light radius of the bulge
(the radius of the isophote that encompasses half of the total bulge
light), obtained by numerically integrating over the profile. 
According to \citet{moll2001} for $\beta > 0.2$ a good
approximation is provided by:
 \vskip 0.25cm 
$r_{eff}$ = (2/$\beta$-0.33)$^{1/\beta}$  $h_b$.
\vskip 0.25cm 

The S\'ersic function can be applied for any
non-axisymmetric component, or alternatively the Ferrers function can
be used (Section 3.2 specifies when these functions are used), described as:
\vskip 0.25cm 

$I_{\rmn{bar}}(r_{\rmn{bar}})  =  I_{\rmn{obar}} (1 - (r_{\rmn{bar}}/a_{\rmn{{bar}}})^2)^{n_{\rmn{bar}}+0.5}$, $r_{\rmn{bar}}<a_{\rmn{bar}} $
\vskip 0.20cm 

$I_{\rmn{bar}}(r_{\rmn{bar}})  = 0, \ \ \ r_{\rmn{bar}}>a_{\rmn{bar}}$,


\vskip 0.25cm 
\noindent where $I_{obar}$ is the central surface brightness of the
bar component, $a_{bar}$ is the component major axis length, and $n_{bar}$ is the exponent
of the model defining the shape of the radial profile.  The
isophotal radius, $r_{bar}$, is defined via the parameters $q_{bar}$, $c_{bar}$ and
$\phi_{bar}$. In the equation the designations are for a bar, but they can refer
to any non-axisymmetric component. 


\vskip 0.25cm 

\subsection{Fitting procedure}
\vskip 0.25cm 

The decompositions will be used for two purposes: (1) for
morphological analysis, and (2) for adjusting the images to face-on
orientation. For the morphological analysis we use the full advantages
of the algorithm, whereas for adjusting the images we need to
assume that the bulges are spherical. This is because it is not
possible to convert a non-circular 2D brightness distribution into 3D space
density in a unique manner. The inclination-corrected
images will be used in a forthcoming paper while analyzing the
properties of bars.

\vskip 0.25cm 

Fitting is done in flux units using a weighting function where the
pixel weights are inversely proportional to Poisson noise. However, as shown by
\citet{lauri2005} the weighting function 
is not critical for the obtained model.  In order to account
for the effects of seeing, the fitted model is convolved with a
Gaussian PSF, based on the stellar FWHM measured for each image. To obtain
reasonable solutions the main structural components need to be
identified first.  The identifications were made based on isophotal
analysis (see SLBK), and on inspection of the original images: the
distinction between bars and ovals is based on the ellipticity of the
structure (ovals have lower ellipticities), whereas the division between
ovals and lenses is not always clear (see Section 5.2 for a definition
of bars, ovals and lenses).  Bars are generally fitted by a Ferrers
function using $n_{bar}$ = 2, whereas for ovals and lenses $n_{bar}$ =
1 or 0 are used, giving flatter light distributions and sharper outer
edges. The small $n_{bar}$-values for the lenses are justified because the
lenses presumably have a small vertical thickness and the flux drops
fairly quickly at the outer edge.  Ovals are assumed to have a larger
vertical thickness, but the division between ovals and lenses is not
always clear. The S\'ersic function generally worked well for the
inner disks, but in a few cases it has also been used for a bar.
\vskip 0.25cm  

The fitting strategy is the same as used by Laurikainen et al. (2004,
2005, 2006). Preliminary solutions were obtained using the images
re-binned by a factor of 4, but the final solutions were found using
the full image resolution.  In simple cases where the galaxies have
only a bulge, an exponential disk and a bar, the solutions were generally found
automatically. However, in more complicated cases (a large majority of
the galaxies) an iterative approach was used. The scale length of the
disk, $h_r$, was found first; if the disk was faint the 1D code was
used to find $h_r$. The other components were found one by one, fixing
some of the parameters and leaving the critical parameters free for
fitting. The parameter $n_{bar}$ in the Ferrers function was not left
free for fitting, otherwise the models for bars, ovals and flattened
bulges become easily degenerate. Galaxies that have a secondary bar
typically also have a bright oval surrounding it. As the oval can be
as bright as the bar, in some cases both components were fitted by a
single Ferrers function.  Once a satisfactory
solution was found, based on inspection of the model image and the fit
to the surface brightness profile, all parameters were left free for
fitting.  However, this last step was not always possible. Finally, we
checked whether the whole surface brightness
profile can be fitted by a single S\'ersic function: this was
done to guarantee that we are not erroneously classifying an
elliptical galaxy as an S0.  The seeing FWHM gives a theoretical lower
limit for the smallest components in the model. However, the central
components fitted in this study are nuclear bars or disks which
clearly exceed this size limit. The outer disk is fitted to a maximum
radial distance where the galaxy is still visible, limited by
the standard deviation of the sky background.

\vskip 0.25cm 
\subsection{Individual cases}
\vskip 0.25cm 

The decompositions are shown in Figure 1.  For each galaxy the right
panel shows the decomposition, whereas in the two left panels the
original image (upper panel) and the model image (lower panel) are
shown. The profiles are shown both in linear and in logarithmic radial
scales.  The y-axis in the right panel is the surface brightness given
in mag arcsec$^{-2}$ and the x-axis is the radius along the sky plane
given in arcseconds, and plotted are the pixel values. In such a
presentation a spherically symmetric component appears as a curve,
while non-axisymmetric components fill an area confined by their major
and minor axis exponential profiles. Exponential disk component fills
a wedge-shaped region when a linear radial scale is used.  A few
representative examples are shown in the paper version, whereas the
decompositions for the rest of the galaxies are available in
electronic form.  

ESO 208-G21 is an example of a sample galaxy which
turned out to be an elliptical galaxy, in agreement with its
classification in the RC3, while ESO 137-G10 is well represented by a
simple bulge/disk decomposition.  For ESO 208-G21 the surface
brightness profile declines smoothly, whereas ESO 137-G10 has a
transition zone between the bulge and the disk. ESO 208-G21 also has a
small inner disk.  IC 5328 and NGC 524 are two examples of non-barred
galaxies with prominent lenses: IC 5328 has one prominent lens,
whereas NGC 524 has at least two lenses, both identified as
exponential sub-sections in the surface brightness profile, indicating
that multi-component decomposition is preferred. In both
cases the lenses are significant enough to affect the parameters of
the bulge. Examples of barred galaxies are NGC 6782 and NGC 2950: for
NGC 6782 two Ferrers functions are used, one for the nuclear bar and
the lens surrounding it, and another for the primary bar and a lens,
ending at an inner ring (B+L).  NGC 2950 has two bars and a lens
extending outside the primary bar (B,B,L). Some of the S0s and S0/a
galaxies in our sample have bulges that are exceptionally small
for their Hubble type (NGC 6654 and NGC 4696): such small bulges can
appear both in barred (NGC 6654) and in non-barred (NGC 4696)
galaxies. NGC 4696 is classified as an elliptical galaxy in the RC3, but
the galaxy clearly has an exponential disk and a prominent lens. In
NGC 6654 the secondary bar is important for making a good fit to the
intermediate zone between the bulge and the primary bar.  The main
parameters of the bulge and the disk, as well as the inclination used, are collected in Table 1.  The
decompositions for the individual galaxies are discussed in Appendix
A.

\vskip 0.35cm
\section{Calculation of the structural parameters}
\vskip 0.35cm

The correlations we wish to investigate need to be based on parameters
corrected for Galactic and internal extinction, as well as for inclination.
For this purpose, we use the following corrections derived by Graham $\&$ Worley (2008, hereafter GW2008) 
and \citet{driver2008}:

\vskip 0.25cm
$h_r^o=h_r/C_1$
\vskip 0.2cm
$\mu_o =\mu_0^{obs}-A_g(K)-C_2$
\vskip 0.2cm
$M_K^o(bulge)=K(bulge)-A_g(K)-C_3-5 \log D-25.0$
\vskip 0.2cm
$M_K^o(disk)=K(disk)-A_g(K)-C_4-5 \log D-25.0$
\vskip 0.2cm
\vskip 0.25cm

\vskip 0.25cm
$C_1=1.02 - 0.13 \ \log (\cos \ i)$
\vskip 0.2cm
$C_2=0.06 + 0.79 \ (2.5 \ \log (\cos \ i)$)
\vskip 0.2cm
$C_3=0.11 + 0.79 \ (1-\cos \ i)^{2.77}$
\vskip 0.2cm
$C_4=0.04 + 0.46 \ (1-\cos \ i)^{4.23}$.
\vskip 0.25cm

\noindent 
In these equations $D$ is the galaxy distance in Mpc, taken either
from the Catalog of Nearby Galaxies \citep{tully1988} or from NED,
based on $H_o$=75 km s$^{-1}$ Mpc$^{-1}$.  
$M_K^o$(bulge) and $M_K^o$(disk) are the absolute magnitudes of the
bulge and the disk, and $\mu_o$ is the central surface brightness of
the disk, all corrected for Galactic and internal extinction.  In addition,
$\mu_o$ has also been corrected to face-on orientation.  The
Galactic extinction, $A_g(K)$, is taken from NED and is based on the
maps of \citet{shleg1998}.  The parameters $C_1$-$C_4$ are the
corrections for internal dust and depend on the inclination $i$ of the
disk. $C_1$ and $C_2$ are based on constants appropriate to the
$K$-band taken from Table 1 of GW2008, while the constants $C_3$ and
$C_4$ are from \citet{driver2008}. Following GW2008, we do not apply
any intrinsic dust corrections for the S\'ersic parameters of the
bulge ($n$, $r_{eff}$) because these corrections are not well
known. However, as we are working in the infrared, the effects of dust
are not expected to be severe.  Also, we don't apply any cosmological
dimming correction because the galaxies studied are fairly nearby
systems.  The magnitude of the bulge is based on the flux of the bulge
model, whereas the magnitude of the disk uses the flux obtained by
subtracting the bulge flux from the total model (e.g., it includes
possible bar/oval/lens components). The fluxes are calculated inside
the maximum radial distance used in the decomposition (not
extrapolated to infinity). The dust corrections in the above equations
by GW2008 and Driver et al. (2008) are based on 3D radiative transfer
models, using the recent determination of the opacity in face-on
nearby disk galaxies, based on the Millennium Galaxy Catalogue of
nearly 10000 galaxies.

\vskip 0.35cm
\section{Analysis of the structure parameters}
\vskip 0.35cm

For the NIRS0S sample the parameters of the bulge and the disk are
compared mainly in three groups of galaxies: (1) S0 vs S0/a-Sa
galaxies, (2) barred vs  non-barred S0s, and (3) galaxies with high
vs low luminosity bulges, using the median value $M_K^o(bulge)$ = $-$22.7 mag as
a dividing line.   In $K$-band the S0s in our sample are on average 0.3 mag
brighter than the S0/a-Sa galaxies, with mean absolute
brightnesses, $M_{tot}$, $-$24.0$\pm$0.1 and $-$23.7$\pm$0.1 mag, respectively.
The uncertainties denote the standard errors of the mean.
However, taking into account the morphological bin-to-bin variations among S0s in NIRS0S,
this difference is insignificant.  
In the following, the OSUBSGS sample is generally divided in
two morphological type bins: Sab-Sbc, and Sc and later.  This is because the
galaxy brightness in maintained nearly constant for S0-Sbc, after
which it drops by almost one magnitude. Compared to optical
surface photometry, the use of near-IR images yields parameters that are
less biased by differences in age and metallicity of the stellar
populations \citep{dejong1996}, and are also less affected by dust
\citep{cardelli1989}.

\vskip 0.25cm 
\subsection{Galaxies included in the statistical analysis}
\vskip 0.25cm 

Based on the presence of the disk, 15 NIRS0S galaxies were included as S0s,
even though they are listed as elliptical galaxies in RC3. These galaxies
are: IC 4889, IC 5328, NGC 584, NGC 1344, NGC 2768, NGC 4696, NGC
4976, NGC 5419, NGC 5846, NGC 5898, NGC 6482, NGC 6958, NGC 7029, NGC
7192 and NGC 7796.  In most cases these galaxies are classified as S0s
also in the Revised Shapley Ames Catalog of Bright Galaxies
\citep{sandage1981} and in the Carnegie Atlas of Galaxies (Sandage
$\&$ Bedke 1994, hereafter CAG).  The fact that many, particularly
late-type ellipticals are mis-classified S0s is consistent with
the recent result by van den Bergh (2009).  In our original sample
truly elliptical galaxies (best fit by a single R$^{1/n}$
profile) appeared to be NGC 439, NGC 3706, NGC 5982 and ESO 208-G21. Of these
NGC 5982 is classified as elliptical galaxy in the RC3, whereas the other 3 galaxies
are classified as S0s.

From the subsequent statistical analysis the following galaxies are eliminated:
(1) the elliptical galaxies indicated above, and (2) dwarf galaxies (having absolute
magnitudes less than $M_B$ = $-$18 mag, see Sandage $\&$ Binggeli
1984), including NGC 4531 and NGC 5206. We also excluded NGC 3900, for
which the decomposition was considered uncertain.  For NGC 4369 and
NGC 4424, the parameters of the bulge were not reliable, mainly because
the number of pixels used in the bulge model was too small and
these two were eliminated from the bulge statistics.  There are also
decompositions in which the bulge model was considered fairly reliable in spite
of the fact that the disk could not be fitted in a reliable manner.
These galaxies are: NGC 1415, NGC 3166, NGC 3998, NGC 4369, NGC 4457,
NGC 4772, NGC 5631, NGC 7029 and NGC 4612. They were eliminated from
the statistics when discussing the disk parameters, but included in
the analysis of the bulges. For the analysis of the bulges we ended up
with 151 galaxies, and for the analysis of the disks with 142 galaxies.

\vskip 0.35cm
\subsection{Identification of the structural components: bars, ovals and lenses}
\vskip 0.35cm

Identification of the structural components was made before starting
the decompositions, which was to ensure that only physical structures
were fitted. The only exceptions are the faint structures that appear
only in the residual images after subtracting the bulge or disk model
functions.  If not otherwise mentioned, we include as barred galaxies
all those systems in which the bar is visually identified in the
image, or is visible in the residual image after subtracting the bulge
model from the original image. Bars identified by eye generally appear
as bars also in the isophotal and Fourier analysis: they appear as a
bump in the ellipticity profile, the position angle being maintained
nearly constant in the bar region (see SLBK). In the bar region there
is also a peak in the radial $A_2$ amplitude profile, where $A_2$ is
the $m$ = 2 Fourier density amplitude normalized to $m$ = 0 (see for
example Buta et al. 2006). There are 10 galaxies in the sample in
which the bar is visible only after subtracting the bulge model from
the original image: NGC 484, NGC 507, NGC 1161, NGC 1351, NGC 2768,
NGC 2902, NGC 3998, NGC 4373, NGC 7377 and IC 5328. For NGC 3998 the
profile was shown in Laurikainen et al. (2009).  For 6 of these
galaxies there is weak evidence for a bar also in the $A_2$ and in the
ellipticity profiles, whereas in the other galaxies the bar is completely
overshadowed by the bulge flux. Whenever a galaxy has a clear bar based
on the $A_2$ and the ellipticity profiles, the galaxy also appeared barred
when inspecting the image visually.

Ovals are global deviations from the axisymmetric shape in galactic
disks (see Kormendy $\&$ Kennicutt 2004). In the isophotal and Fourier
analysis they generally appear qualitatively in a similar manner as
bars.  However, in contrast to bars, they have lower ellipticities,
b/a $>$ 0.85 (Kormendy $\&$ Kennicutt 2004), and therefore lack
significant Fourier terms higher than $m$ = 2 \citep{lauri2007}.

Lenses are defined as components with a shallow or constant surface
brightness profile, and a sharp outer edge (see Kormendy $\&$
Kennicutt 2004).  As lenses can also appear as axisymmetric structures
they cannot be identified in a similar manner as bars and ovals. In addition,
if the bulges are very prominent the lenses may not be directly
visible in the images, but can be identified in the residual images
after subtracting the bulge model. The early papers by \citet{burstein1979}, 
\citet{tsikoudi1980} and Kormendy (1979, 1982) discussed that 
the lenses in some face-on S0 galaxies can appear as nearly exponential
subsections in the surface brightness profiles.  Later, Sandage $\&$
Bedke (1994) used this characteristic, ``three-zone structure''
as part of the classification of S0 galaxies. Recently the exponential
sub-sections in the surface brightness profiles have been used to
identify lenses in S0s in a systematic manner (Laurikainen et al. 2005, 2006, 2009).
However, it is worth noticing that the exponential sub-sections in the
surface brightness profiles are not always interpreted as
lenses: \citet{erwin2005} suggested that they can be 
manifestations of anti-truncated disks inside more spheroidal-like
components such as bulges or halos.

In this study, we use the exponential sub-sections in the surface
brightness profiles as suggestions for lenses. This is justified, because in
most cases the exponential sub-sections can be directly associated with lenses
in the images. In barred galaxies, the lenses often
have a specific orientation with respect to the bar.
In many S0s multiple lenses appear,
extending to the largest visible radial distance in the image, which
rules out the possibility that they were truncated disks inside more
extended spheroidals. In any case, for the present goal of deriving
the scaling relations of S0s, the detailed interpretation
of lenses is not crucial, as long as they are accounted in the
decompositions so that their presence does not affect the
derived parameters of the bulge.

\vskip 0.35cm
\section{Parameters of the bulge and the disk}
\vskip 0.35cm

\vskip 0.35cm
\subsection{Kormendy relation}
\vskip 0.35cm

It has been suggested by \citet{kormendy1977} that both elliptical galaxies
and bulges of S0s show a strong correlation between the central surface
brightness ($\mu_{ob}$) and the effective radius ($r_{eff}$),
subsequently known as the Kormendy relation.  It was shown later that
the surface brightness at the effective radius of the bulge
($\mu_{eff}$) is actually a better estimate of the surface brightness
because it is not as much affected by seeing and is also less model
dependent \citep{djor1987}.  Using the mean surface brightness inside
$r_{eff}$ (denoted by $<\mu>_{eff}$) further minimizes the effects of
the measurement uncertainties. In the early studies the $R^{1/4}$
function was used, but later studies have shown that the S\'ersic
function, $R^{1/n}$, gives a better fit for the elliptical galaxies
\citep{caon1993}, and for the bulges of S0s
\citep{andredakis1995,balcells2003}.  Differences are expected in the
photometric parameters obtained using the $R^{1/4}$ and the $R^{1/n}$
functions, particularly if the images are not very deep
\citep{trujillo2002}, but these variations are not expected to affect
the Kormendy relation \citep{ravikumar2006}.

The Kormendy relations for the NIRS0S and OSUBSGS galaxies are shown
in Figure 2.  In this and in all the following figures the S0 class
consists of the Hubble types S0$^-$, S0$^o$ and S0$^+$. For
comparison, the line for the bright Coma cluster ellipticals is also
shown, taken from Khosroshahi et al. (2000). In Figure 2,
$<\mu>_{eff}$ is used (rather than $\mu_{eff}$) and it is calculated
from the flux calibrated images in a similar manner as in
\citet{khosroshahietal2000} (taking into account the difference in the
Hubble constant $H_0$ used in the two studies).  It appears that
  although S0s are located close to the elliptical galaxies, as a
  group they are offset by 0.3 mag (at a median $r_{eff}$ of 0.58 kpc)
  from the bright Coma cluster ellipticals. For S0/a-Sa galaxies the
  offset is 0.7 mag (at a median $r_{eff}$ of 0.51 kpc).  It is also
  worth noticing that the brightest bulges (those brighter than
  $M_K^o(bulge)$ = $-$22.7 mag), both among the S0s and S0/a-Sa
  galaxies, behave in a similar manner as the elliptical galaxies in
  the Kormendy relation (a similar result is obtained if the S0 sample
  is divided in two bins, based on the total absolute galaxy
  brightness, using $M_{tot}$ = $-$24.5 mag as a dividing line).  Our
sample includes also S0s with $r_{eff}$ smaller than obtained for any
of the Coma cluster ellipticals. Inspite of their small size, they do
not fall in the region of the dwarf elliptical galaxies (dE) with a
similar $r_{eff}$ (see Capaccioli $\&$ Caon 1991; Khosroshahi et
al. 2004; Ravikumar et al. 2006): for a given $r_{eff}$, the dE's have
several magnitudes lower surface brightnesses than the bulges of our
sample galaxies. The offset we had between S0s and ellipticals differs
from previous studies obtained in the optical
\citep{kormendy1984,capaccioli1992} and in the near-IR
\citep{pahre1998,pierini2002,ravikumar2006}, according to which the
Kormendy relations are similar for the S0s and for the bright
elliptical galaxies: hence we find that they are, in fact, offset.

 For spirals, the Kormendy relation has previously been studied by
  many groups. In an early study by \citet{kormendy1984}, the bulges were
  found to follow the same relation as the elliptical galaxies, but
  later studies showed that spiral galaxies actually lie below the
  bright elliptical galaxies, and also have a larger dispersion
(Kent 1985; Kodaira, Watanabe $\&$ Okamura 1986; Khosroshahi et al. 2000; 
Capaccioli, Caon $\&$ D'Onofrio 1992; Hunt, Pierini $\&$ Giovanardi 2004; Ravikumar et al. 2006; GW2008). 
  The dispersion is largest for the late type spirals
 (Ravikumar et al. 2006).  Our study for OSUBSGS and NIRS0S spirals confirms
  the above tendencies: the S0/a-Sa galaxies appear slightly
  below the bright ellipticals (and S0s), whereas the later types are more
  dispersed towards lower surface brightnesses.  \citet{ravikumar2006}
  have shown that faint spirals with $M_K$$>$ $-$22 mag do not
  obey the same Kormendy relation as the brighter galaxies (see also
  Capaccioli $\&$ Caon 1991). Using a limiting absolute magnitude of
  $-$22.7 mag for the bulges, we see a similar effect.

\vskip 0.35cm
\subsection{Photometric plane}
\vskip 0.35cm

The photometric plane for early-type galaxies was first addressed by
\citet{scodeggio1997}, who added a third parameter to the Kormendy relation,
namely the difference between the galaxy magnitude and the
characteristic magnitude of the cluster. Compared to the fundamental
plane using velocity dispersion as the third parameter, it has the
advantage of making the relation achievable with photometry alone.
Compared to the Kormendy relation the dispersion is
smaller, but it can be used only for galaxies in clusters. The
photometric plane for more general use was addressed by
\citet{khosroshahi2000}: it was defined by $\mu_{ob}$, $r_{eff}$ and the
S\'ersic index $n$ which, compared to the Kormendy
relation, likewise reduces the dispersion.  The photometric plane for
bright elliptical galaxies has been studied also by \citet{graham2002}, who 
replaced $\mu_{ob}$ with $\mu_{eff}$. 

In this study, we use
the photometric plane as defined by Khosroshahi Wadadekar $\&$
Kembhavi (2000). In order to study the photometric plane Khosroshahi
et al. (2000) used $K$-band images of a sample of 26 non-barred S0-Sbc
galaxies.  They found that the bulges of these galaxies which are
  mainly spirals (including 6 S0s), occupy the same region in the
photometric plane as the bright Coma cluster ellipticals. On the other
hand, \citet{abreu2008} found that bulges of S0s appear in the same
region as the bulges of early-type spirals, though they did not
investigate whether these galaxies are located also in the same region
as the elliptical galaxies.  Their sample consisted of 148 non-barred
S0-Sb galaxies (mainly S0s, including 58 spirals) and they used
$J$-band images taken from 2MASS. 

The photometric plane for the NIRS0S and OSUBSGS galaxies is shown in
Figure 3, again comparing it with the bright Coma cluster
ellipticals. We can see that in the photometric plane bulges of S0s
typically lie slightly below the elliptical galaxies, indicating that, for
a given $r_{eff}$ or $\mu_{ob}$, they have a smaller S\'ersic index than the elliptical
galaxies. When extrapolating the line for the bright ellipticals to
lower $r_{eff}$, the deviations of S0s from the elliptical galaxies increase.
Compared to spirals, the S0s form a more homogeneous group of
galaxies: spiral  galaxies have a larger dispersion which increases towards
lower $r_{eff}$ and later Hubble types.  
In the photometric plane the barred and non-barred S0s occupy the same region.

Our result for early-type spirals (S0/a-Sa galaxies are not like the
ellipticals) thus deviates from that of Khosroshahi et al. (2000), and
slightly also from that of M\'endez-Abreu et al. (2008) (the
early-type spirals do not behave exactly like the S0s).  A possible
reason for these differences is likely to be our use of deeper images
and the use of multi-component decompositions. It is also worth noticing that
the brighter galaxies in the NIRS0S sample deviate from the elliptical galaxies
more than the fainter galaxies. The spiral galaxies Sab-Sbc are located
below the Coma cluster elliptical, in a similar manner as the S0/a-Sa spirals.

\vskip 0.35cm
\subsection{Bulge-to-total flux ratios}
\vskip 0.35cm

The distributions of bulge-to-total ($B/T$) flux ratios for the
galaxies in the NIRS0S and OSUBSGS samples are shown in Figure 4.  Bulge-to-disk
($B/D$) flux ratios for the OSUBSGS galaxies were taken from
\citet{lauri2004} and converted to $B/T$ flux ratios. The
  $B/T$-values in both samples were corrected for the effects of Galactic
and internal dust
  using the equations in Section 4, applying different corrections 
for the bulge and the disk.  The corrected mean $B/T$ values in
different Hubble type bins are shown in Table 2. The values we find
for S0-S0/a galaxies are consistent with $B/T$ $\sim$ 0.25-0.28 as found
previously by Balcells et al. (2003, 2007) for non-barred galaxies,
and by Laurikainen et al.  (2005, 2006), \citet{gadotti2008}, and
Weinzirl et al. (2009) using a multi-component approach.  For later
type galaxies in the OSUBSGS sample, similar low $B/T$ values have been obtained 
also by Weinzirl et al. (2009).

Fig. 4. shows the characteristic peaks in the $B/T$ distribution for the
different Hubble type bins.  The rather wide spread for each bin 
reflects the fact that the bulge was
not the only criterion in the original Hubble
classification of galaxies. In the original classification the bulges
played a more important role for the early-type galaxies than for the
later types, where the spiral arms were the dominating feature. Using
deep $K_s$-band images, applying a multi-component decomposition
method, and including proper dust corrections, we are in a position to re-evaluate the $B/T$
distribution. 
It appears that the S0 galaxies have a large range of  
$B/T$ extending to very small values.  
In particular, the number of S0s with small $B/T$$\sim$ 0.10
is 3$\%$ (N=3), and 13$\%$ (N=12) have $B/T<$ 0.15.
If S0/a galaxies are also included the values are 13$\%$ and 25$\%$, respectively. 
Small $B/T$ values, comparable to those obtained for Sc galaxies, have been
found previously by Erwin et al. (2003) for two
S0s, using both photometric and kinematic data, and by Laurikainen et
al. (2006) for one S0, based on a morphological analysis.  In our sample,
the barred and non-barred S0s have similar $B/T$ flux ratios 
 ($<B/T> = $0.29$\pm$0.02 vs 0.33$\pm$0.03,
  respectively), which is not fully consistent with Aguerri et al. (2005) and
  Laurikainen et al. (2007), who found indications that $B/T$ is smaller for 
barred galaxies. Most probably the difference between barred
and non-barred S0s is real, but is so weak that it is visible,
at a statistically significant level, only
when strongly barred S0s are studied.

 In Fig. 5 the $B/T$ flux ratio, S\'ersic index $n$, the bulge effective radius scaled
  to the disk scale length $r_{eff}$/$h_r^o$, and the total absolute
  magnitude $M_{tot}$, are shown as a function of Hubble type
  ($r_{eff}$/$h_r^o$ is discussed in Section 6.6).  $M_{tot}$ is the
  total observed $K$-band absolute magnitude corrected for Galactic
  extinction.  For the OSUBSGS galaxies $B/T$ and $n$ as a function of
  Hubble type, have previously been shown by Laurikainen et
  al. (2007), with the difference that their $B/T$ values were not
  corrected for internal dust. The $B/T$ values for
  the OSUBSGS sample has been discussed also by Weinzirl et al. (2009).
  The symbols denote the median values in each Hubble type bin, and
  the error bars indicate the standard deviations of the mean.  For
  comparison, the median values from GW2008 are also shown.  They
  compiled all reliable decompositions from the literature, leading to
  a sample of 400 disk galaxies, including 16 S0s.  In Fig. 5, the
  mean $B/T$ and $n$ are almost constant for S0$^-$, S0$^o$, S0$^+$,
  S0/a and Sa galaxies, except that $B/T$ increases gradually towards
  S0$^-$ types. A small drop in the $B/T$ flux ratio towards the
  early-type S0s found by Laurikainen et al.  (2007) is not seen here
  (their sample consisted of a sub-sample of only 37 S0s of NIRS0S
  representing mainly barred galaxies). Note that a small $B/T$ (and
  $n$) value for S0s is given also by GW2008, however this is based on
  16 galaxies from the NIRS0S sample, belonging to the above-mentioned
  sub-sample.

For spiral galaxies in the OSUBSGS sample the mean $B/T$ flux ratio
decreases from Sab towards later types, in agreement with GW2008,
shown also in many previous studies (see the references in GW2008). In
our study the mean S\'ersic index $n$ is only slightly larger for the
S0-Sa galaxies than for the later types. It appears that our
values for $n$ and to some extent also for $B/T$, are systematically lower than the
values compiled by GW2008. The comparison is meaningful only between
S0/a-Sc types, because the number of galaxies is small among the
later type galaxies in the OSUBSGS sample, and among the S0s in
GW2008.  The systematic shift between GW2008 and the current study is
most likely explained by the different decomposition approaches used:
although GW2008 collected all reliable measurements from the
literature, most of them, particularly for the later type galaxies,
are based on bulge/disk decompositions. It is expected that excluding
the contribution of the bar flux in the decomposition makes the fitted
bulge profile shape less exponential and increases the fitted $B/T$
flux ratio (Laurikainen et al. 2006), and likewise also the estimated
effective radius.

\vskip 0.35cm
\subsection{Other parameters of the bulge}
\vskip 0.35cm

The total luminosity of a bulge can be expressed by the parameters of
the S\'ersic function: $L$ = $k_L$ $I_{eff}$ $r_{eff}^2$, where
$I_{eff}$ is the flux at $r_{eff}$ (corresponding to $\mu_{eff}$) and
$k_L$ is the structural parameter determined by the shape of the light
distribution (i.e., by S\'ersic $n$). This implies that due to the
S\'ersic function alone correlations are expected for $M_K^o(bulge)$
vs $r_{eff}$, and for $M_K^o(bulge)$ vs $\mu_{eff}$.  If
$M_K^o(bulge)$ additionally correlates with $M_K^o(disk)$ (as
discussed in Section 6.6), then $r_{eff}$ and $\mu_{eff}$ are also
expected to correlate with the $B/T$ flux ratio, which is the case for
S0s. The correlations are weaker when $B/T$ is used instead of
$M_K^o(bulge)$, because the contribution of the disk light to the
total light varies.

More fundamental relations are: (1) $n$ vs $M_K^o(bulge)$, (2)
 $n$ vs $B/T$, and (3) $M_K^o(bulge)$ vs $M_{tot}^o$.  
For the spiral galaxies in the OSUBSGS sample, the
S\'ersic index $n$ has a clear correlation with $M_K^o(bulge)$ (see
Fig. 6a), which confirms the previous result for spirals
\citep{graham2001b,moll2001,noordermeer2007}.
However, for the S0 galaxies no such correlation is present, at a statistically
significant level.  On the other hand, there is a
correlation between $n$ and $B/T$, both for the S0 galaxies and
spirals (Fig. 6b), again for spirals confirming the previous results
(Andredakis, Peletier $\&$ Balcells 1995; Graham 2001b; Trujillo et
al. 2002; Balcells, Graham $\&$ Peletier 2007).

To illustrate how for S0s $n$ can be correlated with $B/T$, but not
with $M_K^o(bulge)$ itself, $M_K^o(bulge)$ is shown as a function of
$M_{tot}^o$ in two S\'ersic index bins, using $n$ = 2.0 as a dividing
line (Fig. 7).  Here $M_{tot}^o$ is the total $K$-band absolute
magnitude, corrected for Galactic and internal dust, in a similar
manner as was done for $B/T$ in Section 6.3. Different dust
corrections were applied for the bulge and the disk, with the fluxes
of these components being obtained from the decompositions.  The left panel
shows the NIRS0S galaxies, whereas the right panel shows the OSUBSGS
spirals. The dashed lines indicate three different $B/T$ values of
0.01, 0.1 and 1.0.  Evidently, for S0s both luminous and less luminous
bulges can have a large range of S\'ersic indexes (e.g., no correlation
exists between $M_K^o(bulge)$ and $n$). However, within each
$M_{tot}^o$ bin the S\'ersic index increases with increasing $B/T$ (a
correlation exists between $B/T$ and $n$). Apparently, the OSUBSGS
sample can be divided into two main bins using $M_K^o(bulge)$ = $-$20
mag as a dividing line: the more luminous bulges have much more frequently
higher values of the S\'ersic index and the $B/T$ ratio than the less
luminous bulges. This implies that the correlation between
$M_K^o(bulge)$ and $n$ for spirals largely follows by lumping the
faint and bright bulges together. It seems that the bulges of S0s are
fairly similar to the bulges of spirals  having $M_K^o(bulge)$ $<$ $-$20 mag.

Clearly, the bulges of S0 galaxies (in NIRS0S) and spirals (in OSUBSGS) are
different from the elliptical galaxies
(that bulges of spirals are different from elliptical galaxies
was shown also by Graham 2001b; their Fig. 14). This is
manifested in the $r_{eff}$ vs $M_K^o(bulge)$ diagram (Fig. 8a) where
both appear below the elliptical galaxies (shown as a dashed line,
taken from GW2008), shown also by GW2008 (see their Fig. 13). These
parameters are correlated, at a statistically significant level, for
both S0s and spirals. It is also worth noticing that the bulges of S0s
are slightly more compact than the bulges of spirals of a similar
bulge luminosity. 
For spiral galaxies the correlation between $r_{eff}$ and $M_K^o(bulge)$
was found previously also by Balcells, Graham and Peletier (2007) for
19 (non-barred) galaxies, but not by GW2008 for a sample 400 disk
galaxies (include 16 S0s). A possible reason why this correlation was
found by Balcells, Graham $\&$ Peletier (2007) and by us, but not by
GW2008, is again the decomposition method used: for non-barred
galaxies the method of choice (two-component vs multi-component) is
not critical, but it does matter when barred galaxies are included.
Figure 8 also shows that for a given bulge brightness, the bulges of S0s have typically higher
surface brightnesses than the bulges of spirals.

In conclusion, bulges of S0 galaxies
are fairly similar to those spiral bulges brighter than
$M_K^o(bulge)$ $\sim$ $-$20 mag.

\vskip 0.35cm
\subsection{Parameters of the disk}
\vskip 0.35cm

For both the NIRS0S and the OSUBSGS samples, the main parameters of
  the disk correlate with each other: with an increasing scale length
  of the disk ($h_r^o$ ) the disk luminosity ($M_K^o(disk)$) increases, and
  the central surface brightness ($\mu_o$) becomes fainter (see
  Figs. 9 a,c), shown previously for S0s by \citet{abreu2008}.  
For spirals, the former correlation has been previously demonstrated
  by \citet{balcells2007}, and the latter by \citet{dejong1996},
  Graham $\&$ de Blok (2001) and GW2008.
In addition, the scale 
length for S0s increases with increasing galaxy magnitude
  ($M_{tot}^o$), as shown previously for S0s by \citet{aguerri2005b}, \citet{bedregall2006} and
  \citet{abreu2008}.  Also, the disks of S0s and spirals do not follow
  the Freeman law (Freeman 1970), pertaining a constant central
  surface brightness. Instead, $\mu_o$ has a large dispersion
  (Fig. 9b), as shown previously for spiral
  galaxies M\"ollenhoff $\&$ Heidth (2001), Aguerri et al. (2005b) and Balcells, Graham,
  Peletier (2007), and for S0s by \citet{aguerri2005b} and \citet{abreu2008}.

The Figure 9 further demonstrates that the S0 galaxies are located at the
same regions with the spiral galaxies, including the late-type
spirals. In GW2008 S0s were dispersed to lower $\mu_o$ (see their
Fig. 8).  For S0s in NIRS0S the range in $h_r^o$
is found to be 1-10 kpc, which is very similar to that obtained for
spirals, which is larger than that obtained by \citet{aguerri2005b} and \citet{abreu2008} for S0s (maximum $h_r^o$ 
$\sim$ 6 kpc; see their Fig. 7, based on 2MASS images).  Also, $\Delta \mu_o$
$\sim$ 16-20 mag for S0s is very similar as the range reported for spirals.
To our knowledge the similarity of the disks in S0s and spirals has
not been explicitly shown previously, most probably because deep
images are needed for detecting the faint outer disks in S0s. 

\vskip 0.35cm
\subsection{Parameters of the bulge and the disk cross-correlated}
\vskip 0.35cm

Naively, strong correlations between the parameters of the bulge and
the disk are expected if the disks were formed first and the bulges
emerged from the disk by secular evolution. Indeed, for spiral
galaxies a correlation has previously been found between $h_r$ and
$r_{eff}$
\citep{courteau1996,dejong1996,graham1999,khosroshahi2000,arthur2003,carollo2007,noordermeer2007}.
A similar correlation was found also for 15 barred S0 galaxies by
Aguerri et al.  (2005), and for S0s in the NIRS0S sample by
Laurikainen et al. (2009), the latter galaxies following the relation
  $h_r$ = 0.62 + 0.46 log $r_{eff}$ ($h_r$ and $r_{eff}$ expressed in kpc).  A similar correlation was found
  by \citet{abreu2008} for a sample of S0-Sb galaxies.  For the
  galaxies in OSUBSGS and NIRS0S samples, this correlation is shown in Figure
  10.  Clearly, the correlation is similar for all morphological types
  except that the slope is slightly shallower for the late-type
  spirals.

Having established the existence of correlations between $M_K^o(bulge)$ and
$r_{eff}$, $M_K^o(disk)$
and $h_r^o$, and $h_r^o$ and $r_{eff}$, a correlation between
$M_K^o(bulge)$ and $M_K^o(disk)$ is also expected.  However, this
correlation, shown in Fig. 11, is a direct, and the most illustrative
way to demonstrate how similar or dissimilar the bulges and disks in S0
galaxies and spirals actually are.  For spiral galaxies, this
correlation has previously been reported by \citet{dejong1996} and
\citet{hunt2004}, and marginally also by \citet{noordermeer2007}.
The exact relation between $M_K^o(bulge)$ and $M_K^o(disk)$ naturally depends
on the morphological type, which partly follows from the peaked
$B/D$-distribution in each morphological type bin: a linear
correlation means a constant $B/D$ flux ratio, the slope relates to the
value of that constant, and the dispersion manifests the
variations in $B/D$ for a given morphological type bin.  We find
fairly similar correlations for S0s and S0/a-Sa galaxies, whereas, as
expected, for later types the relative mass of the disk increases. For
S0s (in NIRS0S) and Sab-Sbc spirals (in OSUBSGS) we obtain:
\vskip 0.25cm
$M_K^o(disk)$ = 0.63 $M_K^o(bulge)$ $-$ 9.3  (S0s)
\vskip 0.25cm 
$M_K^o(disk)$ = 0.38 $M_K^o(bulge)$ $-$ 15.5  (Sab-Sbc)
\vskip 0.25cm
Contrary to what was obtained by Hunt, Pierini $\&$ Giovanardi
(2004) for spirals, in our study the correlation between $M_K^o(bulge)$
and $M_K^o(disk)$ is equally good for
galaxies with small and large S\'ersic indexes, where we use $n$=2.0 as a
dividing line (not shown in any figure). It is worth noticing that
S0s and S0/a-Sa galaxies have very similar bulge and disk luminosities.
The relative disk luminosity increases towards late-type galaxies, and
decreases towards galaxies with luminous bulges, which is the case
for all Hubble types. The range of bulge
luminosity is fairly similar for S0-Sbc galaxies, $M_K^o$(bulge) $\sim$ $-$19 - $-$25 mag, whereas for
Sc-Scd types $M_K^o$(bulge) $\sim$ $-$16 - $-$22 mag.
This is related to the total galaxy luminosity which drops for the
Sc-Scd types.

Due to the correlation $M_K^o(bulge)$ vs $M_K^o(disk)$, a correlation
is expected also between $M_K^o(disk)$ and $r_{eff}$. Indeed, such a
correlation is present, both for the S0 galaxies and for spirals in
the OSUBSGS sample (Fig. 10a). This correlation was not found by
\citet{balcells2007}.

Particularly useful is the parameter $r_{eff}$/$h_r^o$, because it can
be directly compared to predictions of simulation models, related to
galaxy formation and evolution.  The median value for
$r_{eff}$/$h_r^o$ for S0s in NIRS0S is 0.20, and for S0/a-Sa galaxies
it is 0.15 (Table 2). These values are nearly a factor of two smaller
than the value of $<r_{eff}$/$h_r^o>$ = 0.32-0.36 obtained for the
early-type disk galaxies by Khosroshahi et al. (2000; 26 mainly
spirals), Noordermeer $\&$ van der Hulst (2007; 19 S0-Sab, mainly
spirals), and by M\'endez-Abreu et al. (2008; 148 S0-Sb, mainly
S0s). For the OSUBSGS spirals we find a median values of 
$r_{eff}$/$h_r^0$ $\sim$ 0.10,
again nearly a factor of two smaller than the median value
$r_{eff}$/$h_r^o$ $\sim$ 0.22-0.24 given by GW2008 for spirals, or
$<r_{eff}$/$h_r^o>$ $\sim$ 0.22 obtained by MacArthur, Courteau $\&$
Holzman (2003) for the late-type spirals. This difference is most 
probably related to the different decomposition methods used: in the 
previous studies the bulge/disk bar decomposition approach was used,
which in barred galaxies is expected to overestimate the flux of the bulge,
and therefore also $r_{eff}$. De Jong (1996) found
$<r_{eff}$/$h_r^o>$ $\sim$ 0.15 in $K$-band, but in this early
study a constant S\'ersic index was used. 

\vskip 0.35cm
\section{Discussion}
\vskip 0.35cm

In the current paradigm of galaxy formation, the hierarchical
$\Lambda$CDM cosmology, simulations have suggested that
mergers of primordial disk galaxies form the first elliptical galaxies
(e.g., Barnes $\&$ Hernquist 1992; Naab, Khochfar $\&$ Burkert 2006).
The most massive ellipticals or
spheroidals are suggested to form in a merger of two gas-poor (dry)
ellipticals of a similar mass (e.g., Naab, Burkert $\&$ Hernquist 1999; Naab
$\&$ Burkert 2003), or through multiple or hierarchical mergers (e.g., Weil
$\&$ Hernquist 1996; Burkert et al. 2008). On the other hand,
intermediate mass spheroidals are suggested to form by mergers of an
elliptical galaxy with a massive spiral \citep{burk2005,naab2006}. In
this picture, a new disk is formed of the gas, which gradually settles
into the disk plane from the dark matter halo
\citep{kauffmann1999,springer2005}. Simulations
\citep{aguerri2001,moral2006,younger2007} also show that bulges can
form by minor mergers, which are much more frequent in the
observed Universe. In these simulations, minor mergers can heat the
original disk and trigger a central starburst, which consumes most of
the gas in the galaxy, leaving a featureless gas-poor remnant with an
outer disk.  The bulges formed by minor mergers can be dynamically
cooler (Younger et al.  2007; Eliche-Moral et al. 2006), but both in
major and minor mergers the formation of bulges and disks is expected
to be decoupled.

Cosmological simulations with merger histories (e.g., D'Onghia et
al. 2006) have, however, faced serious problems which are not yet
fully resolved. For example, they have been unable to unambiguously
produce galaxies without classical bulges, or galaxies without any
bulges at all. This contradicts the observation that up to 30$\%$ of
the total stellar mass in nearby disk galaxies might reside in
bulge-less galaxies (Kautsch et al. 2006; Cameron et al. 2009).  Also,
the merger remnants produced in the cosmological simulations are too
metal rich, by a factor of 4-8, compared to any observed spheroidals,
ellipticals or bulges \citep{naab2009}. This is the case at least if
the mergers appeared at z $>$ 0.3.

As an alternative it is possible that the disks were formed first, and
the bulges were made of the disk material, followed by subsequent star
formation (see Kormendy $\&$ Kennicutt 2004 for a comprehensive
review). The discovery (Illingworth 1981) that bulges can rotate and
are most likely rotation-flattened oblate spheroids, fits to this
picture. This is suggested to be the case for the bulges in low mass
spirals (see MacArthur et al. 2008), but might also be an explanation 
for the bulges of massive disk galaxies, including S0s. Two examples of 
such low mass bulges in S0s
were discussed by Erwin et al. (2003), and more recently,
similar cases were found for a sample of early-type galaxies based on integral-field
spectroscopy by Cappellari et al. (2007). 

\vskip 0.25cm
\subsection{Photometric properties of the bulges in S0s: like ellipticals or more like bulges in spirals?}
\vskip 0.25cm 

If the photometric properties of bulges in S0s are similar to those of
elliptical galaxies, we may assume that also their
formative processes are similar, most probably related to galaxy
mergers. On the other hand, if the bulges in S0s resemble more 
the bulges of spiral galaxies, it is possible that they were formed
by secular evolution. However, it is worth noticing that the formation of bulges
in all kinds of disk galaxies is under debate, and therefore a possible
connection to spirals does not immediately tell what are the physical
mechanisms of bulge formation in S0s.

In this study we have discussed many relations between the photometric
parameters of the bulge and the disk in S0 galaxies, which link their
bulges more tightly to bulges of spirals than to the elliptical
galaxies. In fact, bulges of S0s are fundamentally different from the
elliptical galaxies, which is manifested in the photometric plane
($r_{eff}$, $\mu_{ob}$ and $n$; Fig. 3), in the $M_K^o(bulge)$ vs
$r_{eff}$ diagram (Fig. 8a), and to some extent also in the Kormendy
relation ($r_{eff}$ and $<\mu>_{eff}$; Fig. 2).  For S0s, the
luminosities of the bulge and the disk correlate with the total
absolute galaxy magnitude, and both also correlate with the $B/T$ flux
ratio, in a similar manner as found previously for spirals (see
Trujillo et al. 2002; Balcells et al. 2003; Carollo et al. 2007).

There is also a group of S0s with $r_{eff}$ = 0.05-0.2 kpc, similar to the
values obtained for dwarf elliptical galaxies (dE), but having
surface brightnesses several magnitudes higher.  Simulations suggest
that dwarf galaxies can form from bright galaxies which lose
their disks in dense galaxy clusters.  However, dwarf galaxies show a
large diversity of structures (Lisker et al. 2006), and not all of
them are expected to form by galaxy harassment.  It is particularly
for the disk-like dwarfs (dS0s) that harassment might be at work
(Mastropietro et al. 2005). The possibility of several mechanisms being responsible 
is supported by comparing the photometric
properties of dEs and dS0s with the bulges of bright galaxies in clusters
(Aguerri et al. 2005b), and also, by comparing the ages and masses of
star clusters around pseudobulges in late-type spirals with those
obtained in dwarf galaxies (di Nino et al. 2009). However, the
simulations by \citet{aguerri2009} do not support the view in which also
galaxies with initially large bulge-to-disk ratios (presumably early-type disk galaxies) could evolve into
early-type dwarfs by galaxy harassment. This simulation result is consistent with our
result.

Bulges of S0s have also characteristics
that slightly deviate from those obtained for spirals. The faint
  bulges are slightly more compact than the bulges of spirals of a
  similar bulge luminosity (Fig. 8a). Also, contrary to spirals, S0s show no
  correlation between $M_K^o(bulge)$ and the S\'ersic index $n$.
(Fig. 6a). These characteristics
could manifest environmental effects: the S0s
might have had a more prolonged interaction with the intergalactic
matter, or with small companions, both of which may have 
slightly modified their parameters
\citep{tissera2003,naab2006}.  

We may conclude that the bulges of S0s are photometrically fairly
similar to the bulges of spirals, but different from the
elliptical galaxies.

\vskip 0.25cm
\subsection{Are the formation of bulges and disks in S0s coupled?}
\vskip 0.25cm 

The different formation scenarios of bulges predict different scaling
relations for the photometric parameters of the bulge
and the disk.  The masses and the scale parameters are expected to
correlate if the disks were formed first and the bulges emerged
from the disk material by secular evolution.  On the other hand, in
hierarchical clustering the properties of bulges were established
already during the merger event \citep{burk2005}, and presumably they
changed very little after that. The disks continue to grow only
slightly in mass and size, due to internal dynamical effects,
and due to gas falling into the disk.  

In this study, scaling relations are discussed, suggesting that the
formative processes of bulges and disks in S0 galaxies are coupled.
Most importantly, the absolute brightnesses of the bulge
($M_K^o(bulge)$) and the disk ($M_K^o(disk)$), indicative of their
relative masses, are correlated for S0s (a correlation $M_K^o(disk$)
vs $r_{eff}$ follows from this). For S0s in NIRS0S, a correlation
exists also between the scale parameters of the bulge ($r_{eff}$) and
the disk ($h_r^o$) \citep{lauri2009}. Evidently,
there must be some regulation process leading to the above
correlations.

The scaling relations we obtain for the spiral galaxies in the
  OSUBSGS sample are similar to those found previously for
  spirals in many similar studies.  An exception is the relation between $r_{eff}$ and
$M(disk)$ (Fig. 10) for which \citet{balcells2007} did not find any
correlation.  This lack of correlation, together with the lack of
correlation between the other parameters of the bulge and the disk
($M_K(disk)$ vs $B/T$, or $M_K(disk)$ vs $\mu_o(bulge)$) in their
study was considered as a challenge for the secular origin of bulges
in spiral galaxies: ``bulges do not know how massive their disks
are''. However, the sample by Balcells,
Graham $\&$ Peletier (2007) consisted only of 19 spirals, which might
explain the lack of correlation. With our sample of 175 spirals such
a correlation is evident, and we conclude that the formation
processes of bulges and disks also in spirals galaxies are indeed coupled.

\vskip 0.25cm
\subsection{The formation processes of bulges}
\vskip 0.25cm 

The scaling relations discussed above fit a scenario where the
  disks in a large majority of S0s were formed first, and the bulges
  emerged from the disk material.  In this picture, mass transfer
  towards the central regions of the galaxies is expected
  \citep{atha2002}, leading to central star formation, thus adding to
  the mass of the bulge over a timescale of 1-2 Gyrs (Athanassoula
  Lambert $\&$ Dehnen 2005; Bournaud, Combes $\&$ Semelin 2005;
  Debattista et al. 2006; Sarzi et al. 2007). As a
  consequence, correlations between the masses and the scale
  parameters of the bulge and the disk appear. However, these scaling
  relations cannot be unambiguously used to distinguish between
  various formation processes of bulges. A correlation between $h_r^o$
  and $r_{eff}$ is expected also if S0s were formed by galaxy mergers:
  a small companion swallowed by the main galaxy increases the bulge
  mass, and at the same time the stars in the disk are spread to a
  larger radius.  
For spirals we find median $r_{eff}$/$h_r^o$ $\sim$
    0.10, and for S0 galaxies $r_{eff}$/$h_r^o$ $\sim$ 0.20.
    Compared to the model prediction of $r_{eff}$/$h_r^o$ =
    0.28$-$0.33 by Naab $\&$ Trujillo (2006), both values are
    not well compatible with the merger scenario. Naab $\&$ Trujillo
    constructed simulation models for galaxy mergers using a large
    range of mass ratios between the main galaxy and the companion.
  The transport of baryonic matter to the central regions of galaxies
  by minor mergers increases also the S\'ersic index and the mass of
  the bulge, leading to a correlation between $B/T$ and $n$
  \citep{tissera2003,naab2006}.  Since secular evolution also produces
  such a correlation, it does not yield any unique interpretation for
  the origin of bulges.

In principle, bulges in late-type spirals can be interpreted as
disk-like pseudobulges formed by star formation in the disk.  Along
the same lines, it has been argued that early-type disk galaxies,
due to their larger bulge masses, do not have enough gas to produce
their bulges in a similar manner. It has been estimated that secular
evolution can produce pseudobulges with masses that contain 10$\%$
of the total stellar mass in galaxies (Kormendy $\&$ Kennicutt 2004).
According to Kormendy $\&$ Kennicutt this is comparable to $B/T$
$\sim$ 0.16. This is still smaller than the new estimates of $B/T$ =
0.25-30 for S0s (Laurikainen et al. 2007; Balcells, Graham $\&$ Peletier 2007;
Gadotti 2008; GW2008; Weinzirl et al. 2009). The estimate by Kormendy
$\&$ Kennicutt includes only star formation in circum-nuclear rings,
so that the $B/T$-value would be somewhat larger if accretion of external
gas (Bournaud $\&$ Combes 2002), converted to bulge mass, was also
taken into account.

Another way of evaluating the formation processes of bulges is to look
at their colors. Bulges formed in a hierarchical clustering scenario
are expected to be red, whereas bulges formed recently by disk star
formation are naturally blue. However, for example accretion of external gas or
small blue companions (Aguerri, Balcells $\&$ Peletier 2001) can
rejuvenate the bulges. The close relationship between
the colours of bulges and disks in spiral galaxies (Peletier $\&$
Balcells 1996; MacArthur et al. 2004; Cameron et al. 2009) is
consistent with the idea that bulges in spirals were formed by secular
evolution. But it has also been suggested by Driver et
al. (2006) that the global properties of bulge-disk systems might
result from the mixing of a red bulge stellar population with a blue
disk population, in varying degrees. 
Therefore, based on colors it is not yet completely
clear whether the bulges in spiral galaxies are intrinsically older,
rejuvenated by recent
star formation, or if they really are young bulges, produced by secular
processes.

\vskip 0.25cm
\subsection{Are S0s descendants of spirals?}
\vskip 0.25cm

The problem related to the fairly large bulge masses in S0s would
  vanish if we assume that their bulges were formed by galaxy mergers
  in the distant past.  However, in that case the obtained scaling
  relations for S0s, indicating that the bulges of S0s are not like
  the elliptical galaxies, or that the formative processes of bulges
  and disks in these galaxies most probably are coupled,
would not have any unambiguous explanation. 
On the other hand, there are many factors that
affect the masses of bulges: a) minor mergers can add mass to an
existing bulge, b) dust can obscure massive bulges, particularly in
late-type spirals, and c) the most massive bulges in S0s might 
have been formed at higher redshifts in an environment which is
different from that observed in the nearby Universe. Also, not all
bulges in S0s are massive: even 13$\%$ of them have $B/T$ ratios as small as
typically seen in late-type spirals.
\vskip 0.25cm

In order to evaluate whether the S0s are stripped spirals, the
  following picture can be outlined:
\vskip 0.25cm

(a) The parameters of the bulges and disks for the S0s in the NIRS0S sample
are correlated, in a similar manner as for the spirals in the OSUBSGS
sample, supporting the view that the {\it formation processes of
  bulges and disks in these galaxies are coupled}. This is expected if
the disks were formed first and the bulges emerged from the disks by
secular evolution.

(b) {\it The properties of bulges in S0s are fundamentally different
  from the elliptical galaxies}. This is the case not only for the 
bright S0s, but also for the
faintest bulges, comparable in size with the dwarf ellipticals (dEs),
which implies that the S0s are not likely to be the  
progenitors of
dEs.  On the other hand, the {\it bulges and disks of S0s are similar to
those in S0/a$-$Scd spirals having bulges brighter than $\sim$ $-$20 mag in $K_s$-band}, 
which is compatible with the idea that S0s are stripped spirals.

(c) {\it The bulges of a large majority of disk galaxies were formed by 
star formation in
  the disk}.  This view is consistent with the above two points, and
is supported by the obtained ratio $r_{eff}$/$h_r^o$ for S0s and spirals:
the  obtained small values are not easily compatible with the predictions
of simulations in which bulges were formed by galaxy mergers. 

 (d) {\it Spiral galaxies with bulges brighter than $M_K^o$ $\sim$ $-$20 mag,
can be transformed into S0s by stripping the disk gas, followed by
truncated star formation}. If star formation of the disk in spiral galaxies were truncated
1-6 Gyrs ago (see Begregal, Arag\'on-Salamanca $\&$ Merrifield 2006), it could explain
the observed magnitude difference of $\sim$ 1 mag in the disk luminosity
between S0s and spirals (see our Fig. 11). Spiral galaxies which can evolve into S0s depends largely
on the galaxy luminosity, and therefore also on the luminosity of the bulge,
and not on the spiral Hubble type. 


This scenario, where even late-type spirals could be progenitors of
S0s, reminds of the early ideas by van den Bergh (1976, 1998; see also
Sandage 2005), who suggested that spirals and S0s are spread
throughout the Hubble classification scheme in parallel tuning forks
(S0$_a$, S0$_b$ and S0$_c$).  The S0 galaxies studied by van den Bergh
were fainter than the Sa spirals. Therefore he suggested that those
S0s with small $B/T$ flux ratios are ``anemic'' Sb and Sc galaxies,
stripped of gas, which leaves only a bulge and a featureless
disk. However, at that time no such galaxies were found. The S0
galaxies we study are not ``anemic'', they are not fainter than the
early-type spirals. For a representative sample of disk galaxies
  the $K$-band luminosities of S0s and spirals was studied recently by
  Burstein et al. (2005), showing that S0s and spirals have similar
  total luminosities. Taking into account that fading of star
  formation should lower the luminosity of S0s, this was considered as
  a problem for the stripping scenario.  But, on the other hand,
  increase in the bulge luminosity due to internal dynamical effects
  might partly compensate the loss of the disk luminosity in S0s.
Also, as shown in Figure 11, bright spirals might more easily transform into S0s. Moreover
the disks in most S0 galaxies are not featureless: bars are known to
be prominent in S0s, nuclear bars are frequent (Erwin $\&$ Sparke
2002; Laine et al. 2002; Laurikainen et al. 2009), and even a large
majority of non-barred S0s have ovals or lenses (Laurikainen et
al. 2009).  The large number of bars, rings and lenses in the
early-type galaxies is again consistent with the idea that S0s were
stripped spirals.  Otherwise it would be difficult to understand why
these galaxies, deficient of gas, which should make the disk stable,
have formed such frequently observed, prominent sub-structures in
their disks.

\vskip 0.25cm
The original idea of S0s as stripped spirals where star formation has
ceased comes from the observation by Dressler (1980), who showed that
S0s are common in nearby clusters, whereas spirals are more common in
distant clusters.  This has been recently re-investigated for the
cluster Cl0024+16 at redshift $z$=0.39 by Moran et al. (2007). They
found a large abundance of passive spirals (with only a
small amount of HI gas and a low level of star formation), which were
interpreted as galaxies in a transition state from spirals to S0s. The
number of these galaxies was estimated to be large enough to explain
the fraction of S0s in the nearby clusters.  It was suggested by Moran et
al. (2007) that the processes to form ``passive'' spirals in a low
density interstellar medium could be ram pressure stripping, and in a more dense
environment, galaxy interactions, starvation, and gentle stripping of
the disks. However, it has been questioned whether typical spiral
galaxy luminosities are sufficient to build up the brightest S0s in
local clusters (Poggianti et al. 1999; Gerken et al. 2004). A natural
solution for this problem comes from the recent study of luminous
infrared galaxies (LIRGs) by Geach et al. (2009), who used 24 $\mu m$
band Spitzer observations. They showed that the LIRGs at z$\sim$0.5
can account for the star formation needed to explain the bright end of
the S0 luminosity function. These galaxies have morphological types of
Sab, indicating that the bulges in the most massive S0s might form
already at a fairly early stage of galaxy evolution. Based on an analysis
of colors, Dom\'inguez-Palmero $\&$ Balcells (2009) obtained a similar
conclusion. 

There is also other evidence supporting the view that S0s are stripped spirals.
If interactions of galaxies with the intra-cluster medium,
e.g., gas stripping, are responsible for the transformation of spirals
into S0s, the number of globular clusters in these galaxies will not
be affected. Indeed, this is what was found by \citet{aragon2006} and
\citet{barr2007}. Results for the Tully-Fisher (Luminosity vs maximum
rotation velocity) relation hint in the same direction: Bedregal,
Arag\'on-Salamanca $\&$ Merrifield (2006) explained the location of
the S0 galaxies below the spirals (lower L for a given
$V_{max}$) arising from the luminosity evolution of spiral galaxies,
in which star formation is ceased.  The transformation was assumed to
occur over a range of timescales, with the galaxies passively fading
in the latest stage of the evolution. This is consistent with
the results obtained in this study.

\vskip 0.35cm
\section{Conclusions}
\vskip 0.35cm

New 2D multi-component decompositions are presented for 122 early-type
disk galaxies, as part of the Near-IR Survey S0 galaxy Survey
(NIRS0S), using deep $K_s$-band images. Combined with our previous
decompositions, this leads to a sample of 175 galaxies (117 S0s + 22
S0/a and 36 Sa galaxies). Of these galaxies, reliable decompositions
could be made for 160 galaxies. As a comparison sample we use the Ohio
State University Bright Spiral Galaxy Sample (OSUBSGS) of spirals, for
which similar multi-component decompositions have previously been made
by us.  Particular attention has been paid on correcting the effects
of dust on the parameters of the bulges and disks. In our
multi-component approach, 2D light distributions of galaxies are
decomposed to structure components including, besides bulge and disk,
also bars, ovals and lenses, using either S\'ersic or Ferrers model
functions.  To our knowledge, this is a first attempt to study the
scaling relations for a representative sample of S0s and spirals,
applying the same multi-component decomposition approach for all
galaxies.  

\vskip 0.25cm
Our main conclusions are the following:
\vskip 0.25cm

(1) The photometric properties derived for the NIRS0S and the
  OSUBSGS samples show that the bulges of S0 galaxies are more tightly
  related to bulges of spirals than to elliptical galaxies. This is
  manifested in the $M_K^o(bulge)$ vs $r_{eff}$ diagram, in the
  photometric plane, and in the Kormendy relation (Figs. 8a, 3, and 2,
  respectively). However, the brightest bulges among the S0-Sa galaxies in
  the Kormendy relation are located in the same region as the
  elliptical galaxies.

(2) The bulges in S0s are 
  different from the elliptical galaxies. This is the case, not only for the bright
galaxies, but 
  also for those with sizes similar to the
  dwarf elliptical galaxies (S0s have several magnitudes higher
  surface brightnesses than dEs). This suggests that S0s are
  not likely to evolve into dwarf galaxies by galaxy harassment.  On the other
  hand, the bulges of S0s are fairly similar to those of
  spirals having $M_K^o$(bulge) $<$ $-$20 mag, except that they are slightly more compact (Fig. 8a) and
  do not show a statistically significant correlation 
between $M_K^o(bulge)$ and the S\'ersic  index $n$ (Fig. 6a) .

(3) The properties of bulges and disks of S0s in the NIRS0S
  sample are coupled, in a similar manner as has been found previously for
  spirals.  This is manifested in the $M_K^o(bulge)$ vs $M_K^o(disk)$
  (Fig. 11), $M_K^o(bulge)$ vs $r_{eff}$ (Fig. 8a), as well as in the
  $r_{eff}$ vs $h_r^o$ diagrams (Fig. 10).  For $r_{eff}$ vs $h_r^o$
  we confirm previous results.

(4) We obtained median $r_{eff}$/$h_r^o$ = 0.20, 0.15 and 0.10 for S0, S0/a
  and Sab-Sc galaxies, respectively.  These values are a factor of
  $\sim$2 smaller than obtained previously in the literature. 
Compared to the model predictions by Naab
  $\&$ Trujillo (2006), these values are not well compatible with the
  idea that the bulges of disk galaxies were formed by galaxy mergers,
  not even for S0s.

(5) The properties of the disks are very similar among the S0s and spirals:
they occupy the same regions in the (a) $h_r^o$ vs $M_K^o(disk)$, (b)
$\mu_o$ vs $M_K^o(disk)$, and (c) $\mu_o$ vs $h_r^o$ (Figs. 9a,b,c, respectively)
diagrams.  For both types of galaxies $\mu_o$ varies by 4.5 mag and the
range in $h_r^o$ is 1-10 kpc, in agreement with that obtained
by GW2008 for spirals.

(6) The bulge-to-total flux ratio, $B/T$, for S0s covers the full
range of $B/T$ $<$ 0.7 found in the Hubble sequence. Even 13$\%$ (N=12) have
$B/T$ $<$ 0.15, typically seen in late-type spirals.
\vskip 0.25cm

 Our results support the view according that
  spiral galaxies with bulges brighter than $-$20 mag in $K_s$-band, can
  evolve directly into S0s by stripping the gas from the disk,
  followed by truncated star formation. This is estimated to be the
  case up to $\sim$ 50$\%$ of the Sc-Scd type spirals.  This view
  in consistent with the obtained scaling relations and the properties of
bulges and disks of S0s and spirals, as discussed in our
  study.

\section*{Acknowledgments}

We thank the anonymous referee of carefull reading of the manuscript,
and Dr Alister Graham for constructive criticism. 
We acknowledge the time allocation committees of the
NTT (074.B-0290(A), 077.A-0356(A),081.B-0350(A)), WHT,
TNG, NOT and KPNO telescopes for giving a significant number of observing
nights for this project. The New Technology Telescope (NTT) is operated at the 
European Southern Observatory (ESO) in Chile. The William Herschel 
Telescope (WHT), the  
Italian Telescopio Nazionale Galileo (TNG), and the Nordic Optical  
Telescope (NOT), are operated on the island of La Palma by the Isaac  
Newton Group, the Fundacion Galileo Galilei of the INAF (Istituto  
Nazionale di Astrof\'isica), and jointly by Denmark, Finland, Iceland,
Norway, and Sweden, respectively, in the Spanish Observatorio del  
Roque de los Muchachos of the Instituto de Astrof\'isica de Canarias.
 This publication
makes use of data products from the Two Micron All Sky Survey, which
is a joint project of the University of Massachusetts and the Infrared
Processing and Analysis Center/California Institute of Technology,
funded by the National Aeronautics and Space Administration and the
National Science Foundation. It also uses the NASA/IPAC Extragalactic
Database (NED), operated by the Jet Propulsion Laboratory in Caltech.
EL and HS acknowledge the Academy of Finland of significant financial
support. RB acknowledge the support of NSF grant AST-0507140. JHK
acknowledges support by the Instituto de Astrof\'isica de Canarias
(3I2407).

\newpage

\appendix

APPENDIX A: DISCUSSION OF THE DECOMPOSITIONS FOR THE INDIVIDUAL GALAXIES
    
\vskip 0.25cm In the following decompositions for individual NIRS0S
galaxies are discussed. It is done (1) if the solution is uncertain,
(2) it was complicated to find, (3) the galaxy has a bar although it
is not classified as a barred galaxy in RC3, or (4) the cataloqued
Hubble type needs revision.  For 15 of the 176 galaxies no
decomposition was made, either because the inclination was higher than
65$^0$ (NGC 2549, NGC 4220, NGC 4293, NGC 4429, NGC 5308, NGC 5448,
NGC 5838, NGC 7332), the galaxy formed a strongly interacting pair
(NGC 4105, NGC 6438, NGC 3718, NGC 5353), the image was not deep
enough for detecting the exponential disk (NGC 6482), the galaxy had
practically no bulge (IC 5240), or the decomposition was otherwise
uncertain (NGC 3900).
Weak bars detected in the residual images are
discussed, but they are shown in the data-paper (SLBK) where all the
structural components for the sample galaxies are illustrated.
\vskip 0.25cm 

{\bf IC 4329:} The galaxy has a very dispersed bar-like structure
ending up to a faint lens at the same radial distance ($\sim$
55"). Inside the bar there is a similar bar/lens structure at r $<$
25" appearing also as an exponential subsection in the surface
brightness profile.  The two bar/lens components are visible also in
the ellipticity profile.  In the decomposition the inner bar/lens was
fitted by a Ferrers function.

\vskip 0.25cm 
{\bf IC 4889:} The galaxy is an elliptical galaxy in RC3 but,
as noticed by Sandage $\&$ Bedge in CAG, it has a developed outer envelope which makes a
classification to S0 justified. The inner parts of the galaxy also
show isophotes that deviate from the outer
disk. A weak lens was added to the decomposition, but it does not
affect the parameters of the bulge.
\vskip 0.25cm 

\vskip 0.25cm {\bf IC 5267:} In the near-IR this galaxy looks like
typical S0 with two bright lenses with nearly circular isophotes, and a faint outer
disk. In the decomposition the lenses were fitted by two Ferrers
functions. However, in blue light it is classified as S0/a having also
a partial outer ring. 

\vskip 0.25cm

{\bf IC 5328:} This galaxy is classified as an elliptical galaxy in
RC3, and as an S0 in RCA and in CAG. We
first tried to use a single S\'ersic function, but it failed to model
the surface brightness profile properly. As the galaxy obviously has a
lens, a Ferrers function was added, which considerably improved the fit:
the solution converged even when leaving all the parameters of the bulge and the
disk free for fitting. The residual image, after subtracting the bulge
model, shows a very weak bar. There is no doubt that this is not an
elliptical galaxy.

\vskip 0.25cm {
\bf ESO 137-G10:} In the $K_s$-band image the galaxy
image looks featureless. Our best fitting model includes a fairly
exponential bulge and an exponential disk. In the residual image,
after subtracting the bulge mode, a faint lens is visible, detected
also in CAG in the optical region.
\vskip 0.25cm

{\bf ESO 337-G10:} A simple bulge/disk decomposition was a good fit to the
surface brightness profile, but it left a a residual image where
a lens was visible.

\vskip 0.25cm {\bf NGC 439:} This galaxy is classified as SAB0$^-$(rs)
in RC3, and as E5 in CAG.  Our $K_s$-band image shows no sign of a
bar. In fact, the galaxy looks featureless and the two-dimensional
surface brightness profile can be satisfactorily fitted by a single
S\'ersic function, supporting the GAC classification.

\vskip 0.25cm {\bf NGC 474:} The galaxy has a weak oval bar surrounded
by a lens with circular isophotes, each of which was fitted with a Ferrers
function. The disk is exponential and shows some arc-like structures.
These arcs have been interpreted as merger-created shells, possibly also
influenced by the neighbor NGC 470 \citep{turn1999}.

\vskip 0.25cm 
{\bf NGC 484:} 
This galaxy is classified non-barred in RC3 and at first sight it indeed looks
featureless. But, it has an elongated central component, most probably
a bar, surrounded by a nuclear ring. The galaxy has also a weak lens
extending to a large radial distance. However, these components are
too weak for taking into account in the decomposition. We find two
possible solutions, one obtained with a single S\'ersic function, and
another using a bulge-disk model.  However, as a bar and a lens are
visible in the residual image in the single function solution, the
bulge-disk model was considered more reliable.
\vskip 0.25cm

{\bf NGC 507:} 
NGC 507 is reported to be a non-barred galaxy in
all previous studies, but our $K_s$-band image shows a weak bar in the
unsharp mask image (see SLBK). The bar is too weak to be fitted in the
decomposition, but as it is surrounded by a bright oval, we can fit
both components with a single Ferrers function.
\vskip 0.25cm

{\bf NGC 524:} This galaxy is a typical example of a non-barred S0:
the surface brightness profile shows two lenses, which are directly
visible also in the image. By adding two Ferrers functions with $n$ =
1 gives a good fit to the observed flux distribution. A comparison
with a deep optical image also shows that the component fitted as an
exponential disk is actually a third lens (see SLBK).
\vskip 0.25cm

{\bf NGC 584:} Although an elliptical in RC3, we classify this galaxy
as an S0, which is also in agreement with Buta, Corwin $\&$ Odewahn
(2007, hereafter BCO). It has a bright central oval and a weak bar
inside the oval. The surface brightness profile shows an intermediate
region between the bulge and the disk.  In our decomposition we fitted
the bar/lens using a single Ferrers function.
\vskip 0.25cm


{\bf NGC 890:} In RC3 this galaxy belongs to the family class AB.
However, there is no clear evidence of a bar in the $K_s$-band
image. However, the galaxy has boxy isophotes at r $<$
10''. In the decomposition a Ferrers function was fitted to this boxy
component. By excluding it a boxy oval appears in the residual image,
after subtracting the bulge model from the original image. However, the
oval does not affect the parameters of the bulge or the disk.
\vskip 0.25cm




{\bf NGC 1161:} This galaxy is classified as non-barred S0 in RC3 and
by BCO, but it shows a weak bar in the residual image after
subtracting the bulge model from the original image. We tried first a
simple bulge/disk decomposition which gives a quite good fit. However,
by including a Ferrers function for the bar, less structure appeared
in the residual image.
\vskip 0.25cm

{\bf NGC 1201:} This is again a non-barred galaxy in RC3, but our
$K_s$-band image shows two bars. The secondary bar is also surrounded
by an oval. In our best fitting model the two bars were fitted by
Ferrers functions: for the secondary bar and the oval a single
function was used.
The different structure components were found iteratively in small steps.
The final solution was maintained even after leaving all the key
parameters free for fitting.
\vskip 0.25cm

{\bf NGC 1302:} This galaxy has a nearly round bulge and a bar
surrounded by a lens. The bar and the lens were fitted separately. If
all the parameters were left free for fitting, the flux of the bulge,
the lens, and the bar were degenerated, which was possible to see in
the model image.
A decomposition for this galaxy has been previously made also by
\citet{lauri2004} using the OSUBSGS $H$-band image, leading
to a higher $B/T$ flux-ratio than obtained in this study ($B/T$ = 0.40, and 0.25,
respectively). This difference can be explained by the fairly bright
lens included to the decomposition in this study, but not 
by Laurikainen et al.
\vskip 0.25cm



{\bf NGC 1344:} The galaxy is classified as E5 in RC3, but in our
classification it is an S0.  To the decomposition we included an oval,
fitted with a Ferrers function.


\vskip 0.25cm {\bf NGC 1351:} As the surface brightness profile shows
an exponential disk and a prominent bulge, we started with a
bulge/disk decomposition.  However, after subtracting the bulge model
from the original image a weak bar appears in the residual image.  In
the final decomposition a Ferrers function was used for the bar.
However, the bar is so weak that it does not affect the parameters of
the bulge or the disk.
\vskip 0.25cm

{\bf NGC 1371:} The galaxy has a weak bar with some characteristics of
spiral arms. In the decomposition this component was fitted with
a Ferrers function. The spiral arms seem to be forming an inner ring.
\vskip 0.25cm

{\bf NGC 1380:} The galaxy has a weak bar with ansae at the two
edges. In the central parts of the galaxy there is either a flattened
bulge or an oval. 
\vskip 0.25cm


{\bf NGC 1389:} The galaxy has a small bar oriented along the main
disk (visible in the direct image). The bar is surrounded by a lens
with a different orientation. In the decomposition the lens was fitted
by a Ferrers function, whereas the bar was too weak to be taken into
account.








\vskip 0.25cm 
{\bf NGC 1537:} The galaxy has a bar visible in the
direct image, but it was too weak to be taken into account in the
decomposition.
\vskip 0.25cm

{\bf NGC 1543:} This galaxy has two bars, a nearly circular lens
surrounding the primary bar, and a prominent, broad and diffuse outer
ring. The bulge is nearly spherical. The outer ring dominates the outer  
part of the disk.  In the decomposition the two bars were fitted by
Ferrers functions. The decomposition did not converge, but at the end
the solution changed very little in the iterations.  In order to avoid
too long bar model the length was forced to 75 arcsec. At the
beginning some flattening for the bulge was given, but in the
iterations the bulge turned to spherical.
\vskip 0.25cm 



{\bf NGC 1617:} The galaxy has a bar with some characteristics of
spiral arms. It has also an outer ring, probably forming a
lens. The bar and the lens/outer ring were fitted by Ferrers
functions.


\vskip 0.25cm {\bf NGC 2217:} The galaxy has two bars and a lens
surrounding the primary bar.  These components were fitted with three
Ferrers functions. The secondary bar was fitted together with an oval
surrounding it. Due to the oval, estimating the flattening of the bulge
was not self-evident. However, using a nearly spherical bulge, a well-defined
oval (similar to the observed one) in the model image was
produced. The outer disk is dominated by an outer ring, which makes
the scale length of the disk unreliable.
\vskip 0.25cm



{\bf NGC 2300:} This galaxy is classified as an S0, in RC3 and by
\citet{michard1994}, whereas in CAG it is classified
as an elliptical galaxy. The surface brightness profile shows clearly
an exponential disk which confirms that the galaxy is an S0. It has
also a nearly exponential subsection at $r$ $<$ 20"
associated to a lens, which is visible also in the direct image.  The lens was
fitted by a Ferrers function.
\vskip 0.25cm


{\bf NGC 2655:} This galaxy has a large flattened component and
a faint, extended and non-regular outer envelope. In the decomposition
we assumed that the flattened structure is a bulge. This lead
to an extremely large $B/T$ flux-ratio ($B/T$ = 0.66), which is consistent
with the visual impression of the image and the surface brightness profile. 


\vskip 0.25cm {\bf NGC 2685:} NGC 2685 is classified as barred S0 in
RC3.  \citet{ravikumar2006} have also
detected an inner disk in the $H$-band using the Hubble Space
Telescope. Our image shows that the galaxy has an oval-shaped
component at $r$ $<$ 50", oriented along the outer disk. This oval
component can be either a bar ($\epsilon$ = 0.65) or an inner disk.
In our decomposition this component was fitted by a Ferrers function,
whereas the inner disk was not considered.

\vskip 0.25cm 
{\bf NGC 2768:} This galaxy is classified as E6 in RC3,
and as an S0 in CAG and by BCO. Although the galaxy is fairly
featureless, it has boxy isophotes at a large radial range.  The
surface brightness profile clearly shows an intermediate region
between the bulge and the disk.  In the decomposition we fitted an
oval using a Ferrers function. Leaving the parameters of
the oval free for fitting, the boxy isophotes were automatically
found. Our best fitting model is good everywhere else
except in the innermost pixels. This is because the galaxy has
also a nuclear bar at $r$ $<$ 2'' (see SLBK), which was
not possible to take into account in the decomposition.
\vskip 0.25cm


{\bf NGC 2782:} This galaxy has a small, dusty bar at $r$ $<$ 7'',
and a complex lens-like structure outside the bar. In the
decomposition only the flat component was possible to take into account
(the bar is too weak and knotty). The decomposition
did not converge, but the obtained solution looks reasonable.

\vskip 0.25cm {\bf NGC 2787:} This galaxy has a bar and a bright lens
or oval inside the bar, each of which was fitted with a Ferrers
function ($n$ = 2 and $n$ = 1, respectively).  The central
regions of the image have nearly spherical isophotes. We made first a
bulge/disk/bar decomposition, leaving all the key parameters free for
fitting.  This approach gave a fairly good fit to the surface
brightness profile, but the model image was not comparable with the
observed image. Adding an oval considerably improved the model
image. 
\vskip 0.25cm




{\bf NGC 2902:} This is again a prototypical S0 having a prominent lens,
which is important to take into account in the decomposition.
The galaxy has also a weak bar inside the lens, visible 
only in the residual image after subtracting the bulge model from
the original image (see SLBK).
\vskip 0.25cm


{\bf NGC 2950:} This galaxy has two bars, both surrounded by a lens or
an oval. The secondary bar and the surrounding oval were fitted by a
single Ferrers function, whereas for the primary bar and the lens
surrounding it two different functions were used. The decomposition
was started by finding the parameters of the disk, which were fixed at
the beginning. After that a bulge/disk/bar decomposition was made, and
finally the rest of the components were added to the fit
iteratively. Once all the structure components were found, all
parameters of the bulge and the flux of the other components, were
left free for fitting. The solution did not converge, but at the end
it changed extremely slowly. The found solution looks reasonable.
\vskip 0.25cm


{\bf NGC 3032:} The galaxy has two nearly circular lenses visible both in the
direct image and in the surface brightness profile. In the decomposition 
they were fitted by Ferrers functions. It is possible that the exponential
part of the surface brightness profile fitted as a disk, actually is a lens.
\vskip 0.25cm



{\bf NGC 3166:} This galaxy has three classical bars perpendicular to
each other. The primary bar is surrounded by a nearly circular lens,
and the secondary bar by an other lens. The tertiary bar has boxy
isophotes in the middle part of the bar. Unfortunately the FOV is too
small for detecting the outer disk.  In the decomposition we used
three Ferrers functions: one for the primary bar, one for the lens
surrounding it, and one for the secondary bar together with the
surrounding lens.  An exponential function was used to model the outer
disk, including the tertiary bar.  The components were found
iteratively in small steps. In spite of the fact that the solution for
the outer disk is only an approximation, the inner regions were
possible to fit in a reasonable manner.
\vskip 0.25cm

{\bf NGC 3169:} In RC3 this is a non-barred galaxy, but our $K_s$-band
image shows a weak bar at $r$ $<$ 10''. The bar is visible in the
unsharp mask image (see SLBK). The image shows boxy isophotes at $r$
$<$ 50'', which might form part of a larger bar. However, both
components are too weak to be taking into account in the decomposition.
\vskip 0.25cm

{\bf NGC 3245:} This galaxy has a small inner disk visible in the
direct image, which might be related to a nuclear ring. It is surrounded by a bright lens or an oval, manifested
as a broad bump in the surface brightness profile. In the
decomposition the lens and the outer disk were fitted by a single
exponential function, whereas the inner disk was fitted by a Ferrers function.
\vskip 0.25cm

{\bf NGC 3358:} The galaxy has a secondary bar and a primary bar
surrounded by a weak lens.  The outer disk is very flat and drops
rapidly at the outer edge. The disk has also some spiral structure. In
the decomposition it was possible to model the nuclear bar with one
Ferrers function, and the primary bar, together with the surrounding
lens, with another Ferrers function. The bulge is the small, nearly
round component in the galaxy center. In this case there is no
confusion with the bulge and the other components of the galaxy.
\vskip 0.25cm 

{\bf NGC3384:} The image shows a bright, nearly round bulge, a
weak bar, and an elongated central structure, most probably a nuclear
bar. In our decomposition the two bars were fitted by Ferrers
functions. This solution leads to a considerably smaller $B/T$
flux-ratio than obtained previously by \citet{fisher1996} using a
simple bulge/disk approach ($B/T$ = 0.32 vs. 0.53, respectively).
\vskip 0.25cm

{\bf NGC 3414:} In the CAG this galaxy is given as an example of an S0
having a very bright bulge and one of the faintest disk-to-halo
(bulge) ratios seen in any disk galaxy. De Vaucouleurs et al. (1976)
reports also a faint diffuse bar, but the bar has not been considered
in more recent classifications of this galaxy. In our decomposition a
bar was fitted by a Ferrers function. However, it is also possible
that this is a galaxy seen nearly edge-on.  For the uncertainty in the
galaxy orientation the obtained solution should be considered with
caution.
\vskip 0.25cm

{\bf NGC 3489:} In spite of its apparent simplicity this galaxy has
actually a fairly complicated morphology (see SLBK): it has a weak
bar inside a flattened bulge, and boxy isophotes outside the bar. In
the decomposition the bar was fitted by a Ferrers function.


\vskip 0.25cm {\bf NGC 3516:} This galaxy has a bar and a lens
surrounding it. In the decomposition we used two Ferrers functions to
fit these components. However, the interpretation of the bulge is not
self-evident. The central regions of the galaxy are flattened and are 
assumed to be the bulge. However,
based on the twisted isophotes in that region (see SLBK) it could be
an oval as well. We consider the obtained $B/T$ as an upper limit.
\vskip 0.25cm

{\bf NGC 3607:} This galaxy has a bright, nearly circular lens, and a
slightly flattened bulge.  The lens, fitted by a Ferrers function in
the decomposition 
changes the $B/T$ flux-ratio from 0.31 to 0.36, and the
S\'ersic index from $n$ = 1.5 to $n$ = 1.6.
\vskip 0.25cm


{\bf NGC 3665:} NGC 3665 has a flattened bulge or an oval, as well as
inner and outer disks. However, the inner disk is so weak that it
cannot be taken into account in the decomposition. In the best fitting
decomposition the flat inner component was interpreted as a bulge. We
also tried whether a single S\'ersic function is capable to fit the
whole surface brightness profile, but that was not the case.
\vskip 0.25cm



{\bf NGC 3729:} This galaxy is dominated by a knotty central
component, a bar and an inner ring. We made a bulge/disk/bar
decomposition.  The parameters of the bar were found first. After that
the bar parameters, except for the flux, were fixed and the parameters
of the bulge were found. In this approach a slightly flattened,
extremely small bulge with $B/T$ = 0.03 was found. The found extremely
small $B/T$ flux-ratio is not consistent with the Sa Hubble type given for
this galaxy. 
\vskip 0.25cm

{\bf NGC 3892:} The galaxy has a bright central component, and a
diffuse bar ending up to a nearly spherical lens. Outside the lens a
weak disk appears. In the decomposition two Ferrers functions were used
to fit the bar and the lens.  We first tried a bulge/disk/bar/lens
decomposition, but it left a large residual to the central regions of
the galaxy. The fit in the central regions was considerably improved
by adding another Ferrers function representing another lens inside
the bar.  The obtained $B/T$ = 0.11 is one of the smallest found for
the S0 galaxies in our sample.


 
\vskip 0.25cm {\bf NGC 3945:} This galaxy has a primary and a
secondary bar, both surrounded by a lens. It has also an inner and an
outer ring. In our decomposition different Ferrers functions were used
to model the primary bar, the lens surrounding it, and the inner
bar/lens system. The scale length of the disk, which for this galaxy
is only an approximation, was found using the 1D-version of the code,
by fitting an exponential function to the outer isophotes. The other
components were found iteratively in small steps. At the end all the
parameters of the bulge and the flux of the other components were left
free for fitting. 

\vskip 0.25cm

{\bf NGC 3998:} This is one of the clearest examples of lens dominated
S0 galaxies in our sample. In the surface brightness profile an
exponential sub-section at $r$ $<$ 15'' shows a bright lens. There is
also a weak bump in the surface brightness profile at $r$ $\sim$ 40'',
which hint to an outer lens. In the decomposition the bright inner
lens was fitted with a Ferrers function. The exponential disk in the
decomposition included also the outer lens, but the lens is not expected to
significantly affect the parameters of the disk.
This galaxy has a weak bar at r $<$ 8'', which is visible in the
residual image after subtracting the bulge model. A more detailed discussion
of this galaxy is found in \citet{lauri2009}. 
\vskip 0.25cm

\vskip 0.25cm

{\bf NGC 4138:} This galaxy has an inner and an outer lens, the latter
ending up to an inner ring. In our decomposition we fitted the inner
lens, whereas the outer lens and the ring were included to the 
exponential disk. The bulge is so knotty that this approximation
is not expected to affect the parameters of the bulge. Excluding the
lens in the decomposition barely affects the $B/T$ flux-ratio ($B/T$ =
0.07 and 0.11, respectively), but it changes the S\'ersic index from
$n$ = 2.4 to $n$ = 1.7.
\vskip 0.25cm



{\bf NGC 4203:} This galaxy has a weak bar embedded in a bright oval,
and an almost circular weak outer lens.  The bar is well visible in the
residual image after subtracting the bulge model (see SLBK).  In the
decomposition we fitted the inner lens and the bar with a single
Ferrers function.  The disk and the outer lens were included to the fit of the
exponential disk. 



\vskip 0.25cm

{\bf NGC 4262:} The galaxy has a bright, nearly spherical bulge, a bar,
and a lens extending well outside the bar radius.  We made a
bulge/disk/bar decomposition, which showed the lens in the residual
image, after subtracting the model image from the original image.
However, the lens is so weak that it does not affect the parameters of
the bulge or the disk. This galaxy is reported to have also an inner
disk in the optical region \citep{erwin2004}, but it is not visible in
our $K_s$-band image.
\vskip 0.25cm



\vskip 0.25cm {\bf NGC 4314:} NGC 4314 has a weak nuclear bar
surrounded by a nuclear ring.  It has also a strong primary bar and a
fat inner component elongated along the thin bar, most probably being
part of the bar itself.
In the decomposition we used two Ferrers functions, one for the
primary bar, and one for the nuclear bar/ring system.  In the residual
image, after subtracting the model from the original image, the fat
bar component is visible. 
We tried to add also the fat bar component
to the decomposition, but it disappeared in the iterations.  
We also tried a solution where some flattening was given to the
initial bulge model, and the orientation of the bar, but it did not
lead to a flat bulge model.


\vskip 0.25cm {\bf NGC 4369:} This galaxy has a small bar surrounded
by a lens, which again is surrounded by another lens.  The surface
brightness profile shows that the bulge and the bar have almost
similar surface brightnesses. A bump in the surface brightness profile
at $r$ = 10-15'' corresponds to irregular spiral structure around the
bar. In the decomposition we fitted the bar with a Ferrers function
and the outer lens by an exponential function. No separate disk was
fitted.  However, due to the mall number of pixels above the
exponential part of the surface brightness profile, the obtained
solution is highly uncertain. Most probably the $B/T$ flux-ratio for
this galaxy is extremely small.
\vskip 0.25cm

{\bf NGC 4371:} This galaxy has an ansae-type bar ending up to an inner ring.
Inside the bar there is an inner disk and a nuclear ring. 
In the decomposition Ferrers functions were used to fit the bar and the
inner disk/ring system.  The bulge was forced to be spherical. If the
ellipticity of the bulge was left free for fitting the bulge model
would erroneously consist of the flux of the outer disk. These two
models would give completely different $B/T$ flux-ratios ($B/T$ = 0.19
and 0.53, respectively).

\vskip 0.25cm {\bf NGC 4373:} This galaxy is classified as SAB0$^-$ in
RC3 and as an elliptical galaxy in CAG. In our $K_s$-band image the galaxy
looks featureless, but the surface brightness distribution cannot be
fitted well by a single S\'ersic function. We then fitted a bulge and a
disk: this model did not converge, but at the end the model changed
extremely slowly. In the residual image (observed image - model image)
a weak bar was visible.
\vskip 0.25cm

{\bf NGC 4378:} This is an Sa galaxy having faint, very tightly wound
spiral arms.  The arms seem to be forming two lens-like
components. The weak disk is dominated by an outer pseudo-ring. In the
decomposition we fitted the two lens-like components with Ferrers
functions. These functions affected considerably the $B/T$ flux ratio:
while the simple bulge/disk decomposition gave $B/T$ = 0.54,
including the lenses gave $B/T$ = 0.31.

\vskip 0.25cm {\bf NGC 4424:} This galaxy has a very small bulge or no
bulge at all. We give a bulge/disk/bar decomposition, but for the
small number of pixels in the bulge, only the parameters of the disk
are reliable. This is again one of the early-type galaxies that
have a bulge, similar to those found in late-type galaxies.
\vskip 0.25cm


{\bf NGC 4435:} This galaxy has a bar with a boxy oval inside.
We made a simple bulge/disk/bar decomposition, in which
case the flux of the boxy oval goes to the bulge. However,
the boxy oval might be a manifestation of a nuclear
bar,  for which reason the obtained $B/T$ ratio 
can be considered as an upper limit. 
\vskip 0.25cm

{\bf NGC 4457:} A dominant feature in this galaxy is a lens, which has
a weak embedded bar. The outer disk is dominated by a
weaker lens-like outer ring, whereas in the central regions there is a
secondary bar surrounded by an oval.  In the decomposition the
secondary bar and the surrounding oval were fitted by a single Ferrers
function.  In the decomposition we fitted the primary bar/lens system
with an exponential function and ignored the outer ring. This solution
is expected to be only an approximation for the disk, but presumably
does not significantly affect the parameters of the bulge.
\vskip 0.25cm

{\bf NGC 4459:} This galaxy has a bright, nearly round central
region, and an exponential disk where also a lens appears. The lens
is visible both in the direct image and in the surface brightness
profile. In the decomposition it was fitted by a Ferrers function. A
good solution was found using an image rebinned by a factor of 4, but
a similar solution using the full resolution image did not
converge. Therefore, in the final solution the number of iterations
was limited.
\vskip 0.25cm

{\bf NGC 4477:} The galaxy has a classical bar surrounded by multiple 
shell-like or ring-like structures, which form part of an exponential disk. In
the surface brightness profile the shell-like structures appear as
flux slightly above the exponential light distribution. A good
fit was obtained using a Ferrers function for the bar.

\vskip 0.25cm
{\bf NGC 4531:} This is a dwarf galaxy having a spiral-like bar embedded inside a
lens.  The surface brightness profile shows that the bulge is very 
small and has a low surface brightness. 
All the components of the disk were fitted by a single exponential
function.  A simple bulge/disk decomposition, leaving all the key
parameters free for fitting, leads to an extremely large S\'ersic index
($n$ = 5.3). In this solution the bulge extends throughout the galaxy.
We consider this solution unrealistic and find an other solution with
a smaller bulge (shown in Fig. 1). It was found by selecting the
initial parameters more carefully. We also limited the number of
iterations otherwise, after a certain number of iterations, a sudden
jump to the large bulge solution occurs.
\vskip 0.25cm

{\bf NGC 4546:} In our best fitting solution a bulge, a disk and a bar
were fitted.  The galaxy has also a weak secondary bar visible in the
residual image, after subtracting the bulge model from the original
image.  The secondary bar appears also as a small peak in the
ellipticity profile.  However, it is not expected to affect the
parameters of the bulge or the disk.
\vskip 0.25cm



{\bf NGC 4612:} The galaxy has a weak bar surrounded by a prominent
lens.  The outer disk is weak and it is dominated by an outer ring. In
the decomposition we fitted the bar with a Ferrers function and the
lens with a S\'ersic function.  Our best fitting decomposition looks a
reasonable fit, but the solution did not converge. For this galaxy the
bar affects the parameters of the bulge: a simple bulge/disk
decomposition would considerably overestimate the flux of the bulge
leading to $B/T$ = 0.63 whereas the bulge/disk/bar/lens model gives
$B/T$ = 0.21. The S\'ersic index in the two models also differ
considerably ($n$ = 4.1 and 2.2, respectively).  The image is not deep
enough for obtaining the scale length of the disk in a reliable
manner.
\vskip 0.25cm


{\bf NGC 4694:} The galaxy has a bright oval-like component and a
peculiar nuclear bar or an inner disk.  In the decomposition the
central component was fitted by a nuclear exponential function.

\vskip 0.25cm {\bf NGC 4696:} This galaxy is classified as an
elliptical galaxy both in RC3 and in RSA, whereas in CAG it is
S0$_3(0)$ galaxy.  We found that it is not possible to fit the 2D
surface brightness profile by a single S\'ersic function, thus
supporting the S0 classification. The S0 nature is manifested also by
several exponential subsections in the surface brightness profile,
associated to lenses.  Also, the outer part of the surface brightness
profile is exponential.
\vskip 0.25cm

{\bf NGC 4754:} The galaxy has two bars, of which the secondary bar is
embedded inside an oval.  In the decomposition the bars were fitted by
Ferrers functions. 
\vskip 0.25cm

{\bf NGC 4772:} This galaxy is reported in the literature as an Sa
galaxy in the optical (RC3, BCO, CAG), and as an S0 in the near-IR
\citep{esk2002}. The galaxy has a bright bulge surrounded by a disk, 
which might be seen nearly edge-on. It is surrounded by a faint outer disk. 
\vskip 0.25cm

{\bf NGC 4880:} This galaxy has a very small bulge, a small peculiar
bar, and a lens extending well outside the bar radius. In the
decomposition the bar and the lens were fitted by two Ferrers
functions. Due to the small number of pixels in the bulge model it is
not considered to be uncertain.

\vskip 0.25cm {\bf NGC 4976:} Again, a controversy exists in the
classification given in RC3, where the galaxy is an elliptical, and in
the CAG where it is classified as a prototypical S0$_1$.  Indeed, the
$K_s$-band surface brightness profile shows a deviation from a typical
profile of an elliptical galaxy: the central region is too bright and
also, the flux density does not decrease smoothly. As a confirmation
check we tried to fit the 2D flux distribution with a single S\'ersic
function, but the model failed to fit the central regions of the
galaxy. However, a good fit in the central regions was obtained by
adding a Ferrers function, representing a lens, to the model. This
component (r $\sim$ 17") can be associated also as a bump in the
ellipticity profile (see SLBK).

\vskip 0.25cm {\bf NGC 4984:} NGC 4984 has a primary and a secondary
bar, both embedded inside lenses.  In the decomposition the two
bar+lens systems were fitted by Ferrers functions. While finding the
parameters of the primary bar a series of decompositions were made
until the model image looked similar to the observed image. The galaxy
has a weak outer disk having also an outer ring, which makes the
parameters of the disk somewhat uncertain.
\vskip 0.25cm


{\bf NGC 5026:} The galaxy has a bar surrounded by a weak lens, ending
up to the the same radius with the bar.  Spiral arms, forming an inner
ring, also start from the two ends of the bar. A flattened bulge
appears inside the bar. In our decomposition the bar was fitted by a
Ferrers function, whereas the lens was not fitted.  The exponential
function for the outer disk is not perfect, because the disk is fairly
flat. It is possible that the component fitted as an exponential disk
is actually a lens.
\vskip 0.25cm

{\bf NGC 5087:} This galaxy is classified as an S0 both in RC3 and in CAG.
The image shows a weak bar which was fitted by a Ferrers function.
\vskip 0.25cm

{\bf NGC 5121:} The galaxy has a bright, nearly circular bulge, a
small central component, and a bright lens at $r$ $<$ 40''. The lens
is manifested also as a broad bump in the surface brightness
profile. The lens and the central elongated component were fitted by
Ferrers functions.
\vskip 0.25cm

{\bf NGC 5206:} The galaxy has a bright nucleus and a very faint
dispersed bar at $r$ $<$ 80''. The galaxy has also a faint circular
component at $r$ $<$ 15", which most probably is a lens. In the
decomposition we fitted the nucleus with a Gaussian function, the
bulge with a S\'ersic function, and the bar with a Ferrers function. The
lens was not modeled, but it is also so weak that it does not affect
the decomposition. The found solution gives one of the smallest $B/T$
flux-ratios ($B/T$ = 0.08) for S0s.
\vskip 0.25cm

{\bf NGC 5266:} The galaxy has either a bright oval or a slightly
flattened bulge, surrounded by a faint disk. The dust lane crossing
the flattened component is visible in the $K_s$-band image. In the
decomposition the flattened component was fitted by a S\'ersic function
and it was interpreted as a bulge. A single S\'ersic function does not 
give a good fit to the surface brightness distribution.

\vskip 0.25cm {\bf NGC 5273:} This galaxy has a nuclear bar and a lens
extending to a much larger radius than the bar. In the decomposition
only the lens was fitted. The bar is so weak that it does not affect
the model.
\vskip 0.25cm


{\bf NGC 5333:} NGC 5333 has an oval-bar, having also an elongated
inner component inside the bar, aligned nearly in the same orientation
with the bar.  These two components were fitted by Ferrers functions in the
decomposition.


\vskip 0.25cm {\bf NGC 5365:} This galaxy has a primary bar embedded
in a weak lens, and a secondary bar embedded in a bright oval. The
outermost structure in the image is probably also a lens ending up to
an outer ring. This galaxy is discussed in detail by \citet{lauri2009}, 
where the image is also shown in different
scales, demonstrating the different structure components.  The primary
bar and the lens were fitted by a single Ferrers function, whereas the
nuclear bar and the oval were fitted by a single S\'ersic's
function. The outer part of the image was fitted by an exponential
function.
\vskip 0.25cm

{\bf NGC 5377:}  The galaxy has a strong bar with boxy or 
x-shaped isophotes inside the bar. The galaxy has also a nuclear bar.
In the decomposition the two bars were fitted with Ferrers functions.

\vskip 0.25cm {\bf NGC 5419:} This galaxy is classified as an
elliptical galaxy in RC3, and as S0 in SGC and in CAG. The $K_s$-band
surface brightness profile clearly shows an exponential disk,
characteristic for S0s.  There is also a lens, manifested in the in
surface brightness profile.  Also, the 2D flux distribution cannot be
fitted by a single S\'ersic function.  The best fit was obtained with a
three component model where a Ferrers function was used for the lens.


\vskip 0.25cm
{\bf NGC 5473:} This galaxy has a classical bar inside a bright oval. 
The bar/oval structure is embedded in a weak lens, oriented
perpendicular to the oval. We made a simple bulge/disk/bar
decomposition where a single Ferrers function was used for the bar/oval 
system. The lens was not fitted, but it is so weak that it does
not affect the parameters of the bulge or the disk.
\vskip 0.25cm

{\bf NGC 5485:} The galaxy image looks fairly featureless, but more
careful inspection shows a boxy oval or a lens, manifested also in the
surface brightness profile.  We made first a simple bulge/disk
decomposition leaving all the key parameters free for fitting, but it left
a lens in the residual image. In the final decomposition the oval 
was fitted by a Ferrers function.
\vskip 0.25cm

{\bf NGC 5493:} The galaxy has a long bar and also an inner elongated
component along the bar.  Due to its orientation it is most probably
part of the bar.  The bar was fitted by a Ferrers function.


\vskip 0.25cm {\bf NGC 5631:} This galaxy looks fairly featureless,
except that it has a lens, manifested both in the galaxy image and in
the surface brightness profile. The outer disk looks exponential, but
a comparison to an optical image shows that it is actually another
lens (see SLBK). In the decomposition the lens was fitted
by a Ferrers function, although it does not affect the parameters of
the bulge.  

\vskip 0.25cm

{\bf NGC 5636:} NGC 5636 is classified as an elliptical galaxy in RC3.
However, the galaxy has an exponential surface brightness distribution
at a large radial range, which hints to an S0. Also, a decomposition
using a single de Vaucouleurs type S\'ersic function did not work for
this galaxy. Leaving all the parameters of the bulge and the disk free
for fitting results to $B/T$ = 0.48. A slightly smaller $B/T$
flux-ratio of 0.41 is obtained by adding to the decomposition a lens.
The decompositions with and without a lens give equally good fits.

\vskip 0.25cm {\bf NGC 5728:} This galaxy has a primary and a
secondary bar, both surrounded by a lens. The inner bar/lens system is
also surrounded by a nuclear ring.  In our decomposition we fitted the
primary bar and the surrounding lens with two Ferrers functions, and
the secondary bar/lens system with a third Ferrers function. The
exponential disk is very weak so that $h_r$ was found using the
1D-version of the code. In order to avoid degeneracy of the different
components we started with a spherical bulge and found the parameters
of the two outer Ferrers functions first. These parameters were then
fixed. After that the bulge (which was still assumed to be spherical)
was fitted. Then the parameters of the nuclear component were found
and finally, except for the flux, were fixed. At the end the key
parameters of the bulge (including the flattening) were fitted
together with the other components whose fluxes were left free for
fitting. 
\vskip 0.25cm

{\bf NGC 5750:} A prominent feature in this galaxy is the inner ring
surrounding a weak bar. There is also an oval inside the bar. In our
decomposition only the bar was fitted with a Ferrers function.  The
outer disk was found by fitting an exponential function to the surface
brightness profile outside the ring, using the 1D-version of the
code. The final 2D solution did not converge, but at the end it
changed extremely slowly. Finally the number of iterations was
limited.
\vskip 0.25cm


{\bf NGC 5846:} This galaxy has three lenses with nearly circular isophotes, the
first showing as a bump in the surface brightness profile at $r$ $<$
4'', whereas the other two are manifested as exponential sub-sections
in the profile. The outermost lens at $r$ $<$ 30'' is obvious in the
direct image, but is less clear in the surface brightness
profile.  In the decomposition all three lenses were fitted with flat
Ferrers functions, although only the two inner lenses affected the bulge/disk
solution.
\vskip 0.25cm

{\bf NGC 5898:} This galaxy, with its multiple lens structure, is again
a prototypical S0 (see SLBK).  We tried to fit the lenses
that appear at $r$ = 7'' and 20'', but only the outer lens was
maintained in the decomposition. We found that including it to the
decomposition is important, otherwise we would have $B/T$ = 0.66,
instead of 0.44, and a S\'ersic index 3.1 instead of 2.2.
\vskip 0.25cm


{\bf NGC 5953:} The galaxy has a nearly circular bulge, flocculent
spiral arms in the inner part of the disk forming a lens, and a
featureless outer disk. The lens appears as a bump in the surface
brightness profile, and in the decomposition it was fitted by a flat
Ferrers function. This galaxy also forms part of a closely interacting
pair for which reason the decomposition was limited to a radial
distance of 33". In the adopted portion of the surface brightness
profile there is still some flux from the companion, visible above the
exponential disk, but that is not expected to affect
the scale length of the disk.

\vskip 0.25cm {\bf NGC 5982:} NGC 5892 is again classified as an
elliptical galaxy in RC3, which is consistent with the observation that it has
shell-like structures in the $K_s$-band image.  In our decomposition
it was possible to fit the surface brightness profile both by a single
S\'ersic function, and using two S\'ersic functions, where the inner
component represents a small inner disk. We confirm that this galaxy
is an elliptical galaxy.
\vskip 0.25cm

{\bf NGC 6340, NGC 6407:} These galaxies are non-barred systems, but have
evidence of a prominent lens, manifested as exponential sub-section in
the surface brightness profile. If a simple bulge/disk decomposition 
is made for these galaxies, and the bulge model subtracted from the
original image, the lenses are clearly visible in the residual images.


\vskip 0.25cm


{\bf NGC 6654:} This galaxy is classified as a barred S0/a in RC3. It
is a double barred system, the secondary bar being embedded inside a
bright oval. In the decomposition both bars were fitted by Ferrers
functions. For the secondary bar the bar model includes also the flux
of the surrounding oval.
\vskip 0.25cm

{\bf NGC 6684:} This galaxy has a primary bar ending up to an inner
ring, and a secondary bar embedded inside an oval. The scale length of
the disk was found using the 1D-version of the code. In the final 2D
decomposition both bars were fitted, using a single Ferrers function
for the secondary bar and the surrounding oval.  However, the nuclear
bar+oval did not significantly affect the parameters of the bulge. The
solution did not converge, but at the end changed very slowly. Finally
the number of iterations were limited.
\vskip 0.25cm

{\bf NGC 6703:} This galaxy has two lenses, at $r$ = 20" and at $r$ = 45".
In our decomposition we fitted only the inner lens, manifested as an
exponential sub-section in the surface brightness profile. The outer
lens is too weak to be taken into account in the decomposition.  
\vskip 0.25cm


\vskip 0.25cm

{\bf NGC 6861:} 
$K_s$-band image shows an elongated structure which can be an inner
disk or a bar. Taking into account that it is oriented along the disk
major axis, it is most probably a disk.  In our decomposition it was
fitted by a S\'ersic function.

\vskip 0.25cm

{\bf NGC 6958:} In RC3 NGC 6958 is classified as an elliptical galaxy,
but in CAG it appears as an S0. Based on the fact that the galaxy has
an exponential outer disk and a prominent lens, manifested as an
exponential sub-section in the surface brightness profile, the S0
classification is confirmed in this study.

\vskip 0.25cm {\bf NGC 7029:} This galaxy has a bulge and an inner
disk, surrounded by a boxy oval.  Most probably there is an
exponential disk outside the boxy oval, but the FOV of the image is
too small to confirm that.  For this galaxy we made a simple
bulge/disk decomposition: we were not able to fit the inner disk,
which disappeared during the iterations. However, the inner disk is so
weak that it barely affects the parameters of the bulge. The obtained
solution needs to be used critically, because the flux that was fitted
as an exponential disk actually corresponds the flux of the boxy
oval. In the statistics we don't use the obtained parameters for the
disk.


\vskip 0.25cm

{\bf NGC 7049:} This galaxy is classified as an S0 in RC3.
We made a simple bulge/disk decomposition, which gave one of the 
highest $B/T$ flux-ratios in our sample. It was not possible to fit
the 2D surface brightness distribution by a single S\'ersic function.
\vskip 0.25cm

{\bf NGC 7079:} NGC 7079 has a weak primary bar embedded in a lens,
and an oval inside the bar. In the decomposition we fitted the bar and
the lens with a single Ferrers function. We tried to fit also the
oval, but it did not survive in the iterations. The slightly curved
surface brightness profile at $r$ $>$ 20'' hints to the possibility that the
section we call an exponential disk is probably an outer lens.


\vskip 0.25cm {\bf NGC 7192:} This galaxy is dominated by two lenses,
the inner lens manifested as an exponential sub-sections at $r$ =
10''-17'' in the surface brightness profile, and the outer lens as a
broad bump at $r$ = 17''-47''. The outer disk is exponential.  In the
decomposition the 2D surface brightness distribution is well fitted
using Ferrers functions for the lenses.
\vskip 0.25cm

{\bf NGC 7213:} This galaxy is again dominated by two lenses. The
bright inner lens, which appears as a weak bump in the surface
brightness profile at $r$ $<$ 20'', includes a double ring.  The
outer lens is manifested as a weak exponential sub-section at $r$ =
20''-35''.  In our decomposition the two lenses were fitted by
Ferrers functions. For this galaxy the typically used Ferrers index
$n$ = 1 gave too sharp outer edges for the lenses.  Therefore, the
Ferrers index was left free for fitting, leading to $n$ = 2.6 and 6 for
the inner and outer lenses, respectively.  Our solution did not
converge, but at the end it changed very slowly.
\vskip 0.25cm



{\bf NGC 7371:} This galaxy has a weak bar, whereas outside the bar
flocculent spiral arms dominate the disk. It is due to the spiral arms
that the exponential function is not a perfect fit to the outer disk.
Using a Ferrers function for the bar in the decomposition gave a good fit
to the observed image.  Notice the extremely small $B/T$
flux-ratio of 0.10 for this galaxy.

\vskip 0.25cm

{\bf NGC 7377:} This galaxy has flocculent spiral arms in the optical
region, but in our near-IR image the spirals arms are absent. The
galaxy has a bright lens, manifested as an exponential sub-section in
the surface brightness profile. In the decomposition a Ferrers
function was used for the lens which considerably improved the model.
After subtracting the bulge model from the original image a weak bar
was visible in the residual image (see SLBK).
\vskip 0.25cm


{\bf NGC 7585:} NGC 7585 is classified as a peculiar S0 galaxy in RC3.
Indeed, the peculiar disk is dominating this galaxy.  The galaxy
has also a lens with slightly boxy isophotes. The lens is visible in
the direct image and manifested also in the surface brightness
profile. In the decomposition the lens was fitted by a Ferrers
function. The residual image, after subtracting the bulge model, shows
a nuclear disk with spiral arms at $r$ $<$ 4" (see SLBK).
\vskip 0.25cm

{\bf NGC 7727:} : The galaxy has a small classical bar and a perturbed
peculiar disk.  The dominant feature in the disk is a lens-like
component with some spiral structure.  In the decomposition the
bar was fitted by a Ferrers function. We added to the decomposition
also another Ferrers function, representing the flux in the complicated
structural components outside the bar, but it did not affect the
parameters of the bulge or the disk.

\vskip 0.25cm {\bf NGC 7742:} This galaxy is classified Sb in RC3, but
in the $K_s$-band image it has no spiral structure. The galaxy has two
lenses: an inner lens at $r$ $<$ 10'', characterized also by a double
ring, and an outer lens, manifested as a bump at $r$ = 15-25'' in the
surface brightness profile. In our decomposition we first tried to fit
both lenses with Ferrers functions, but after a few iterations the
inner lens disappeared. 

\vskip 0.25cm

{\bf NGC 7796:} This galaxy is classified as an elliptical galaxy both
in RC3 and in CAG. However, our $K_s$-band surface brightness profile
shows that the galaxy has a disk, which is exponential at a large radial
range.  Also, it was not possible to fit the surface brightness
profile by a single de Vaucouleurs type S\'ersic function, thus
confirming that the galaxy is an S0.  A good solution was found by
fitting a bulge and a disk.
\vskip 0.25cm

\clearpage
\newpage

\begin{table}
\small
\centering
\caption{Parameters of the bulge and the disk, derived from 2D-decompositions. All parameters are uncorrected for
the effects of dust, except for $B/T$ (corr) in which both Galactic and internal dust effects are corrected.}
 \begin{tabular}{@{}lllllllll@{}}
   & & & &  & & & &  \\
  \hline

 Galaxy           &  inc   &  $n$ & $r_{eff}$ & $M_{K}(bulge)$  & $h_r$ & $\mu_o^{obs}$   & $B/T$    & $B/T$ (corr) \\ 
                  &  [deg] &      &  [kpc]    &              & [kpc] & [mag \ arcsec$^{-2}$] &   &     \\

  \hline

ESO 137-G10     &  45.0   &   1.4    &  1.14    &  -23.24   &  4.15   & 16.90   &  0.25   &  0.26   \\
ESO 208-G21$^d$ &  43.8   &   4.2    &  0.62    &  -22.60   &  0.08$^c$   & 13.81$^c$   &  0.97   &  0.97  \\
ESO 337-G10     &  37.6   &   1.7    &  1.52    &  -24.33   &  6.03   & 17.52   &  0.41   &  0.43 \\
IC 1392         &  47.2   &   1.6    &  0.79    &  -23.80   &  3.94   & 16.65   &  0.28   &  0.30  \\
IC 4214         &  47.6   &   2.9    &  0.60    &  -22.97   &  4.96   & 18.16   &  0.28   &  0.30  \\
IC 4329         &  56.0   &   1.5    &  0.88    &  -23.71   &  9.97   & 17.20   &  0.13   &  0.14  \\
IC 4889         &  42.3   &   2.3    &  0.99    &  -23.52   &  3.06   & 16.71   &  0.37   &  0.39  \\
IC 4991         &  44.6   &   1.6    &  2.83    &  -24.90   &  9.94   & 17.65   &  0.36   &  0.38  \\
IC 5267         &  33.7   &   2.2    &  0.82    &  -22.84   &  6.26   & 18.22   &  0.20   &  0.21  \\
IC 5328         &  49.9   &   2.7    &  1.37    &  -23.80   &  5.34   & 17.51   &  0.37   &  0.40  \\
NGC 439$^d$     &  49.2   &   3.7    & 17.9    &  -26.20   &         &         &         &       \\
NGC 474         &  23.6   &   1.8    &  0.57    &  -22.70   &  4.01   & 17.66   &  0.25   &  0.27  \\
NGC 484         &  45.1   &   2.8    &  1.69    &  -24.63   &  5.89   & 17.94   &  0.62   &  0.64  \\
NGC 507         &  27.6   &   2.7    &  2.36    &  -24.81   &  9.48   & 17.65   &  0.31   &  0.33  \\
NGC 524         &  17.4   &   2.7    &  1.39    &  -24.10   &  4.12   & 16.62   &  0.28   &  0.29  \\
NGC 584         &  51.3   &   2.6    &  1.27    &  -23.93   &  4.37   & 17.75   &  0.58   &  0.61  \\
NGC 718$^a$     &  31.6   &   1.7    &  0.22    &  -21.37   &  2.30   & 17.58   &  0.21   &  0.22  \\
NGC 890         &  46.1   &   3.2    &  2.53    &  -24.55   &  4.91   & 16.80   &  0.42   &  0.44  \\
NGC 936$^a$     &  42.3   &   1.5    &  0.29    &  -21.90   &  3.06   & 16.80   &  0.12   &  0.13  \\
NGC 1022$^a$    &  21.6   &   2.2    &  0.13    &  -20.48   &  1.80   & 17.12   &  0.10   &  0.11  \\
NGC 1079$^b$    &  53.9   &   2.2    &  0.42    &  -21.34   &  2.81   & 18.42   &  0.25   &  0.28  \\
NGC 1161        &  33.6   &   2.5    &  1.77    &  -24.20   &  4.45   & 18.19   &  0.65   &  0.67  \\
NGC 1201        &  53.1   &   3.4    &  0.82    &  -22.82   &  2.85   & 16.98   &  0.37   &  0.40  \\
NGC 1302        &  15.1   &   2.2    &  0.48    &  -22.24   &  3.34   & 17.77   &  0.24   &  0.26  \\
NGC 1317$^b$    &  17.6   &   2.1    &  0.46    &  -22.21   &  2.12   & 17.44   &  0.34   &  0.35  \\
NGC 1326$^b$    &  38.7   &   3.0    &  0.57    &  -22.55   &  3.00   & 17.64   &  0.34   &  0.36  \\
NGC 1344        &  44.7   &   3.1    &  1.28    &  -22.88   &  3.71   & 18.37   &  0.52   &  0.54  \\
NGC 1350$^b$    &  54.2   &   2.2    &  0.68    &  -22.56   & 11.5   & 19.58   &  0.25   &  0.27  \\
NGC 1351        &  55.1   &   3.3    &  0.77    &  -22.02   &  3.02   & 18.47   &  0.53   &  0.56  \\
NGC 1371        &  43.7   &   1.4    &  0.26    &  -20.83   &  2.55   & 16.83   &  0.08   &  0.08  \\
NGC 1380        &  42.3   &   2.1    &  0.97    &  -23.19   &  2.87   & 16.56   &  0.31   &  0.33  \\
NGC 1387$^b$    &   3.5   &   1.8    &  0.38    &  -22.80   &  2.40   & 17.30   &  0.39   &  0.41  \\
NGC 1389        &  58.8   &   1.7    &  0.40    &  -21.53   &  1.81   & 17.22   &  0.36   &  0.39  \\
NGC 1400$^a$    &  23.8   &   2.5    &  0.15    &  -19.98   &  0.46   & 16.83   &  0.50   &  0.52  \\
NGC 1411$^b$    &  40.6   &   2.0    &  0.04    &  -19.36   &  0.98   & 16.60   &  0.08   &  0.08  \\
NGC 1415$^a$    &  67.0   &   1.3    &  0.16    &  -20.60   &         &         &  0.11   &  0.13  \\
NGC 1440$^a$    &  39.0   &   1.4    &  0.21    &  -21.06   &  1.82   & 16.86   &  0.14   &  0.15  \\
NGC 1452$^a$    &  45.7   &   1.6    &  0.44    &  -21.76   &  3.37   & 18.01   &  0.22   &  0.24  \\
NGC 1512$^b$    &  52.8   &   1.2    &  0.27    &  -20.49   &  1.69   & 17.04   &  0.19   &  0.20  \\
NGC 1533$^b$    &  19.6   &   1.5    &  0.25    &  -21.53   &  1.63   & 17.20   &  0.25   &  0.26  \\
NGC 1537        &  51.9   &   2.9    &  0.61    &  -22.62   &  2.10   & 17.19   &  0.52   &  0.55  \\
NGC 1543        &   0.0   &   2.5    &  0.99    &  -22.68   &  9.94   & 19.71   &  0.32   &  0.34  \\
NGC 1553        &  41.5   &   2.1    &  0.63    &  -22.85   &  3.23   & 16.87   &  0.21   &  0.23  \\
NGC 1574$^b$    &  16.8   &   2.8    &  0.42    &  -22.52   &  2.22   & 17.33   &  0.38   &  0.40  \\
NGC 1617        &  61.2   &   1.6    &  0.27    &  -21.46   &  2.06   & 16.14   &  0.15   &  0.17  \\
NGC 2196$^a$    &  45.8   &   2.8    &  1.55    &  -23.12   &  3.78   & 17.14   &  0.30   &  0.32  \\
NGC 2217        &  29.5   &   2.6    &  0.89    &  -23.24   & 10.1   & 18.59   &  0.33   &  0.34  \\
NGC 2273$^a$    &  53.6   &   1.8    &  0.36    &  -22.18   &  3.49   & 17.49   &  0.24   &  0.26  \\
NGC 2300        &  41.4   &   1.3    &  0.87    &  -23.53   &  3.92   & 16.64   &  0.26   &  0.28  \\
NGC 2460$^b$    &  41.4   &   2.6    &  0.71    &  -21.91   &  1.22   & 15.76   &  0.27   &  0.29  \\
NGC 2655        &  36.9   &   3.8    &  2.40    &  -24.42   &  3.59   & 17.38   &  0.66   &  0.68  \\
NGC 2681$^a$    &  24.2   &   2.2    &  0.12    &  -21.72   &  2.18   & 17.45   &  0.24   &  0.25  \\
NGC 2685        &  50.3   &   2.8    &  0.26    &  -21.29   &  1.85   & 17.53   &  0.27   &  0.29  \\
NGC 2768        &  66.4   &   2.5    &  1.28    &  -23.38   &  4.89   & 16.78   &  0.30   &  0.35  \\
NGC 2781$^a$    &  59.1   &   2.9    &  0.77    &  -22.60   &  7.39   & 19.23   &  0.30   &  0.33  \\
NGC 2782        &  34.2   &   3.2    &  0.57    &  -22.66   &  3.45   & 17.31   &  0.25   &  0.26  \\
NGC 2787        &  55.6   &   1.3    &  0.21    &  -21.51   &  1.49   & 15.95   &  0.18   &  0.20  \\
NGC 2855$^a$    &  33.1   &   2.8    &  2.10    &  -23.68   &  4.16   & 18.04   &  0.57   &  0.59  \\
NGC 2859$^a$    &  40.5   &   1.3    &  0.63    &  -22.99   &  9.76   & 19.35   &  0.28   &  0.29  \\
NGC 2880        &  52.0   &   2.3    &  0.70    &  -22.58   &  3.07   & 17.62   &  0.42   &  0.45  \\
NGC 2902        &  23.9   &   2.0    &  0.31    &  -22.05   &  1.89   & 17.25   &  0.32   &  0.33  \\

 \hline
\end{tabular}
\end{table}
\clearpage
\newpage

\begin{table}
\small
\centering
 \begin{tabular}{@{}lllllllll@{}}
....continue   
& & & &  &   & & &  \\
  \hline
 Galaxy           &  inc   &  $n$ & $r_{eff}$ & $M_{K}(bulge)$  & $h_r$ & $\mu_o^{obs}$   & $B/T$    & $B/T$ (corr) \\ 
                  &  [deg] &      &  [kpc]    &              & [kpc] & [mag \ arcsec$^{-2}$] &   &     \\

  \hline

NGC 2911$^a$      &  40.4   &   3.1    &  1.44    &  -23.70   &  8.57   & 18.36   &  0.34   &  0.35  \\
NGC 2950          &  53.2   &   2.4    &  0.49    &  -23.07   &  3.62   & 17.77   &  0.42   &  0.45  \\
NGC 2983$^a$      &  54.9   &   1.3    &  0.29    &  -21.45   &  3.71   & 17.42   &  0.11   &  0.12  \\
NGC 3032          &  29.9   &   2.4    &  0.10    &  -20.70   &  1.51   & 17.27   &  0.19   &  0.20  \\
NGC 3081$^b$      &  30.6   &   2.1    &  0.24    &  -21.29   &  3.30   & 17.93   &  0.10   &  0.11  \\
NGC 3100$^b$      &  52.9   &   1.8    &  0.96    &  -22.52   &  4.95   & 18.06   &  0.30   &  0.32  \\
NGC 3166          &  60.7   &   0.8    &  0.43    &  -22.85   &         &         &  0.23   &  0.26  \\
NGC 3169          &  51.9   &   2.9    &  0.94    &  -23.62   &  3.16   & 17.22   &  0.58   &  0.60  \\
NGC 3245          &  51.6   &   2.4    &  0.40    &  -22.69   &  2.29   & 16.49   &  0.33   &  0.36  \\
NGC 3358$^b$      &  54.5   &   1.7    &  0.48    &  -22.57   & 10.8   & 18.26   &  0.11   &  0.12  \\
NGC 3384          &  59.7   &   1.5    &  0.21    &  -21.50   &  1.37   & 16.22   &  0.32   &  0.35  \\
NGC 3414$^a$      &  31.1   &   2.6    &  0.54    &  -22.95   &  3.02   & 17.07   &  0.32   &  0.33  \\
NGC 3489          &  49.7   &   2.1    &  0.07    &  -20.00   &  0.52   & 15.54   &  0.23   &  0.24  \\
NGC 3516          &  38.7   &   3.4    &  0.26    &  -23.54   &  4.90   & 18.11   &  0.40   &  0.42  \\
NGC 3607          &  23.3   &   1.5    &  0.63    &  -23.22   &  1.99   & 15.98   &  0.32   &  0.33  \\
NGC 3626$^a$      &  45.9   &   1.9    &  0.34    &  -22.58   &  3.18   & 17.11   &  0.25   &  0.27  \\
NGC 3665          &  39.1   &   2.7    &  1.98    &  -24.36   &  4.61   & 17.55   &  0.58   &  0.60  \\
NGC 3706$^b$$^d$  &  50.7   &   3.5    &  6.79    &  -25.23   &           &         &  0.92   &  0.93  \\
NGC 3729          &  49.6   &   1.3    &  0.15    &  -18.97   &  1.71   & 16.61   &  0.03   &  0.04  \\
NGC 3892          &  13.1   &   2.9    &  0.40    &  -21.63   &  3.46   & 17.44   &  0.10   &  0.11  \\
NGC 3941$^a$      &  44.4   &   1.6    &  0.22    &  -22.17   &  2.03   & 16.07   &  0.17   &  0.18  \\
NGC 3945$^a$      &  47.9   &   2.9    &  0.77    &  -23.15   & 12.7   & 19.96   &  0.37   &  0.39  \\
NGC 3998          &  38.0   &   2.0    &  0.40    &  -23.23   &         &         &  0.36   &  0.38  \\
NGC 4138          &  52.4   &   1.7    &  0.08    &  -20.18   &  1.08   & 15.36   &  0.08   &  0.08  \\
NGC 4143          &  49.6   &   1.1    &  0.23    &  -21.88   &  1.01   & 15.21   &  0.27   &  0.29  \\
NGC 4150          &  47.5   &   2.8    &  0.14    &  -20.11   &  0.65   & 16.59   &  0.38   &  0.40  \\
NGC 4203          &  28.7   &   2.7    &  0.27    &  -21.46   &  1.55   & 17.31   &  0.36   &  0.38  \\
NGC 4245          &  34.6   &   1.7    &  0.22    &  -20.11   &  1.22   & 17.52   &  0.22   &  0.24  \\
NGC 4262          &  28.8   &   1.6    &  0.25    &  -22.07   &  1.13   & 16.86   &  0.54   &  0.55  \\
NGC 4267          &  18.6   &   1.5    &  0.28    &  -22.04   &  2.06   & 17.18   &  0.30   &  0.31  \\
NGC 4314          &  27.1   &   1.5    &  0.22    &  -20.40   &  1.52   & 17.34   &  0.16   &  0.17  \\
NGC 4340$^a$      &  56.2   &   3.4    &  0.78    &  -21.91   &  3.62   & 17.65   &  0.26   &  0.29  \\
NGC 4369          &  13.4   &          &          &  -21.80   &  1.54   & 17.11   &  0.34   &  0.36  \\
NGC 4371          &  59.2   &   2.6    &  0.53    &  -21.78   &  2.54   & 16.75   &  0.20   &  0.22  \\
NGC 4373          &   0.0   &   2.7    &  2.40    &  -24.95   & 10.1   & 18.66   &  0.53   &  0.54  \\
NGC 4378          &  37.2   &   2.4    &  0.83    &  -23.17   &  4.90   & 18.05   &  0.32   &  0.34  \\
NGC 4424          &  68.4   &          &          &  -19.70   &  2.39   & 17.37   &  0.10   &  0.12  \\
NGC 4435          &  45.4   &   1.3    &  0.30    &  -22.08   &  1.64   & 16.49   &  0.28   &  0.30  \\
NGC 4457          &  24.4   &   1.5    &  0.22    &  -22.07   &         &         &  0.28   &  0.29  \\
NGC 4459          &  45.6   &   3.0    &  0.49    &  -22.76   &  2.04   & 16.41   &  0.35   &  0.37  \\
NGC 4477          &  34.2   &   2.5    &  0.34    &  -21.93   &  2.04   & 16.41   &  0.17   &  0.18  \\
NGC 4546          &  58.0   &   1.8    &  0.32    &  -22.85   &  2.02   & 15.79   &  0.29   &  0.32  \\
NGC 4596$^a$      &  44.3   &   1.6    &  0.29    &  -21.67   &  3.32   & 17.34   &  0.13   &  0.14  \\
NGC 4608$^a$      &  31.1   &   1.4    &  0.27    &  -21.14   &  3.03   & 18.06   &  0.14   &  0.15  \\
NGC 4612          &  46.4   &   2.2    &  0.24    &  -20.98   &         &         &  0.21   &  0.22  \\
NGC 4643          &  38.5   &   0.8    &  0.39    &  -22.68   &  3.86   & 17.06   &  0.15   &  0.15  \\
NGC 4694          &  58.9   &   1.9    &  1.16    &  -22.88   &  2.15   & 17.98   &  0.65   &  0.68  \\
NGC 4696          &  43.9   &   1.4    &  1.56    &  -23.81   &  8.42   & 16.78   &  0.11   &  0.12  \\
NGC 4754          &  60.1   &   1.9    &  0.24    &  -21.81   &  2.65   & 16.79   &  0.18   &  0.20  \\
NGC 4772          &  60.8   &   2.8    &  0.61    &  -21.86   &         &         &  0.39   &  0.43  \\
NGC 4880          &  41.4   &   3.2    &  0.24    &  -17.51   &  1.63   & 17.64   &  0.02   &  0.02  \\
NGC 4976          &  55.6   &   2.0    &  0.24    &  -22.45   &  2.91   & 15.84   &  0.12   &  0.14  \\
NGC 4984          &  55.8   &   2.3    &  0.59    &  -22.99   &  2.41   & 16.64   &  0.41   &  0.44  \\
NGC 5026          &  55.7   &   2.7    &  1.28    &  -23.68   &  6.46   & 17.06   &  0.21   &  0.23  \\
NGC 5087$^d$      &  21.9   &   4.1    &  1.40    &  -24.58   &         &         &         &        \\
NGC 5121          &  37.0   &   2.2    &  0.24    &  -21.92   &  1.85   & 17.06   &  0.28   &  0.29  \\
NGC 5206          &  24.6   &   3.0    &  0.18    &  -17.40   &  1.10   & 18.16   &  0.09   &  0.20  \\
NGC 5266          &  54.2   &   2.8    &  1.84    &  -24.79   &  7.94   & 17.44   &  0.45   &  0.48  \\
NGC 5273          &  30.8   &   1.6    &  0.19    &  -20.63   &  1.90   & 16.90   &  0.11   &  0.11  \\

  \hline
\end{tabular}
\end{table}

\clearpage
\newpage

\begin{table}
\small
\centering
 \begin{tabular}{@{}lllllllll@{}}
continue   
& & & &  & & & & \\
  \hline

 Galaxy           &  inc   &  $n$ & $r_{eff}$ & $M_{K}(bulge)$  & $h_r$ & $\mu_o^{obs}$   & $B/T$    & $B/T$ (corr) \\ 
                  &  [deg] &      &  [kpc]    &              & [kpc] & [mag \ arcsec$^{-2}$] &   &     \\

  \hline

NGC 5333       &  59.3   &   3.8    &  0.59    &  -23.67   &  3.23   & 16.92   &  0.47   &  0.51  \\
NGC 5365       &  52.1   &   2.0    &  0.52    &  -23.08   &  9.38   & 18.48   &  0.17   &  0.19  \\
NGC 5377       &  57.4   &   1.9    &  0.52    &  -22.35   &  2.94   & 16.69   &  0.19   &  0.21  \\
NGC 5419       &  43.9   &   1.4    &  1.74    &  -24.60   &  8.57   & 17.09   &  0.24   &  0.25  \\
NGC 5473       &  36.2   &   1.9    &  0.40    &  -22.77   &  2.61   & 16.62   &  0.26   &  0.27  \\
NGC 5485       &  46.4   &   2.3    &  1.66    &  -23.51   &  4.47   & 17.89   &  0.48   &  0.51  \\
NGC 5493       &  29.8   &   2.1    &  1.02    &  -23.77   &  2.40   & 16.76   &  0.50   &  0.52  \\
NGC 5631       &  20.8   &   3.2    &  0.90    &  -23.30   &         &         &  0.45   &  0.46  \\
NGC 5636       &  25.2   &   2.1    &  0.90    &  -23.06   &  2.59   & 17.11   &  0.41   &  0.42  \\
NGC 5728       &  41.3   &   0.6    &  0.69    &  -23.08   & 10.05   & 18.81   &  0.17   &  0.18  \\
NGC 5750       &  61.0   &   4.2    &  1.11    &  -22.80   &  3.31   & 16.73   &  0.31   &  0.34  \\
NGC 5846       &  24.0   &   1.7    &  2.34    &  -24.41   &  4.74   & 17.32   &  0.46   &  0.47  \\
NGC 5898       &  28.8   &   2.2    &  0.92    &  -23.78   &  4.17   & 17.57   &  0.43   &  0.45  \\
NGC 5953       &  33.7   &   2.7    &  0.24    &  -21.99   &  1.36   & 16.32   &  0.26   &  0.28  \\
NGC 5982       &  36.5   &   2.7$^d$     &  0.30    &  -20.14   &  1.67$^c$   & 18.58$^c$   &  0.26   &  0.28  \\
NGC 6340       &  15.0   &   2.6    &  0.45    &  -21.91   &  2.62   & 17.17   &  0.19   &  0.20  \\
NGC 6407       &  35.7   &   1.5    &  1.21    &  -23.66   &  5.69   & 17.04   &  0.20   &  0.21  \\
NGC 6646       &  23.4   &   2.2    &  2.16    &  -24.16   &  7.76   & 17.59   &  0.25   &  0.26  \\
NGC 6654       &  38.1   &   1.5    &  0.33    &  -21.93   &  4.96   & 17.68   &  0.11   &  0.12  \\
NGC 6684       &  49.2   &   3.4    &  0.63    &  -21.98   &  2.15   & 17.26   &  0.36   &  0.39  \\
NGC 6703       &  14.1   &   2.5    &  0.73    &  -23.47   &  3.58   & 17.13   &  0.30   &  0.32  \\
NGC 6782       &  22.3   &   2.1    &  0.60    &  -22.95   &  4.62   & 18.11   &  0.20   &  0.21  \\
NGC 6861       &  43.0   &   2.0    &  0.96    &  -24.09   &  5.22   & 17.54   &  0.36   &  0.38  \\
NGC 6958       &  35.7   &   3.3    &  0.96    &  -23.85   &  7.89   & 18.72   &  0.43   &  0.45  \\
NGC 7029       &  53.0   &   1.8    &  0.80    &  -23.24   &         &         &  0.37   &  0.40  \\
NGC 7049       &  39.1   &   2.7    &  2.42    &  -24.81   &  6.15   & 18.81   &  0.79   &  0.81  \\
NGC 7079       &  46.0   &   1.4    &  0.39    &  -22.34   &  2.11   & 16.08   &  0.21   &  0.23  \\
NGC 7098       &  55.1   &   2.2    &  0.63    &  -22.83   &  4.99   & 17.56   &  0.25   &  0.27  \\
NGC 7192       &  15.9   &   2.7    &  1.38    &  -23.68   &  5.85   & 18.71   &  0.44   &  0.45  \\
NGC 7213       &   0.0   &   2.9    &  0.33    &  -22.76   &  2.61   & 16.56   &  0.17   &  0.18  \\
NGC 7371       &  20.7   &   2.1    &  0.49    &  -21.05   &  2.81   & 17.40   &  0.10   &  0.11  \\
NGC 7377       &  37.3   &   2.5    &  2.19    &  -23.98   &  8.21   & 17.79   &  0.26   &  0.28  \\
NGC 7457       &  57.1   &   3.1    &  0.48    &  -20.48   &  1.54   & 16.83   &  0.20   &  0.22  \\
NGC 7585       &  36.3   &   2.4    &  1.47    &  -24.00   &  6.17   & 17.47   &  0.29   &  0.31  \\
NGC 7727       &  43.1   &   3.2    &  0.94    &  -23.16   &  3.22   & 17.23   &  0.38   &  0.40  \\
NGC 7742       &  10.6   &   3.9    &  0.23    &  -21.42   &  1.15   & 16.20   &  0.22   &  0.23  \\
NGC 7743       &  40.3   &   2.5    &  0.20    &  -21.48   &  3.24   & 17.20   &  0.11   &  0.12  \\
NGC 7796       &  27.9   &   2.2    &  1.56    &  -24.26   &  5.69   & 17.84   &  0.48   &  0.49  \\

 \hline
& & & & & & & & \\
$^a$ from Laurikainen et al. 2005 & & & & & & & & \\

$^b$ from Laurikainen et al. 2006  & & & & & & & & \\

$^c$ inner disk  & & & & & & & &\\

$^d$ elliptical galaxy  & & & & & & & &\\
\end{tabular}
\end{table}

\clearpage
\newpage
\begin{table}
\small
\centering
  \caption{Mean total absolute magnitudes and the parameters of the bulge (median and mean) in different Hubble type bins.
Except for $n$ and $r_{eff}$, the values are corrected for Galactic and internal extinction.
The uncertainties are standard deviations in the bins.}
    \begin{tabular}{@{}llllll@{}}

  \hline
 & & & & & \\ 
    Type &   N  &  $<M_{tot}^o>$ & $<B/T>(corr)$                     &   $<n>$                    &   $<r_{eff}/h_r^o>$         \\   
         &      &             &  mean (median)            &  mean (median)           &  mean (median)              \\ 
  \hline 
 & & & & & \\ 
 NIRSOS: & & & & & \\ 
    -3   &    35 & -24.44      &  0.39$\pm$0.13 (0.39) &  2.230$\pm$0.62 (2.18)  &  0.24$\pm$0.100 (0.23) \\
    -2   &    35 & -24.00      &  0.33$\pm$0.14 (0.33) &  2.265$\pm$0.72 (2.38)  &  0.19$\pm$0.103 (0.18) \\
    -1   &    26 & -23.82      &  0.29$\pm$0.13 (0.29) &  2.227$\pm$0.59 (2.13)  &  0.19$\pm$0.108 (0.19) \\
     0   &    21 & -24.13      &  0.27$\pm$0.15 (0.23) &  2.074$\pm$0.74 (2.11)  &  0.20$\pm$0.139 (0.14) \\
     1   &    26 & -23.53      &  0.25$\pm$0.12 (0.26) &  2.146$\pm$0.64 (2.22)  &  0.18$\pm$0.122 (0.15) \\

         & & & &  & \\
  OSUBSGS: & & & & & \\ 
     2  &    17 & -23.91  &    0.25$\pm$0.13 (0.23)  &   1.69$\pm$0.60 (1.48)  &   0.15$\pm$0.15 (0.09) \\
     3  &    20 & -23.85  &    0.14$\pm$0.09 (0.12)  &   1.49$\pm$0.83 (1.29)  &   0.12$\pm$0.07 (0.09) \\
     4  &    38 & -23.89  &    0.11$\pm$0.08 (0.09)  &   1.44$\pm$0.77 (1.27)  &   0.15$\pm$0.13 (0.11) \\
     5  &    30 & -23.22  &    0.11$\pm$0.13 (0.06)  &   1.42$\pm$0.70 (1.31)  &   0.16$\pm$0.16 (0.08) \\
     6  &    13 & -22.03  &    0.05$\pm$0.10 (0.01)  &   1.45$\pm$1.15 (1.14)  &   0.16$\pm$0.24 (0.06) \\
     7  &     6 & -22.55  &    0.09$\pm$0.13 (0.06)  &   1.53$\pm$0.95 (2.06)  &   0.46$\pm$0.39 (0.39) \\
     8  &     2 & -22.63  &    0.05$\pm$0.06 (0.09)  &   1.15$\pm$1.23 (2.02)  &   0.12$\pm$0.10 (0.19) \\
     9  &     3 & -21.00  &    0.04$\pm$0.05 (0.00)  &   2.01$\pm$0.22 (2.07)  &   0.30$\pm$0.43 (0.06) \\
 & & & & & \\ 
 \hline
\end{tabular}
\end{table}



\begin{figure}
\includegraphics[]{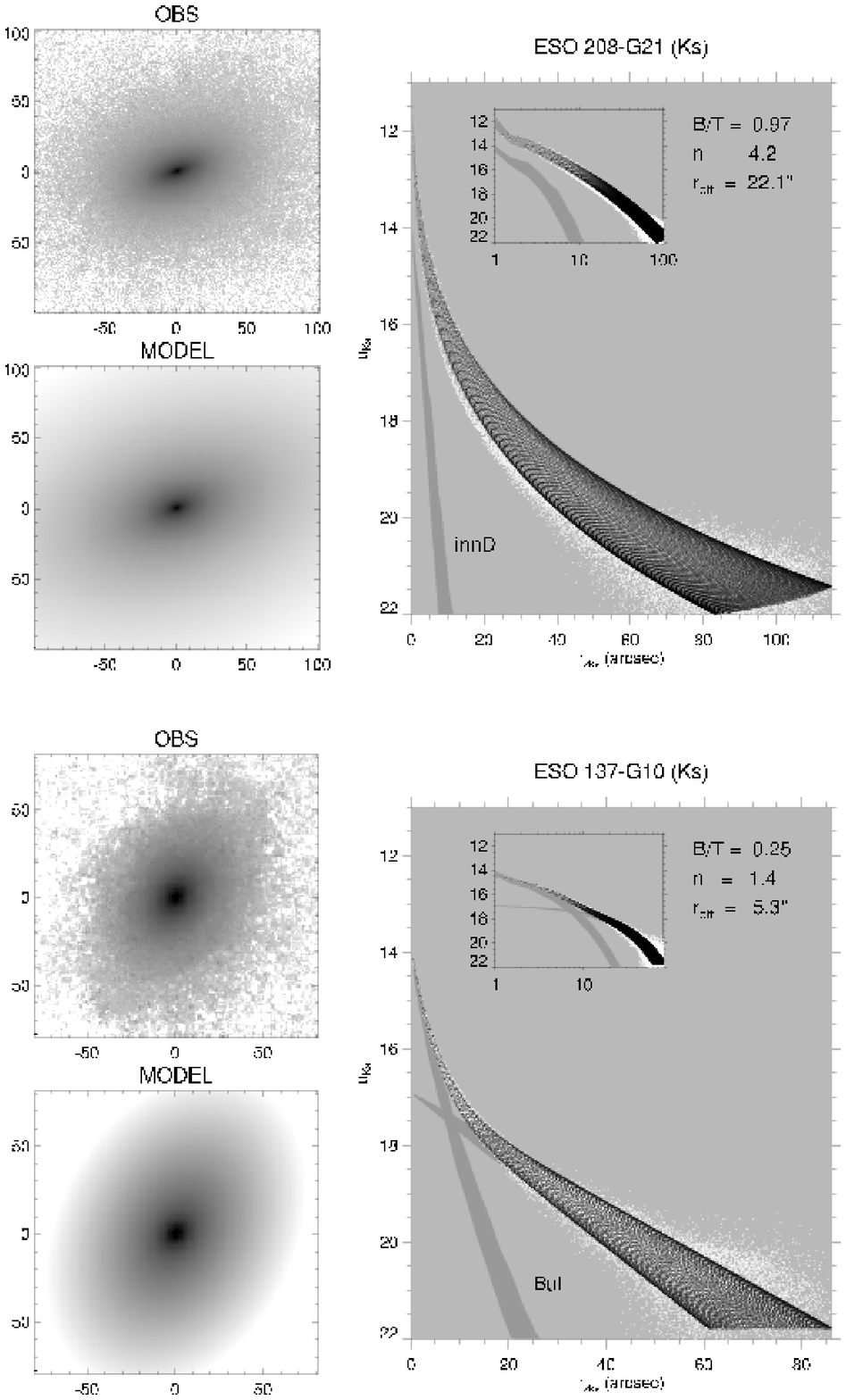}
\label{sample-figure}
\end{figure}
\newpage
\clearpage

\begin{figure}
\includegraphics[]{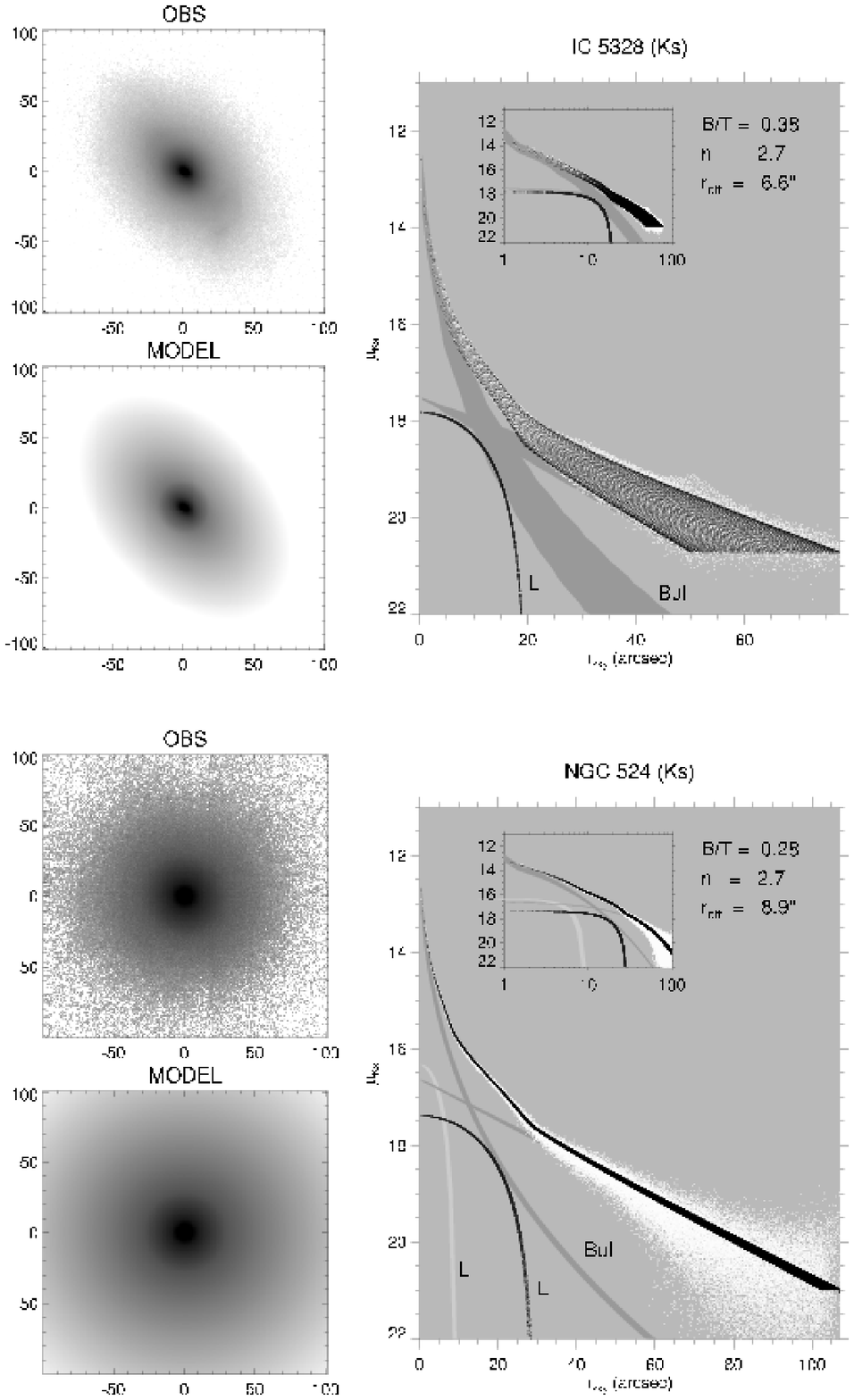}
\end{figure}
\newpage
\clearpage

\begin{figure}
\includegraphics[]{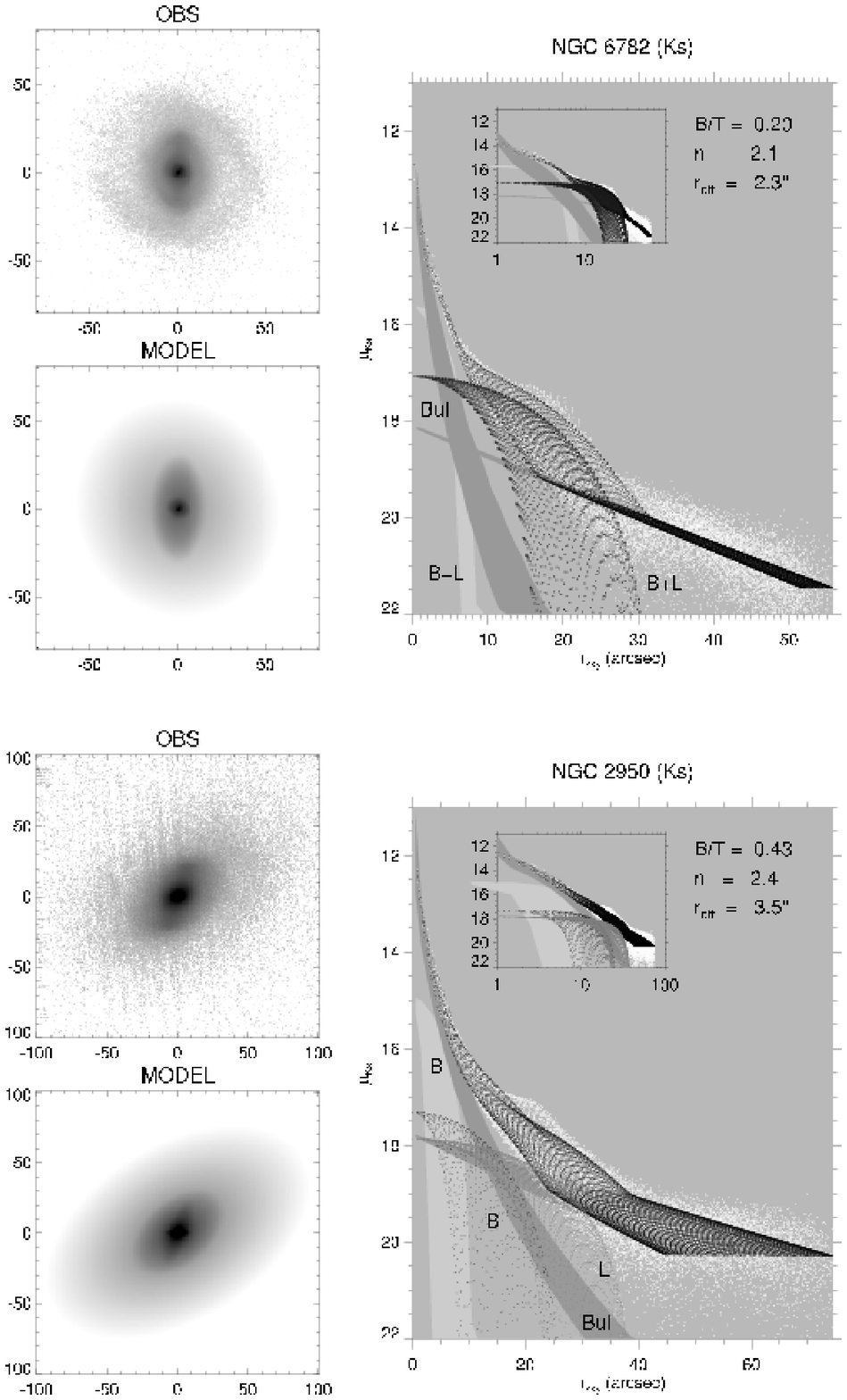}
\end{figure}

\newpage
\clearpage

\begin{figure}
\includegraphics[width=84mm]{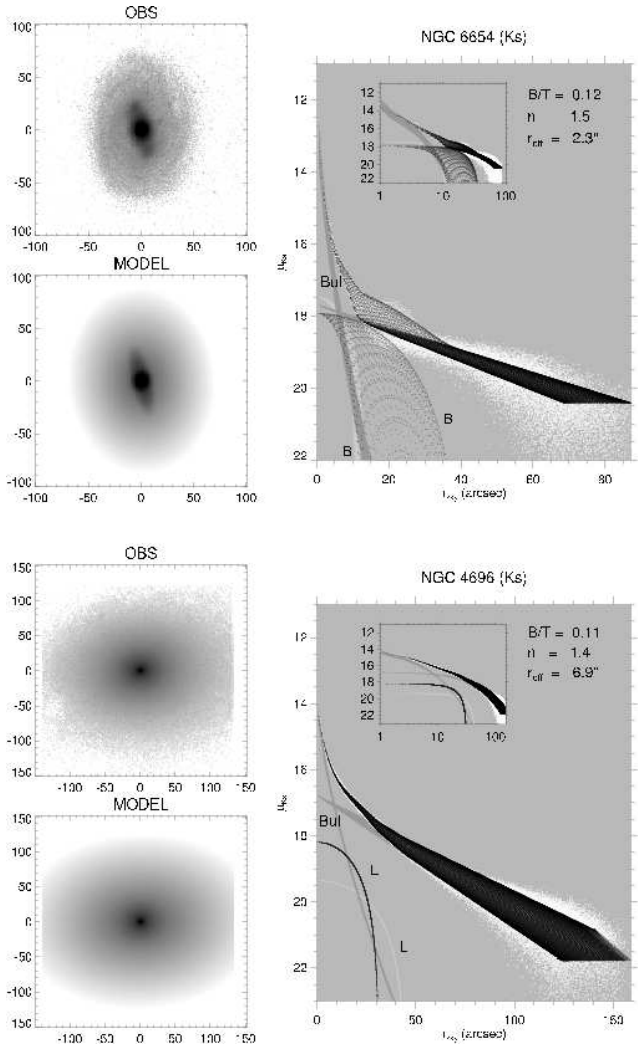}
\caption{Multi-component structural decompositions for 160 of the 175
  galaxies of our sample. In the figure representative examples for 8
  galaxies are shown (the rest of them are available in
  electronic form). {\it Right panel} shows the decomposition, whereas
  the two left panels show the observed image (above) and the model
  image (below).  White dots in the right panel show the data points
  of the 2-dimensional surface brightness distribution (brightness of
each pixel as a function of sky-plane radius from the galaxy center), and the black
  and grey colors show the model components
 The uppermost black dots show the total model.
In the figures a bulge is indicated by ``Bul'', a lens by ``L'', a bar 
by ``B''. In some cases a bar and lens are fitted by a single
function in which case the designation is ``B+L''. The small inserts
show the same decompositions in a logarithmic radial scale. 
}
\label{fig1c}
\end{figure}

\newpage
\clearpage

\begin{figure}
\includegraphics[width=120mm]{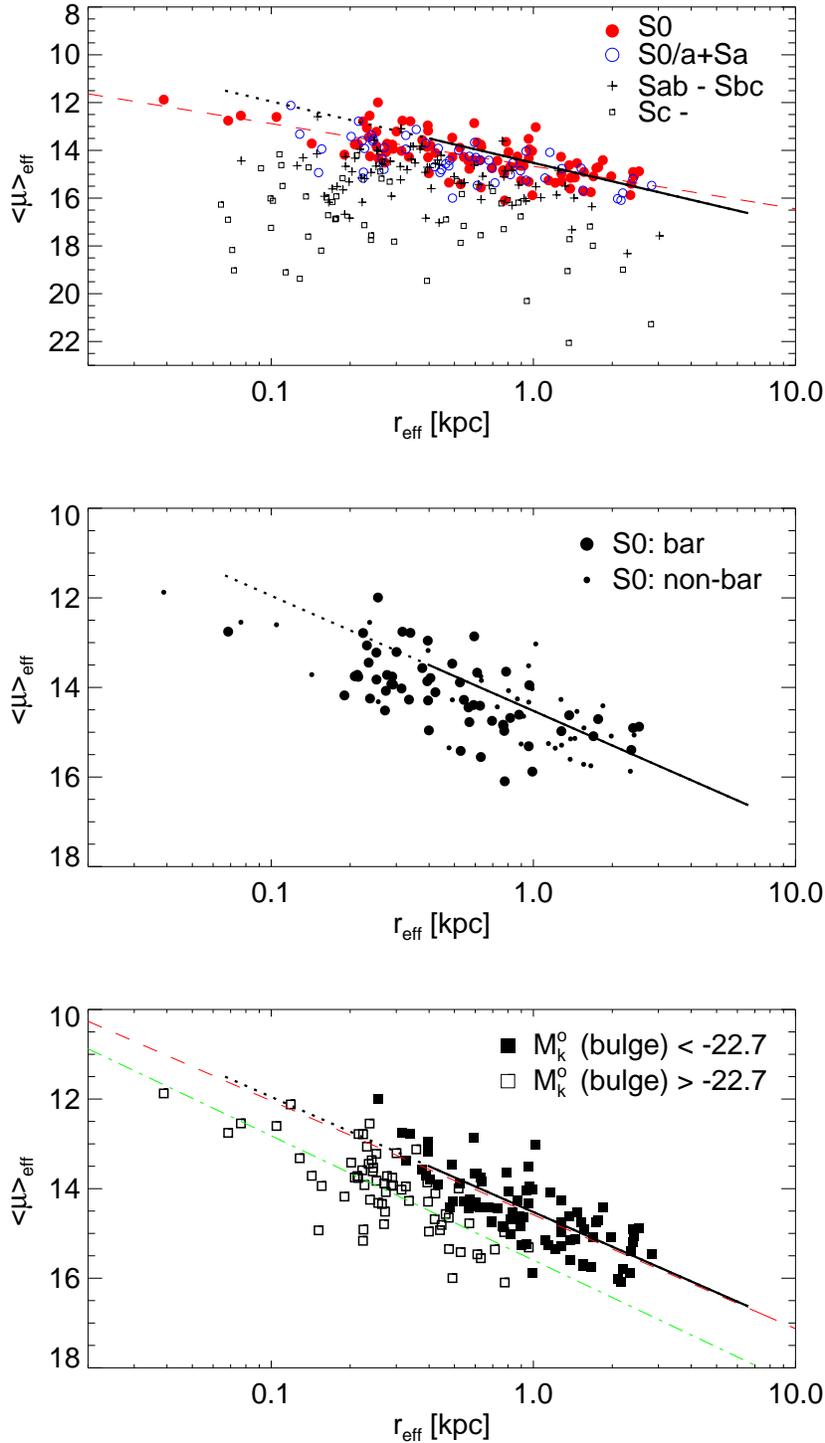} 
\caption{The Kormendy relation, which is an empirical relation between the
  effective radius of the bulge, $r_{eff}$, and the mean surface
  brightness of the bulge inside $r_{eff}$, denoted by $<\mu>_{eff}$.
  {\it The top panel} shows the relation separately for S0s (filled red
  circles), S0/a+Sa spirals (open blue circles), Sab-Sbc spirals (crosses)
and Sc and later-type spirals (squares). 
 {\it The middle panel} shows the
  barred (large filled circles) and non-barred (small dots) S0s,
  whereas {\it The lower panel} shows separately the bright (filled symbols) and dim
  (open symbols) bulges (S0-Sa) in NIRS0S. The solid line in all panels indicates the
  location of the Coma cluster ellipticals, taken from Khosroshahi et
  al. (2000): $<\mu>_{eff}$ = 2.57 $\log$ $r_{eff}$ + 14.07.  The
  dotted section of the line is an extrapolation of the Coma cluster
  ellipticals to smaller $r_{eff}$. The difference in $H_0$ between
  Khosroshahi et al. ($H_o$=50 km sec$^{-2}$ Mpc$^{-1}$) and this work
  ($H_o$=75 km sec$^{-2}$ Mpc$^{-1}$) has been taken into account in
  this figure and in Fig. 3.  In the upper panel the red dashed line
  is a fit to the S0 data points. In the lower panel the red dashed and green dash-dotted lines are
  linear fits for the bright and dim bulges, respectively.
 Note the different scale in the uppermost frame.
The same symbols are consistently used in
  the subsequent type. }
\end{figure}

\newpage
\clearpage
\begin{figure}
\includegraphics[width=120mm]{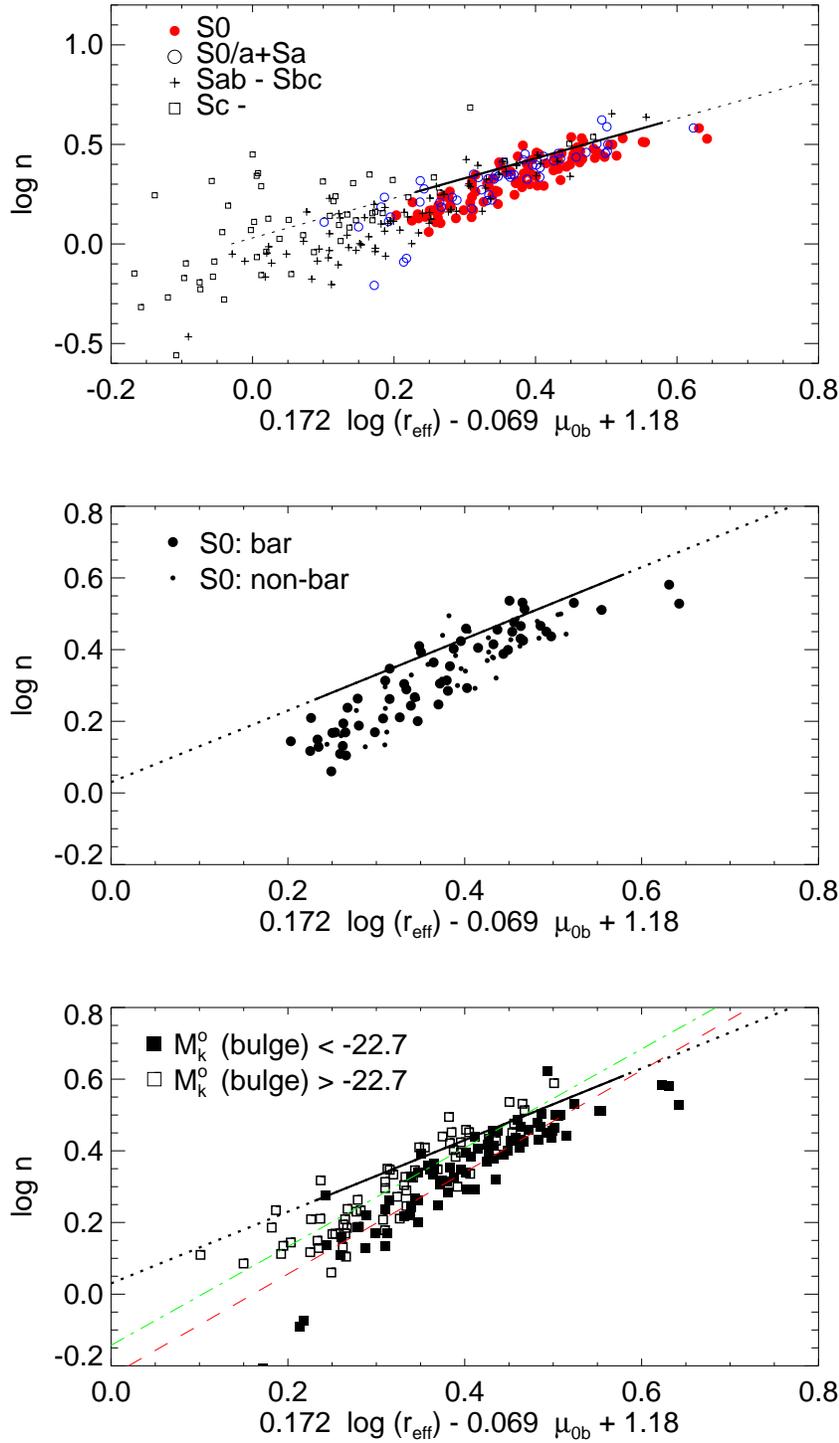}
\caption{Projections of the photometric plane, which is a relation
  between three parameters of the bulge, the S\'ersic index $n$, the
  effective radius $r_{eff}$ in [kpc], and the central surface
  brightness $\mu_{ob}$ of the bulge.  The full line indicates the best-fit plane for
  the Coma cluster ellipticals, taken from Khosroshahi et al. (2000).
  The Coma cluster ellipticals does not go through the origin
  because of the correction made to Hubble constant. The dotted line
  shows an extrapolation outside the Coma cluster ellipticals. The
  symbols are the same as in Fig. 2.  In the lower panel the red dashed and
  green dash-dotted lines indicate the best fits for the bright and 
the dim bulges in NIRS0S, respectively.  }

\end{figure}

\newpage
\clearpage
\begin{figure}
\includegraphics[width=120mm]{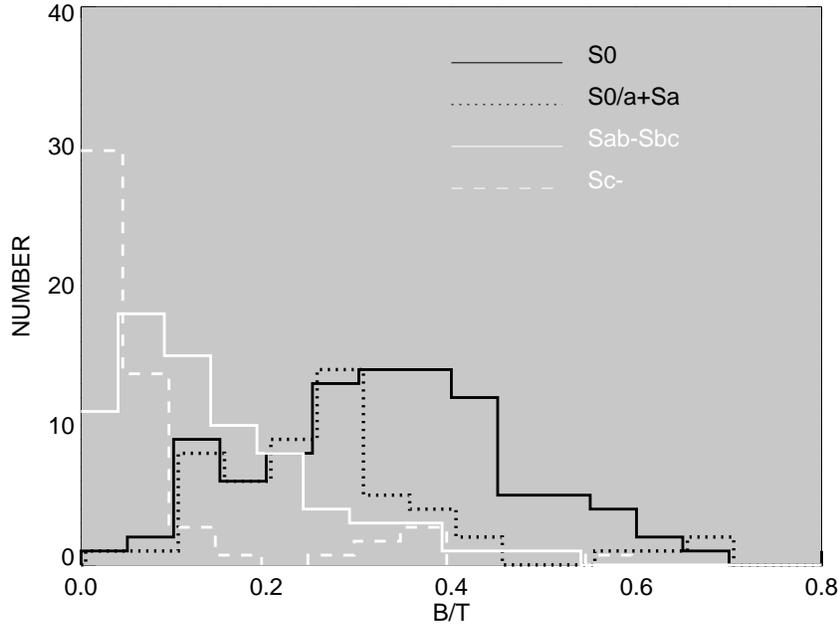}
\caption{Histograms of the $B/T$ flux ratios, corrected for
  Galactic and internal dust, as explained in Section 4 .The histograms 
are shown for the S0s
  in NIRS0S (full black line), S0/a+Sa  galaxies in NIRS0S (dotted black
  line), Sab-Sbc galaxies in OSUBSGS (white full line), and Sc and
  later types in OSUBSGS (white dashed line).  The decompositions for
  the OSUBSGS galaxies are made using $H$-band images, and for the
  NIRS0S galaxies using $K_s$-band images. However, using the
  overlapping Sa-galaxies we checked that there
  is no offset between the two samples.  }

\end{figure}

\newpage
\clearpage
\begin{figure}
\includegraphics[width=84mm]{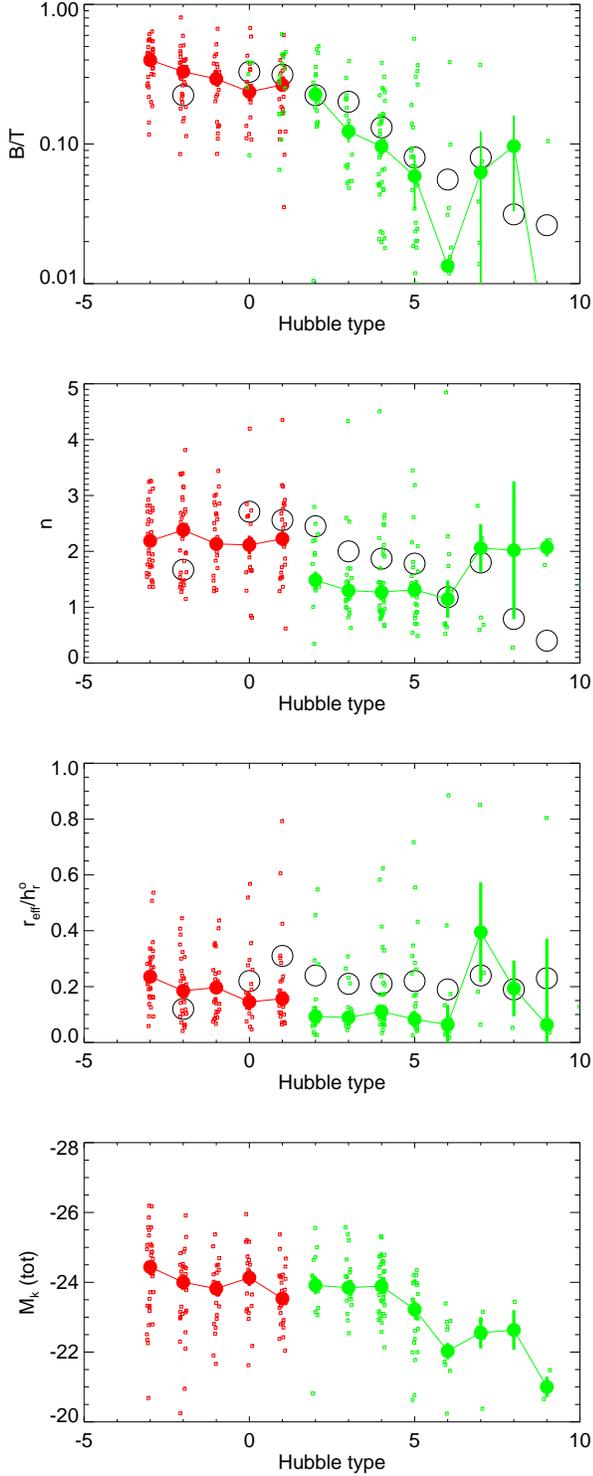}
\caption{ As a function of Hubble type are shown: (a) the dust
  bulge-to-total flux ratio ($B/T$) corrected for Galactic and internal dust, (b) S\'ersic index $n$,
  (c) the ratio of the effective radius of the bulge ($r_{eff}$)
  divided by the scale length of the disk ($h_r^o$), and (d) the total
  absolute galaxy magnitude corrected for Galactic extinction
  ($M_{tot}^o$). The {\it small symbols} show the measurements for the
  individual galaxies (with a small random horizontal offset), {\it filled circles} show the
  median values in each Hubble type bin, and {\it large open circles}
  are the median values from Graham $\&$ Worley (2008; their value
for S0s refers to types -3, -2, -1, and is here drawn at T=-2).  The error
  bars are taken to be the standard errors of the mean (= sample
  variance in each bin divided by square root of the number of galaxies
in each bin), which stresses the large number of
  galaxies in the bin.  Red colour is for the galaxies in NIRS0S and
  green for those in OSUBSGS.
}

\end{figure}

\newpage
\clearpage
\begin{figure}
\includegraphics[width=94mm]{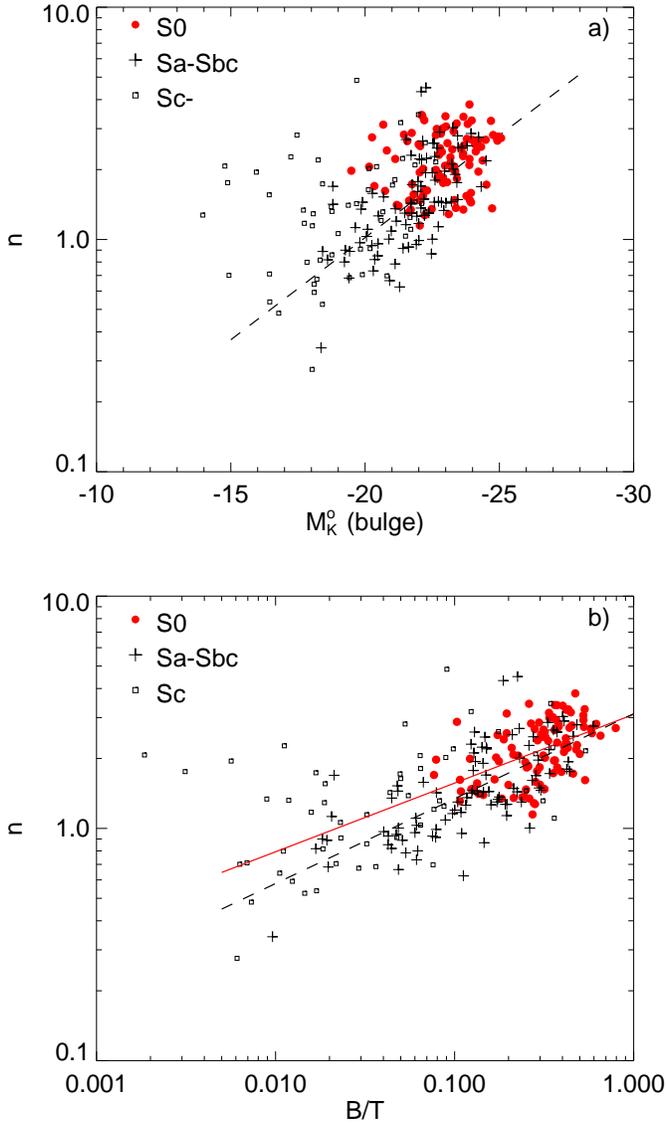}
\caption{ a) The S\'ersic index $n$ is plotted against the absolute brightness of the bulge 
corrected for Galactic and internal extinction ($M_K(bulge)$):
S0s (red filled circles) are from NIRS0S,
whereas Sa-Sbc (crosses) and Sc and later types (squares) are from
OSUBSGS. Dashed line indicates the linear fit for Sa-Sbc galaxies; for S0s
there is no statistically significant correlation (significance level of rank correlation coefficient is 0.03)
b) Dust corrected bulge-to-total flux ratio ($B/T$) is correlated with the S\'ersic
index.  The continuous and dashed lines are the linear fits for S0s and
for Sa-Sbc galaxies, respectively. Both correlations are statistically
significant. S0/a galaxies from NIRS0S fall largely on top
of S0s, and for clarity are not shown.}

\end{figure}

\newpage
\clearpage
\begin{figure}
\includegraphics[width=164mm]{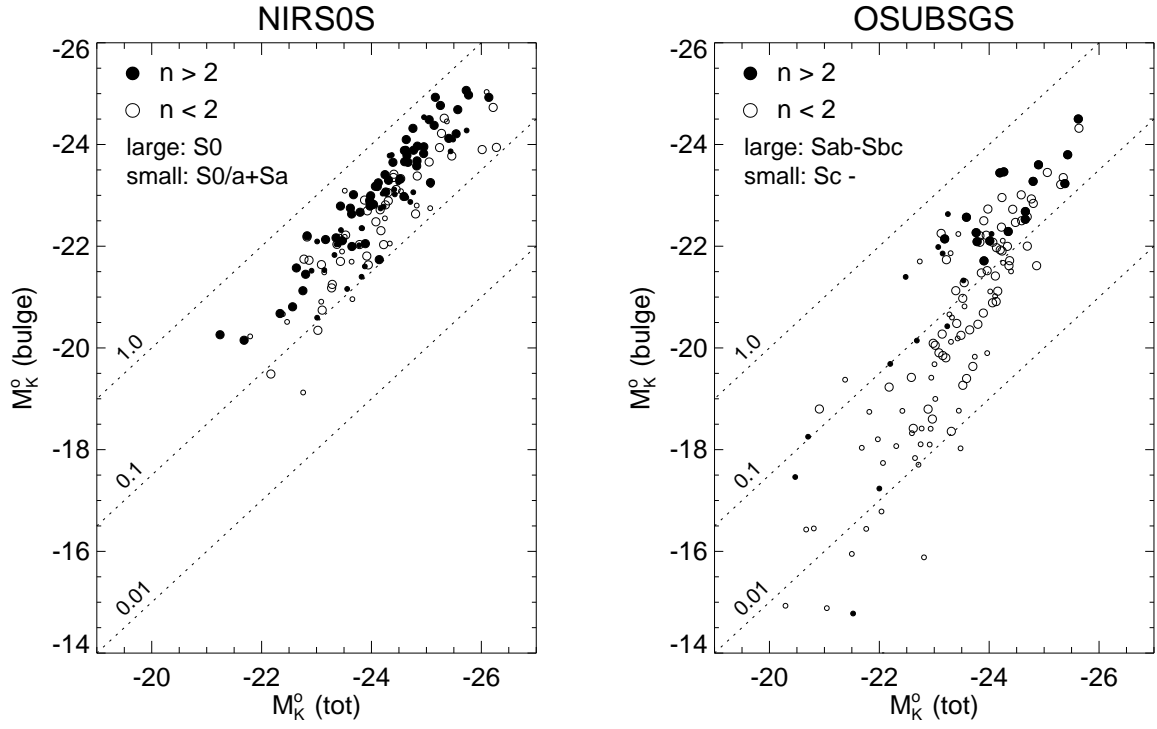}
\caption{The absolute brightness of the bulge ($M_K^o(bulge)$) is
shown as a function of the total absolute galaxy brightness ($M_{tot}^o$ ), 
both being corrected for Galactic and internal dust. The dashed lines denote 
$B/T$ = 0.01, 0.1 and 1.0. {\it Left
panel:} S0s (large symbols) and S0/a-Sa galaxies (small
symbols) in NIRS0S, both divided in two S\'ersic index bins (filled
symbols for n$>$2, open symbols for n$<$2). {\it Right panel:} 
Sa-Sbc (large symbols) and Sc and later types (small symbols), again
in two S\'ersic index bins.}
\end{figure}

\newpage
\clearpage
\begin{figure}
\includegraphics[width=94mm]{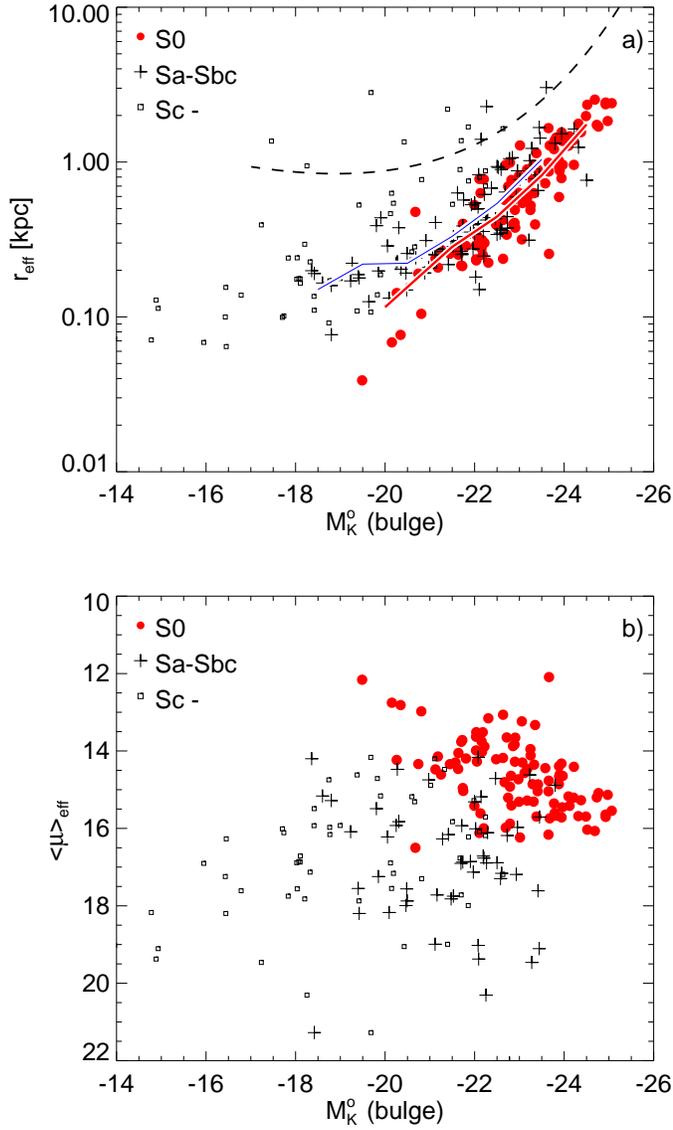}
\caption{ (a) The effective radius of the bulge ($r_{eff}$) plotted
  against the absolute brightness of the bulge ($M_K^o(bulge)$). Shown
  separately are: S0s (red filled symbols), Sa-Sbc spirals (crosses),
  and Sc and later types (squares).  The dashed line shows the
  location of the Coma cluster ellipticals, taken from Graham $\&$
  Worley (2008): log $r_{eff}$[kpc] = 1.137 + 0.217$b_n$ + $M_B$/10 +
  0.5 log [$b_n^{2n}$ / ($n\Gamma$ (2$n$)e$^{b_n}$)], where $b_n$
  $\sim$ 1.9992$n$ - 0.3271, for 0.5 $<$ $n$ $<$ 10; the value of $n$
  is expressed as: $n$ = 10$^{-(14.30+M_B)/9.4}$. The continuous 
  lines show the mean values of the data points in one magnitude bins:
  red (thicker) line is for S0s and blue (thinner) line for Sab-Sbc spirals.  (b) 
the mean effective surface brightness of the bulge ($<\mu>_{eff}$)
plotted against the absolute brightness of the bulge ($M_K^o(bulge)$). 
  }
\end{figure}

\newpage
\clearpage
\begin{figure}
\includegraphics[width=94mm]{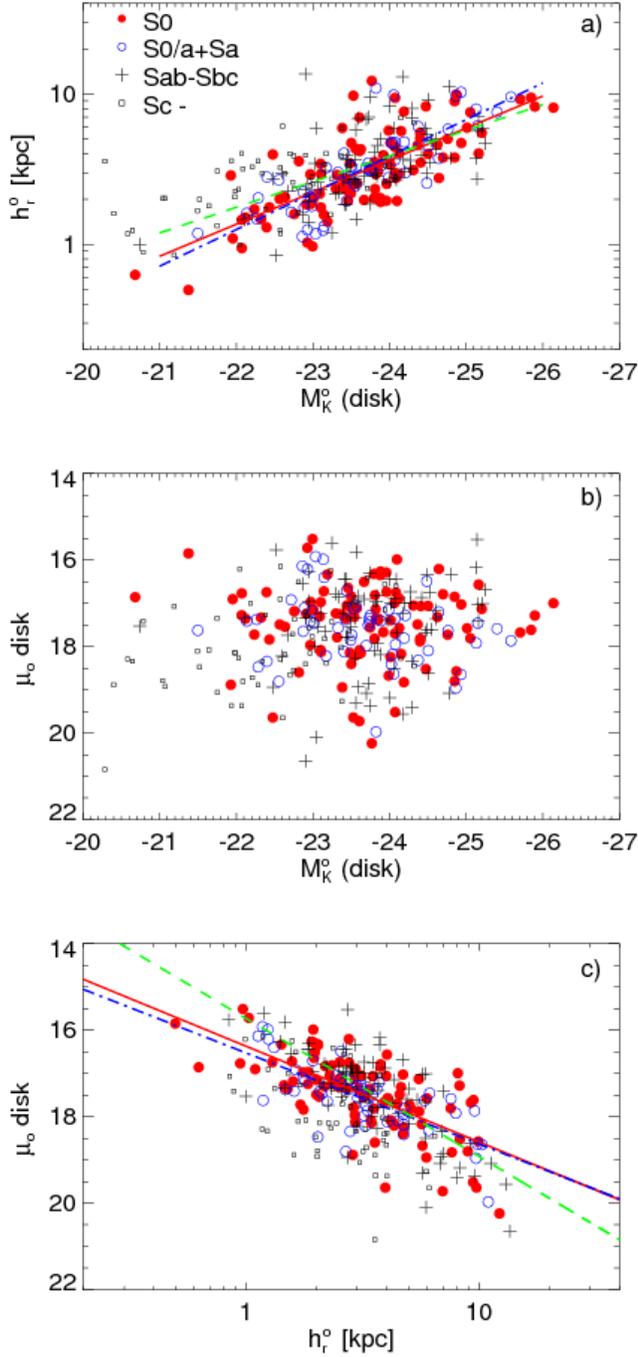}
\caption{The main parameters of the disk are
shown separately for different Hubble type bins: S0s (filled red circles), 
and S0/a-Sa galaxies (open blue circles) in NIRS0S, and Sab-Sbc (crosses)
and Sc and later types (squares) in OSUBSGS.  (a) scale length of the disk
($h_r^o$) vs. disk mass ($M_K^o(disk)$), (b) central surface brightness 
of the disk ($\mu_o$) vs. disk mass ($M_K^o(disk)$), and (c) central surface brightness of
the disk ($\mu_o$) vs. scale length of the disk ($h_r^o$).
The solid, dash-dotted and dashed lines   
in panels a) and c) indicate the best fits for S0s, S0/a+Sa, and Sab-Sbc galaxies,  
all having statistically significant rank correlation coefficients.
 }
\end{figure}

\newpage
\clearpage
\begin{figure}
\includegraphics[width=94mm]{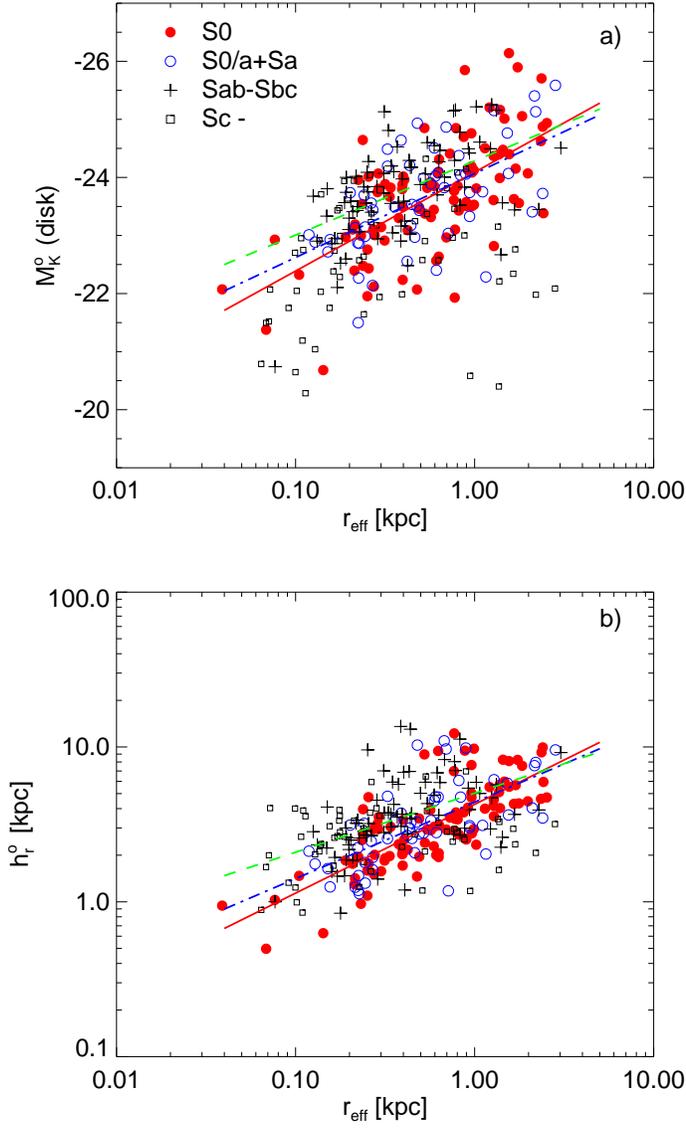}
\caption{(a) Absolute magnitude of the disk ($M_K^o(disk)$), indicative of its mass, is plotted against the  
effective radius of the bulge ($r_{eff}$),
shown in different Hubble type bins. The symbols are as in the previous figures,
and the lines are the best fits to the data points in different Hubble type bins:
S0s (red full line), S0/a-Sa (blue dash-dotted line), Sab-Sbc (green dashed line).
(b) Scale length of the disk ($h_r^o$) plotted against the effective radius 
of the bulge ($r_{eff}$).
The correlations are significant for all the other bins, except for spirals
Sc and later.
}
\end{figure}

\newpage
\clearpage
\begin{figure}
\includegraphics[width=94mm]{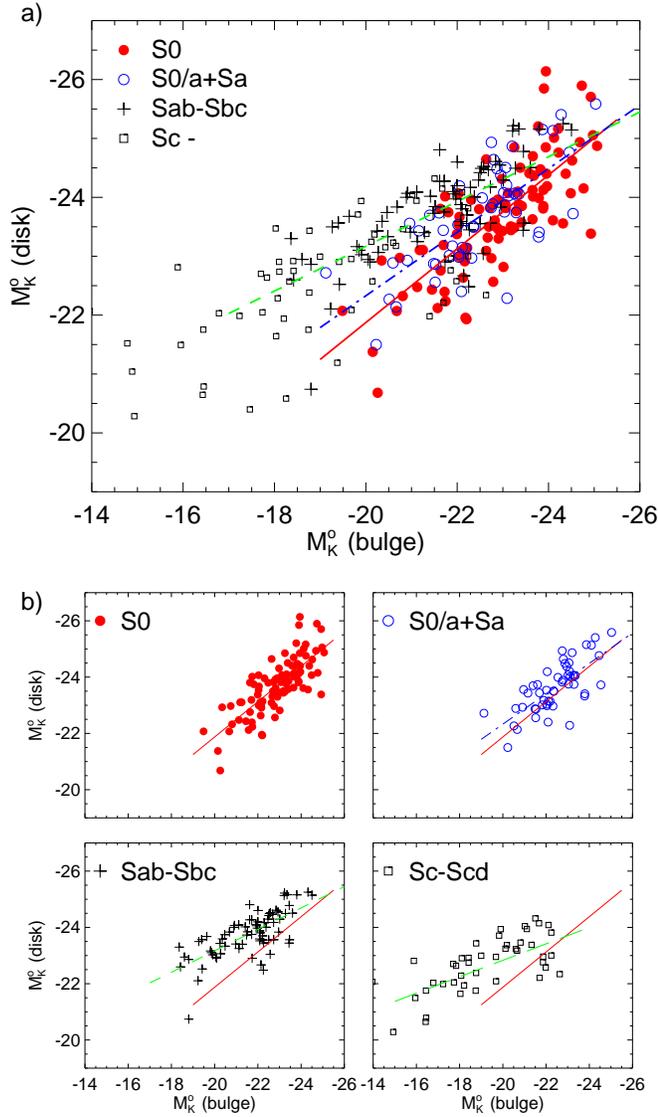}
\caption{Absolute brightness of the disk ($M_K^o(disk)$) is plotted against the
  absolute brightness  of the bulge ($M_K^o(bulge)$). (a) S0s (red
  filled circles), S0/a+Sa (open blue circles), Sab-Sbc (crosses) and
  Sc and later types (squares). Solid, dash-dotted and dashed  lines show linear fits
  for S0s, S0/a+Sa and Sab-Sbc galaxies, respectively, all correlations being 
statistically significant.  The correlation for
  the latest type galaxies is also statistically significant, but for
  clarity the fit is not shown. (b) the correlation
  is shown separately in different Hubble type bins. 
In the lower panels the solid lines indicate
  the linear fit for the S0 galaxies.}
\end{figure}


\begin{thebibliography}{99}

\bibitem[\protect\citeauthoryear{Aguerri, Balcells \& Peletier}{2001}]{aguerri2001} Aguerri J.A.L., Balcells M., Peletier R.F., 2001, AA, 367, 428
\bibitem[\protect\citeauthoryear{Aguerri et al.}{2005a}]{aguerri2005} Aguerri J.A.L., Elias-Rosa N., Corsini E., Mun\'oz-Tu\'non C., 2005a, AJ, 434, 109 
\bibitem[\protect\citeauthoryear{Aguerri et al.}{2005b}]{aguerri2005b} Aguerri, J. A. L., Iglesias-P\'aramo, J., V\'ilchez, J. M., Munoz-Tun\'on, C., S\'anchez-Janssen, R., 2005b, AJ, 130, 475
\bibitem[\protect\citeauthoryear{Aguerri \& Gonz\'alez-Garc\'ia}{2009}]{aguerri2009} Aguerri, J. A. L.; Gonz\'alez-Garc\'ia, A. C., 2009, AA, 494, 891	
\bibitem[\protect\citeauthoryear{Andredakis \& Sanders}{1994}]{andredakis1994}  Andredakis Y. C., Sanders R. H, 1994,  MNRAS, 267, 283
\bibitem[\protect\citeauthoryear{Andredakis, Peletier \& Balcells}{1995}]{andredakis1995} Andredakis Y.C., Peletier R.F., Balcells M., 1995, MNRAS, 275, 874
\bibitem[\protect\citeauthoryear{Arag\'on-Salamanca, Bedregal \& Merrifield}{2006}]{aragon2006} Arag\'on-Salamanca A., Bedregal A.G., Merrifield M.R., 2006, AA, 458, 101
\bibitem[\protect\citeauthoryear{Athanassoula \& Misioritis}{2002}]{atha2002} Athanassoula E., Misioritis A., 2002, MNRAS, 330, 35
\bibitem[\protect\citeauthoryear{Athanassoula, Lambert \& Dehnen}{2005}]{atha2005} Athanassoula E., Lambert J.C., Dehnen W., 2005, MNRAS, 363, 496
\bibitem[\protect\citeauthoryear{Balcells et al.}{2003}]{balcells2003} Balcells M., Graham A. W., Dom\'inguez-Palmero L., Peletier R.F., 2003, ApJ, 582, 79
\bibitem[\protect\citeauthoryear{Balcells, Graham \& Peletier}{2007}]{balcells2007} Balcells M., Graham A.W., Peletier R.F., 2007, ApJ, 665, 1104
\bibitem[\protect\citeauthoryear{Barnes \& Hernquist}{1992}]{barnes1992}  Barnes, J.E., Hernquist L., 1992, ARA\&A, 30, 705
\bibitem[\protect\citeauthoryear{Barr et al.}{2007}]{barr2007} Barr, J.M., Bedregal A.G., Arag\'on-Salamanca A., Merrifield M.R., Bamford S.P., 2007, AA, 470, 173
\bibitem[\protect\citeauthoryear{Bedregal, Arag\'on-Salamanca \& Merrifield}{2006}]{bedregall2006} Bedregal A., Arag\'on-Salamanca A., Merrifield M., 2006, MNRAS, 373, 1125
\bibitem[\protect\citeauthoryear{Bekki, Couch \& Shioya}{2002}]{bekki2002} Bekki K., Couch W.J., Shioya Y., 2002, ApJ, 577, 651
\bibitem[\protect\citeauthoryear{van den Bergh}{1976}]{bergh1976} van den Bergh S., 1976, ApJ, 206, 883 
\bibitem[\protect\citeauthoryear{van den Bergh}{1998}]{bergh1998} van den Bergh S., 1998, ``Galaxy Morphology and Classification''. Cambridge, Cambridge Univ. Press
\bibitem[\protect\citeauthoryear{van den Bergh}{2009}]{bergh2009} van den Bergh S., 2009, ApJ, 694, 120
\bibitem[\protect\citeauthoryear{Bournaud \& Combes}{2002}]{bournaud2002} Bournaud F., Combes F., 2002, AA, 392, 83
\bibitem[\protect\citeauthoryear{Bournaud, Combes \& Semelin}{2005}]{bournaud2005} Bournaud F., Combes F., Semelin B., 2005, MNRAS, 364, 18
\bibitem[\protect\citeauthoryear{Burstein}{1979}]{burstein1979} Burstein D., 1979, ApJ, 234, 829
\bibitem[\protect\citeauthoryear{Burstein}{2005}]{burstein2005} Burstein D., Ho, L. C., Huchra, J. P., Macri, L. M., 2005, ApJ, 621, 246
\bibitem[\protect\citeauthoryear{Buta et al.}{2006}]{buta2006} Buta R., Laurikainen E., Salo H., Block D. L., Knapen J. H., 2006, AJ, 132, 1859
\bibitem[\protect\citeauthoryear{Buta, Corwin \& Odewahn}{2007}]{buta2007} Buta R., Corwin H., Odewahn S., 2007, ``The de Vaucouleurs Atlas of Galaxies'', Cambridge, Cambridge University Press (BCO)
\bibitem[\protect\citeauthoryear{Burkert et al.}{2008}]{burkert2008} Burkert A., Naab T., Johansson P.H., Roland J., 2008, ApJ, 685, 897
\bibitem[\protect\citeauthoryear{Cameron et al.}{2009}]{cameron2009}Cameron E., Driver S. P., Graham A. W., Liske J., 2009, ApJ, 699, 105
\bibitem[\protect\citeauthoryear{Caon, Capaccioli \& D'Onofrio}{1993}]{caon1993} Caon N., Capaccioli M., D'Onofrio M., 1993, MNRAS, 265, 1013
\bibitem[\protect\citeauthoryear{Capaccioli \& Caon}{1991}]{capacc1991} Capaccioli M., Caon, N., 1991, MNRAS, 248, 523
\bibitem[\protect\citeauthoryear{Capaccioli, Caon \& D'Onofrio}{1992}]{capaccioli1992} Capaccioli M., Caon N., D'Onofrio M., 1992, MNRAS, 259, 323
\bibitem[\protect\citeauthoryear{Cardelli, Clayton \& Mathis}{1989}]{cardelli1989} Cardelli J.A., Clayton G.C., Mathis, J.S., 1989, ApJ, 345, 245
\bibitem[\protect\citeauthoryear{Cappellari et al.}{2007}]{cappellari2007} Cappellari, M. et al., 2007, MNRAS, 379, 418 
\bibitem[\protect\citeauthoryear{Carollo et al.}{1997}]{carollo1997} Carollo C. M., Stiavelli M., de Zeeuw P. T., Mack J., 1997, AJ, 491, 434
\bibitem[\protect\citeauthoryear{Carollo \& Stiavelli}{1998}]{carollo1998} Carollo C. M., Stiavelli M., 1998, AJ, 115, 2306
\bibitem[\protect\citeauthoryear{Carollo et al.}{2007}]{carollo2007} Carollo C.M., Scarlata C., Stiavelli M., Wyse R.F.G., Mayer L., 2007, ApJ, 658, 960
\bibitem[\protect\citeauthoryear{Courteau, de Jong \& Broeils}{1996}]{courteau1996} Courteau, S., de Jong R. S., Broeils A.H., 1996, ApJL, 457, 73
\bibitem[\protect\citeauthoryear{Dalcanton, Spergel \& Summers}{1997}]{dalcanton1997} Dalcanton, J. J., Spergel, D. N., Summers, F. J., 1997, ApJ, 482, 659 
\bibitem[\protect\citeauthoryear{Debattista et al.}{2006}]{debattista2006} Debattista V.P., Mayer L. Carollo C.M., Moore B. Wadsley J., Quinn T., 2006, ApJ, 645, 209
\bibitem[\protect\citeauthoryear{Djorgovski \& Davis}{1987}]{djor1987} Djorgovski S., Davis M., 1987, ApJ, 313, 59 
\bibitem[\protect\citeauthoryear{Dom\'inguez-Palmero \& Balcells}{2009}]{palmero2009} Dom\'inguez-Palmero, L., Balcells M., 2009, ApJ, 694, 69
\bibitem[\protect\citeauthoryear{D'Onghia et al.}{2006}]{ong2006} D'Onghia E., Burkert A., Murante G., Khochfar S., 2006, MNRAS, 372, 1525
\bibitem[\protect\citeauthoryear{Dressler}{1980}]{dressler1980} Dressler A., 1980, ApJ, 236, 351
\bibitem[\protect\citeauthoryear{Dressler et al.}{1987}]{dressler1987} Dressler A., Lynden-Bell D., Burstein D., Davies R.L., Faber S.M., Terlevich R., Wegner G., 1987, ApJ, 313, 42 
\bibitem[\protect\citeauthoryear{Driver et al.}{2006}]{driver2006} Driver, S. P., Allen, P. D., Graham, A. W., Cameron, E., Liske, J., Ellis, S. C., Cross, N. J. G., de Propris, R., P., S., Couch, W. J., 2006, MNRAS, 368, 414
\bibitem[\protect\citeauthoryear{Driver et al.}{2008}]{driver2008} Driver S.P., Popescu C.C., Tuffs R. J., Graham A. W., Liske J., Baldry I., 2008, ApJ, 678, L101
\bibitem[\protect\citeauthoryear{Eliche-Moral et al.}{2006}]{moral2006} Eliche-Moral M.C., Balcells M., Aguerri J.A.L., Gonz\'alez-Garc\'\i a A.C., 2006, AA, 457, 91
\bibitem[\protect\citeauthoryear{Erwin \& Sparke}{2002}]{erwin2002} Erwin P., Sparke, L.S., 2002, AJ, 124, 65
\bibitem[\protect\citeauthoryear{Erwin et al.}{2003}]{erwin2003} Erwin P., Beltr\'an, J.C.V., Graham A.W., Beckman J.E., 2003, ApJ, 597, 929
\bibitem[\protect\citeauthoryear{Erwin}{2004}]{erwin2004} Erwin P., 2004, AA, 415, 941
\bibitem[\protect\citeauthoryear{Erwin, Beckman \& Pohlen}{2005}]{erwin2005} Erwin P., Beckman J.E., Pohlen M., 2005, ApJ, 626, 81
\bibitem[\protect\citeauthoryear{Eskridge et al.}{2002}]{esk2002} Eskridge P.B. et al., 2002, ApJS, 143, 73
\bibitem[\protect\citeauthoryear{Freeman}{1970}]{freeman1970} Freeman K.C., 1970, ApJ, 160, 811
\bibitem[\protect\citeauthoryear{Firmani \& Avila-Rees}{2000}]{firmani2000} Firmani C., Avila-Reese V., 2000 MNRAS, 315, 457 
\bibitem[\protect\citeauthoryear{Fisher, Franx \& Illingworth}{1996}]{fisher1996} Fisher D., Franx M., Illingworth G., 1996, ApJ, 459, 110
\bibitem[\protect\citeauthoryear{Gadotti}{2008}]{gadotti2008} Gadotti D., 2008, MNRAS, 384, 420 
\bibitem[\protect\citeauthoryear{Geach et al.}{2009}]{geach2009} Geach J.E., Smail I., Moran S.M., Treu T., Ellis R., 2009, ApJ, 691, 783
\bibitem[\protect\citeauthoryear{Gerken et al.}{2004}]{gerken2004} Gerken B., Ziegler B., Balogh M., Gijbank D., Friz A., Jager K., 2004, AA, 421, 59
\bibitem[\protect\citeauthoryear{Graham \& Prieto}{1999}]{graham1999} Graham A., Prieto M., 1999, ApJSS, 269, 653
\bibitem[\protect\citeauthoryear{Graham}{2001a}]{graham2001a} Graham A., 2001a, MNRAS, 326, 543
\bibitem[\protect\citeauthoryear{Graham}{2001b}]{graham2001b} Graham A., 2001b, AJ, 121, 820
\bibitem[\protect\citeauthoryear{Graham \& de Blok}{2001}]{grahamblok2001} Graham A., de Blok W.J., 2001, ApJ, 556, 177
\bibitem[\protect\citeauthoryear{Graham}{2002}]{graham2002} Graham A., 2002, MNRAS, 334, 859
\bibitem[\protect\citeauthoryear{Graham \& Worley}{2008}]{graham2008} Graham A., Worley C.C., 2008, MNRAS, 388, 1708 (GW2008) 
\bibitem[\protect\citeauthoryear{Gunn \& Gott}{1972}]{gun1972} Gunn J.E., Gott J.R., 1972, ApJ, 176, 1
\bibitem[\protect\citeauthoryear{Hubble}{1926}]{hubble1926} Hubble E., 1926, ApJ, 64, 321
\bibitem[\protect\citeauthoryear{Hunt, Pierini \& Giovanardi}{2004}]{hunt2004} Hunt L.K., Pierini D., Giovanardi C., 2004, AA, 414, 905
\bibitem[\protect\citeauthoryear{Illingworth}{1981}]{ill1981} Illingworth, G. 1981,  in ``The structure and evolution of normal galaxies'', Proceedings of the Advanced Study Institute, Cambridge, England, August 3-15, 1980. (A82-11951 02-90) Cambridge and New York, Cambridge University Press, 1981, p. 27-41. 
\bibitem[\protect\citeauthoryear{de Jong}{1996}]{dejong1996} de Jong R.S., 1996, AA, 313, 45
\bibitem[\protect\citeauthoryear{Kauffman et al.}{1999}]{kauffmann1999} Kauffmann G., Golberg J., Diaferio A., White S., 1999, MNRAS, 303, 188
\bibitem[\protect\citeauthoryear{Kautsch et al.}{2006}]{kautsch2006} Kautsch S.J., Grebel E.K., Barazza F D., Gallagher J.S., 2006, AA, 445, 765
\bibitem[\protect\citeauthoryear{Kent}{1985}]{kent1985} Kent S.M., 1985, ApJS, 59, 115
\bibitem[\protect\citeauthoryear{Khochfar \& Burkert}{2005}]{burk2005} Khochfar S., Burkert A., 2005, MNRAS, 359, 1319
\bibitem[\protect\citeauthoryear{Khochfar \& Silk}{2006}]{burk2006} Khochfar S., Silk J., 2006, MNRAS, 370, 902
\bibitem[\protect\citeauthoryear{Khosroshahi, Wadadekar \& Kembhavi}{2000}]{khosroshahi2000} Khosroshahi H., Wadadekar Y., Kembhavi A., 2000, ApJ, 533, 162
\bibitem[\protect\citeauthoryear{Khosroshahi et al.}{2000}]{khosroshahietal2000} Khosroshahi H., Wadadekar Y., Kembhavi A., Mobasher, B., 2000, ApJL, 531, 103
\bibitem[\protect\citeauthoryear{Khosroshahi et al.}{2004}]{khosroshahietal2004} Khosroshahi H., Raychaudhury, S., Ponman T.J. et al., 2004, MNRAS, 349, 527
\bibitem[\protect\citeauthoryear{Kodaira, Watanabe \& Okamura}{1986}]{kodaira1986} Kodaira K., Watanabe M., Okamura S., 1986, ApJS, 62, 703
\bibitem[\protect\citeauthoryear{Kormendy}{1977}]{kormendy1977} Kormendy J., 1977, ApJ, 218, 333
\bibitem[\protect\citeauthoryear{Kormendy}{1979}]{kormendy1979} Kormendy J., 1979, ApJ, 227, 714
\bibitem[\protect\citeauthoryear{Kormendy}{1982}]{kormendy1982} Kormendy J., 1982, ApJ, 257, 75
\bibitem[\protect\citeauthoryear{Kormendy}{1984}]{kormendy1984} Kormendy J., 1984, ApJ, 286, 116
\bibitem[\protect\citeauthoryear{Kormendy \& Kennicutt}{2004}]{kormendy2004} Kormendy J.,  Kennicutt R., 2004, ARA\&A, 42, 603
\bibitem[\protect\citeauthoryear{Laine et al.}{2002}]{laine2002} Laine S., Shlosman I., Knapen J.H., Peletier, Reynier F., 2002, ApJ, 567, 97
\bibitem[\protect\citeauthoryear{Laurikainen \& Salo}{2000}]{lauri2000} Laurikainen E., Salo H., 2000, AA, 141, 103
\bibitem[\protect\citeauthoryear{Laurikainen et al.}{2004}]{lauri2004} Laurikainen E., Salo H., Buta R., Vasylyev S. 2004 MNRAS, 355, 1251
\bibitem[\protect\citeauthoryear{Laurikainen, Salo \& Buta}{2005}]{lauri2005} Laurikainen E., Salo H., Buta R., 2005, MNRAS, 362, 1319 
\bibitem[\protect\citeauthoryear{Laurikainen et al.}{2006}]{lauri2006} Laurikainen E., Salo H., Buta R., Knapen J. H., Speltincx T., Block D., 2006, AJ, 132, 2634
\bibitem[\protect\citeauthoryear{Laurikainen et al.}{2007}]{lauri2007} Laurikainen E., Salo H., Buta R., Knapen J. H., 2007, MNRAS, 381, 401
\bibitem[\protect\citeauthoryear{Laurikainen et al.}{2009}]{lauri2009} Laurikainen E., Salo H., Buta R., Knapen J. H., 2009, ApJL, 692, L34
\bibitem[\protect\citeauthoryear{Lisker et al.}{2006}]{liske2006} Lisker T., Grebel E. K., Binggeli B., 2006, AJ, 132, 497 
\bibitem[\protect\citeauthoryear{MacArthur, Courteau \& Holtzman}{2003}]{arthur2003}  MacArthur L.A., Courteau S., Holtzman J.A., 2003, ApJ, 582, 689
\bibitem[\protect\citeauthoryear{MacArthur et al.}{2004}]{arthur2004} MacArthur L.A.,Courteau, S., Bell, E., Holtzman, J.A., 2004, ApJS, 152, 175
\bibitem[\protect\citeauthoryear{MacArthur et al.}{2008}]{arthur2008} MacArthur L.A., Ellis R.S., Treu T.U, Vivian U., Bundy K. Moran S., 2008, ApJ, 680, 70
\bibitem[\protect\citeauthoryear{Mastropietro et al.}{2005}]{mastro2005} Mastropietro C., Moore B., Mayer L., Debattista V. P., Piffaretti R., Stadel J., 2005, MNRAS, 364, 607
\bibitem[\protect\citeauthoryear{M\'endez-Abreu et al.}{2008}]{abreu2008} M\'endez-Abreu J., Aguerri J.A.L., Corsini E.M., Simonneau E., 2008, AA, 487, 555
\bibitem[\protect\citeauthoryear{Michard \&  Marchal}{1994}]{michard1994} Michard R., Marchal J., 1994, AAS, 105, 481
\bibitem[\protect\citeauthoryear{Moran et al.}{2007}]{moran2007} Moran S.M., Loh B.L., Ellis R.S., Treu T., Bundy K., MacArthur L.A., 2007, ApJ, 665, 1067
\bibitem[\protect\citeauthoryear{Moore et al.}{1996}]{moore1996} Moore B., Katz N., Lake G., Dressler A., Oemler A., 1996, Nature, 379, 613
\bibitem[\protect\citeauthoryear{M\"ollenhoff \& Heidt}{2001}]{moll2001} M\"ollenhoff C., Heidt J., 2001, AA, 368, 16 
\bibitem[\protect\citeauthoryear{Naab, Burkert \& Hernquist}{1999}]{naab1999} Naab T., Burkert A., Hernquist L., 1999, ApJ, 523, 133 
\bibitem[\protect\citeauthoryear{Naab \& Burkert}{2003}]{naab2003} Naab T., Burkert A., 2003, ApJ, 523, 133
\bibitem[\protect\citeauthoryear{Naab \& Trujillo}{2006}]{naabtrujillo2006} Naab T., Trujillo I., 2006, MNRAS, 369, 625
\bibitem[\protect\citeauthoryear{Naab, Khochfar \& Burkert}{2006}]{naab2006} Naab T., Khochfar S., Burkert A., 2006, ApJ, 636, 81
\bibitem[\protect\citeauthoryear{Naab \& Ostriker}{2009}]{naab2009} Naab T., Ostriker J.P., 2009, ApJ, 690, 1452
\bibitem[\protect\citeauthoryear{di Nino et al.}{2009}]{nino2009} di Nino D., Trenti M., Stiavelli M., Carollo C. M., Scarlata C., Wyse R. F. G., 2009, AJ, 138, 1296
\bibitem[\protect\citeauthoryear{Noordermeer \&  van der Hulst}{2007}]{noordermeer2007} Noordermeer E., van der Hulst J.M., 2007, MNRAS, 376, 1480
\bibitem[\protect\citeauthoryear{Pahre, Djorgovski \& Carvalho}{1998}]{pahre1998} Pahre M.A., Djorgovski S., de Carvalho R.R., 1998, AJ, 116, 1591
\bibitem[\protect\citeauthoryear{Peletier \& Balcells}{1996}]{pel1996} Peletier R. F., Balcells M., 1996, AJ, 111, 2238
\bibitem[\protect\citeauthoryear{Peng}{2002}]{peng2002} Peng, C. 2002, AJ, 124, 294
\bibitem[\protect\citeauthoryear{Pierini et al.}{2002}]{pierini2002} Pierini D., Gavazzi G., Franzetti P., Scodeggio M., Boselli A., 2002, MNRAS, 332, 422
\bibitem[\protect\citeauthoryear{Poggianti et al.}{1999}]{poggianti1999} Poggianti, B.M., Smail I., Dressler A., Couch W.J., Barger W.J., Butcher H., Ellis R.C., Oemler A.J., 1999, ApJ, 518, 576
\bibitem[\protect\citeauthoryear{Prieto et al.}{1997}]{prieto1997} Prieto M., Gottesman S.T., Aguerri J.L., Varela A.M., Munoz-Tunon C., 1997, AJ, 114, 1413
\bibitem[\protect\citeauthoryear{Prieto et al.}{2001}]{prieto2001} Prieto M.,  Aguerri J.L., Varela A.M., Munoz-Tunon C., 2001, AA, 367, 405
\bibitem[\protect\citeauthoryear{Ravikumar et al.}{2006}]{ravikumar2006} Ravikumar C.D., Barway S., Kembhavi A., Mobasher B., Kuriakose V.C., 2006, AA, 446, 827
\bibitem[\protect\citeauthoryear{Reese et al.}{2007}]{reese2007} Reese A.S., Williams T.B., Sellwood J.A., Barnes E.I., Powell B.A., 2007, AJ, 133, 2846
\bibitem[\protect\citeauthoryear{Salo et al.}{2009}]{salo2009} Salo H., Laurikainen E., Buta R., Knapen J.H., 2009 (SLBK, in preparation)
\bibitem[\protect\citeauthoryear{Sandage}{1961}]{sandage1961} Sandage A., 1961, ``Hubble Atlas of Galaxies'', Washington, DC: Carnegie Inst. Wash.Publ. 618
\bibitem[\protect\citeauthoryear{Sandage \& Tammann}{1981}]{sandage1981} Sandage A. R., Tammann G.A., 1981, ``Revised Shapley Ames Catalog of Bright Galaxies'', Carnegie Institute of Washington (RSA)
\bibitem[\protect\citeauthoryear{Sandage \& Bingelli}{1984}]{sandage1984} Sandage A., Bingelli B., 1984, AJ, 89, 919
\bibitem[\protect\citeauthoryear{Sandage}{2005}]{sandage2005} Sandage A, 2005, ARA\&A, 43, 581
\bibitem[\protect\citeauthoryear{Sandage \& Bedke}{1994}]{sandage1994} Sandage A., Bedke J., 1994, ``The Carnegie Atlas of Galaxies''. Washington, DC: The Carnegie Inst. Washington (CAG)
\bibitem[\protect\citeauthoryear{Sarzi et al.}{2007}]{sarzi2007} Sarzi, M.; Allard, E. L.; Knapen, J. H.; Mazzuca, L. M., 2007, MNRAS, 380, 949  
\bibitem[\protect\citeauthoryear{Scannappieco \& Tissera}{2003}]{tissera2003} Scannapieco C., Tissera P.B., 2003, MNRAS, 338, 880
\bibitem[\protect\citeauthoryear{Schlegel, Finkbeiner \& Davis}{1998}]{shleg1998} Schlegel D.J., Finkbeiner D.P., Davis M., 1998, MNRAS, ApJ, 500, 525
\bibitem[\protect\citeauthoryear{Scodeggio, Giovanelli \& Haynes}{1997}]{scodeggio1997} Scodeggio M., Giovanelli R., Haynes M.P., 1997, AJ, 113, 2087
\bibitem[\protect\citeauthoryear{S\'ersic}{1963}]{sersic1963} S\'ersic J.L., 1963, in  Boletin de la Asociacion Argentina de Astronomia, vol. 6, p.41
\bibitem[\protect\citeauthoryear{S\'ersic}{1968}]{sersic1968} S\'ersic J.L., 1968, in Atlas de Galaxies Australes (Cordoba: Observatorio Astronomico)
\bibitem[\protect\citeauthoryear{Skrutskie et al.}{2006}]{skrutskie2006} Skrutskie M.F., Cutri R.M., Stiening R., Weinberg M.D., Schneider S., Carpenter J.M., Beichman C., Capps R., Chester T., Elias J., and 21 coauthors, 2006, AJ, 131, 1163
\bibitem[\protect\citeauthoryear{Somerville \&  Primack}{1999}]{somerville1999} Somerville R.S., Primack J.R., 1999, MNRAS, 310, 1087
\bibitem[\protect\citeauthoryear{Springel \& Hernquist}{2005}]{springer2005} Springel V., Hernquist L., 2005, ApJ, 622, L9
\bibitem[\protect\citeauthoryear{Steinmetz \& Navarro}{2002}]{stein2002} Steinmetz M., Navarro J.F., 2002, NewA, 7, 155 
\bibitem[\protect\citeauthoryear{Trujillo et al.}{2002}]{trujillo2002} Trujillo I., Asensio Ramos A., Rubino-Mart\'in J.A., Graham A.W., Aguerri J.A.L., Cepa J., Guti\'errez C. M., 2002, MNRAS, 333, 510
\bibitem[\protect\citeauthoryear{Tsikoudi}{1980}]{tsikoudi1980} Tsikoudi V., 1980, ApJS, 43, 365
\bibitem[\protect\citeauthoryear{Tully}{1988}]{tully1988} Tully R.B., 1988, ``The Nearly Galaxy Catalog'', Cambridge University Press
\bibitem[\protect\citeauthoryear{Turnbull, Bridges \& Carter}{1999}]{turn1999} Turnbull A.J., Bridges T.J., Carter D., 1999, MNRAS, 307, 967 
\bibitem[\protect\citeauthoryear{de Vaucouleurs}{1959}]{devauc1959} de Vaucouleurs G., 1959, in Handbuch der Physik, Volume 53, p.275
\bibitem[\protect\citeauthoryear{de Vaucouleurs, de Vaucouleurs \& Corwin}{1976}]{devauc1976} de Vaucouleurs G., de Vaucouleurs A., Corwin H.G., 1976, Second Reference Catalogue of Bright Galaxies, 1976, Austin, University of Texas press
\bibitem[\protect\citeauthoryear{de Vaucouleurs et al.}{1991}]{devauc1991} de Vaucouleurs G., de Vaucouleurs A., Corwin H.G. Jr., Buta R., Paturel G., Fouque P., 
                                    1991, Third Reference Catalogue of Bright Galaxies, New York, Springer (RC3)
\bibitem[\protect\citeauthoryear{Weil \& Hernquist}{1996}]{weil1996}  Weil M. L., Hernquist L., 1996, ApJ, 460, 101
\bibitem[\protect\citeauthoryear{Weinzirl et al.}{2009}]{weinzirl2008} Weinzirl T., Jogee S., Khockfar S., Burkert A., Kormendy J., 2009, ApJ, 696, 411
\bibitem[\protect\citeauthoryear{Yoshizawa \& Wakamatsu}{1975}]{yoshizawa1975} Yoshizawa M., Wakamatsu K., 1975, AA, 44, 363
\bibitem[\protect\citeauthoryear{Younger et al.}{2007}]{younger2007} Younger J.D., Cox T.J., Seth A.C., Hernquist L., 2007, ApJ, 670, 269
\bibitem[\protect\citeauthoryear{de Zeeuw \& Franx}{1991}]{zeeuw1991} de Zeeuw T., Franx M., 1991, ARA\&A, 29, 239
\end{thebibliography}
\end{document}